\documentclass[10pt]{iopart}
\usepackage{iopams}
\usepackage{hyperref}
\usepackage{graphicx}
\usepackage{cite}
\newcommand{\D}{{\rm d}}
\newcommand{\bea}{\begin{eqnarray}}
\newcommand{\eea}{\end{eqnarray}}
\usepackage{color}

\expandafter\let\csname equation*\endcsname=\relax
\expandafter\let\csname endequation*\endcsname=\relax
\usepackage{amsmath}

\makeatletter
\renewcommand\@appendixstar{\@@par
 \ifnumbysec 
 \@addtoreset{table}{section}
 \@addtoreset{figure}{section}\fi
 \setcounter{section}{0}
 \setcounter{subsection}{0}
 \setcounter{subsubsection}{0}
 \setcounter{equation}{0}
 \setcounter{figure}{0}
 \setcounter{table}{0}
 \def\thesection{\Alph{section}} 
 \def\theequation{\ifnumbysec
      \Alph{section}.\arabic{equation}\else
      \Alph{section}\arabic{equation}\fi}
 \def\thetable{\ifnumbysec
      \Alph{section}\arabic{table}\else
      A\arabic{table}\fi}
 \def\thefigure{\ifnumbysec
      \Alph{section}\arabic{figure}\else
      A\arabic{figure}\fi}}
\makeatother

\renewcommand{\theequation}{\arabic{section}.\arabic{equation}} 

\begin{document}

\topical{Stochastic Resetting and Applications}

\author{Martin R. Evans$^1$, Satya N. Majumdar$^2$ and Gr{\'e}gory Schehr$^2$}

\address{$^1$ SUPA, School of Physics and Astronomy, University of Edinburgh, 
Peter Guthrie Tait Road, Edinburgh EH9 3FD, UK\\
$^2$ LPTMS, CNRS, Univ. Paris-Sud, Universit\'e Paris-Saclay, 91405 Orsay, France
}
\ead{m.evans@ed.ac.uk,satya.majumdar@u-psud.fr,gregory.schehr@u-psud.fr}

\begin{abstract}
In this Topical Review we consider stochastic processes under resetting, which have attracted a lot of attention in recent years.
We begin with the simple example of a diffusive particle whose position is reset randomly in time with a  constant  rate $r$, which corresponds to Poissonian resetting, to some fixed point (e.g. its initial position). This simple  system already exhibits the  main  features of interest  induced by resetting: (i) the system reaches
a nontrivial nonequilibrium stationary state (ii) the mean time for the particle to  reach a target is finite and has a minimum, optimal, value as a function of the resetting rate $r$.
We then generalise to an arbitrary stochastic process (e.g. L\'evy flights or fractional Brownian motion) and non-Poissonian resetting (e.g. power-law waiting time distribution for intervals between resetting events).
We go on to discuss multiparticle systems as well as extended systems, such as fluctuating interfaces, under resetting.
We also consider resetting with memory which implies resetting the process to some randomly selected previous time.
Finally we give an overview of recent developments and applications in the field.
\end{abstract}

\pacs{05.40.-a, 05.70.Fh, 02.50.Ey, 64.60.-i}

\pagestyle{plain}

\maketitle

\tableofcontents

\section{Introduction}

\subsection{Search processes}

Sometime it's best just to give up and start all over again! Imagine some mundane task such as locating one's keys in the morning.  After a fruitless, haphazard search, which has taken one to areas far away from where the keys should normally be, it is perhaps best to go back to the starting point of the search and try again.  Similarly, in visual search \cite{WH04}, where one tries to locate a face in a crowd,  
the eye typically flicks back to some chosen starting point after darting around in the vicinity of this point.  In both cases one has a search process that entails a local and  to a greater or lesser  extent random  search procedure interspersed with resetting or restart events.

Generally search processes are ubiquitous in nature and  human behaviour \cite{Bell,AD68}: from the search for the  holy grail  and the Higgs boson
all the way to animals searching for food \cite{BC09,VLRS} and biomolecules searching for a binding site  such as proteins on 
DNA \cite{BWH81,CBVM04,GMKC18,Chowdhury19}.
Depending on the specific search problem there are different protocols, but what is common to these problems is
to find an optimal search strategy.
Different classes of search strategies  have been identified, see e.g. \cite{MZ02,Gelenbe,Snider12,AG13,CBR17} 
and prominent among them is {\em intermittent search} strategy wherein there is a mixture of local steps and long-range moves  \cite{BCMSV05,BMSV07,BLMV11}.
During the local step actual searching takes place whereas  during the long relocation move the searcher
moves but is not actively searching.
Such strategies have been shown to be advantageous in a variety of contexts such as
animal foraging and the target search of proteins on DNA molecules \cite{LKMK08,BKSV09}.

\subsection{From stochastic algorithms to chemical reactions}
Another example of such an intermittent search strategy is realised in computer simulations of  dynamics on complex 
(free) energy landscapes, such as in simulated annealing. Here one starts from some initial configuration and tries to locate the global minimum of the landscape.  However  at low temperature the system may  get trapped in a metastable, local minimum for a long time. To speed up the search it has been observed empirically that it helps to  halt the process and restart from the initial configuration, the rationale being that this allows the exploration of new pathways on the landscape.
More generally the advantage of restarting has been  exploited in various stochastic algorithms. The  idea is that a stochastic algorithm may get stuck before completing the intended task  and therefore it is advantageous to simply
restart the algorithm \cite{VAVA91,LSZ93,TFP08,APZ13,Lorenz18}.
Some variants of these problems have been studied in the probability \cite{JP12,APZ18} and combinatorics \cite{BW17} literature.

We also mention chemical reactions where it has been pointed out \cite{RUK14} that a complex chemical process
to produce some product is much like a complex stochastic process which may benefit from restarting.
In this case the restarting can be effected by the unbinding of an enzyme which forms the  initial
catalyst for the process.

\subsection{Nonequilibrium states}
Resetting a stochastic process is also of interest as a paradigm which  stops a system attaining an equilibrium state (as it is continually returned to its initial condition).
However the system will still attain a stationary state which will be off-equilibrium in nature, i.e. there will be probability
currents in the system which would vanish if the system were allowed to relax to thermal equilibrium \cite{KRB10}.
The resetting moves dynamically generate an effective potential which drives the system out of equilibrium.

\subsection{Catastrophes in population dynamics}

Perhaps the first instances of stochastic processes with resetting appeared  in  literature on birth-death processes in which there is an absorbing state (for example, when the population size reaches zero). In the case where the population 
is always absorbed, a restart  process to a finite population size generates interesting stationary properties
\cite{Levikson77,Pakes78,Pakes97}. On the other hand if the population tends  to 
increase exponentially in time,  resetting to a finite initial population, from which there is a finite probability of absorption, renders the mean time to absorption finite \cite{BGR82,Brockwell85}. More generally one can consider the effect of catastrophes
with a distribution of sizes on  growing populations and study  various  stationary properties \cite{Kyriakidis94,EF03,VAME10,DDCGN15}.
The same applies to queueing systems where  catastrophes reset the length of the queue to zero \cite{KA00,DCGNR03,KRMSKL05}.

\subsection{Purpose of this review}
In recent years there has been a surge in the study of stochastic processes subject to resetting
(for general formulations see for example \cite{MZ99,EM1,EM2,MV13,EM14,KMSS14,CM15,CS15,MSS15a,MSS15b,MV16,MC16,EM16,PKE16,NG16,Reuveni16,RLSTG16,PR17,CS18,EM18,VM18,GGC18,MPCM19a,MPCM19b,MPCM19c,Gupta19,LD19,MM19}---we refer the reader to \cite{MMV17} for an historical perspective).
This is a very general problem as resetting to the initial condition can be applied to any stochastic process.
The purpose of this review is to describe these developments in a pedagogical manner focussing on simple models and
the derivation of quantitative analytical results.

We will begin by considering a single diffusing particle  subject to reset in one or higher dimensions \cite{EM1,EM2,EM14}.
In this example we first  show how a nontrivial nonequilibrium stationary state emerges. Then by introducing a  target
for the diffusing particle to search for, we show how the mean time for the particle to locate the target (the mean first passage time or mean time to absorption of the target) is minimised for an optimal choice of the resetting rate. These features turn out to be very general and hold for various 
other stochastic processes, which may correspond to extended, many-particle systems.
We shall also explore various reset protocols, beginning with the simplest one which is Poissonian resetting (with a constant rate) to a fixed initial configuration. We then generalise to non-Poissonian resetting and resetting which uses memory of the past history.
We also give an overview of recent extensions of the subject in various directions.

\section{Single particle process}
\setcounter{equation}{0}

\subsection{Diffusion with Poissonian resetting}
\label{sec:1ddiff}
First let us define diffusion with Poissonian resetting in one space dimension.
We consider a single particle on the real line 
with initial position $ x_0$ at $t=0$ and resetting with rate $r$ to position  $X_r$.
We stress here that the initial position  $x_0$ and resetting position
$ X_r$ are in general distinct, although at the end of some calculations it is convenient to set them to be equal.

\begin{figure}[ht]
  \begin{center}
\includegraphics[width = 0.6\linewidth]{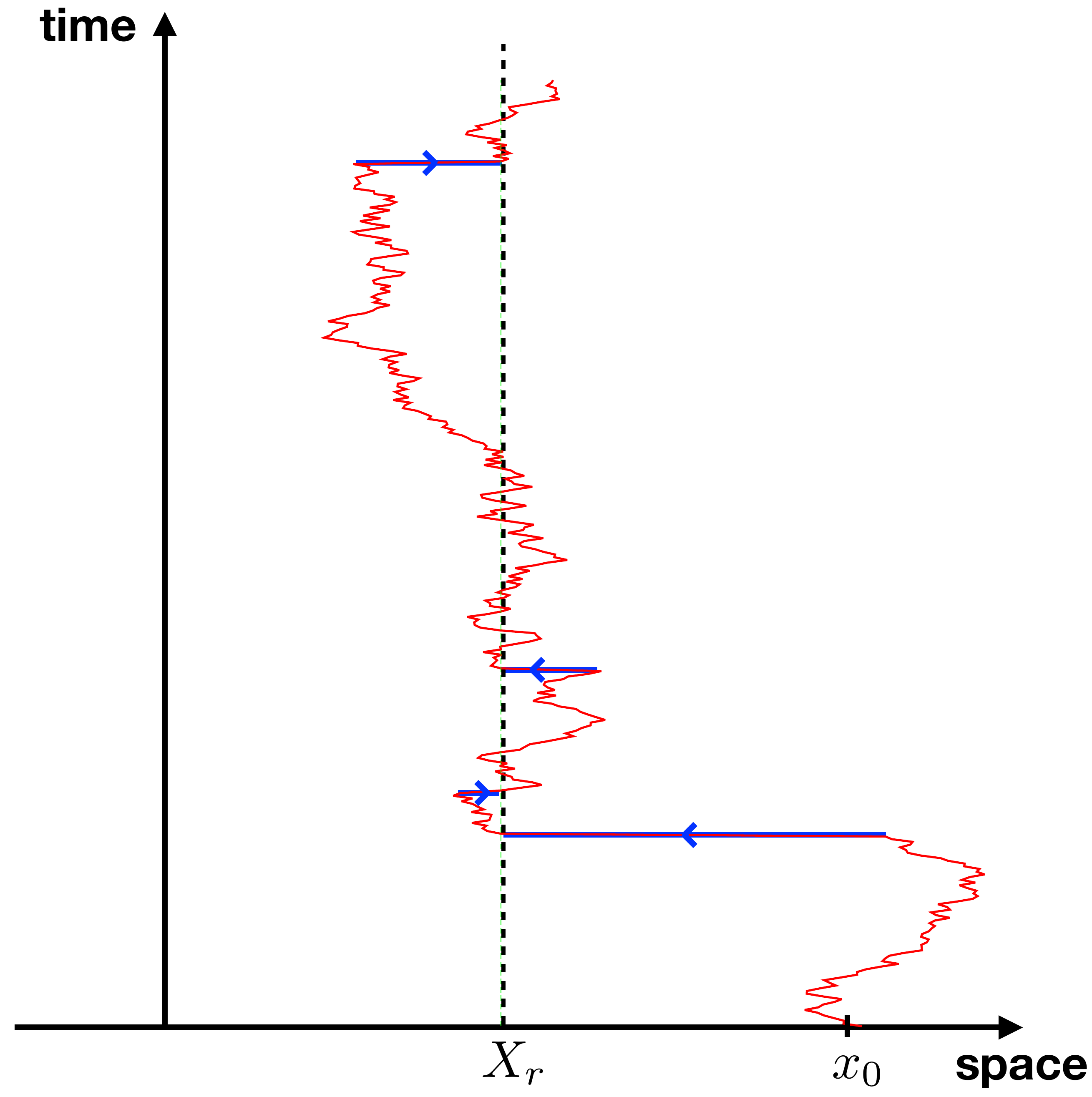}
  \end{center}
  \caption{Illustration in $d=1$ of the diffusion with resetting process: the particle starts at initial position $ x_0$ and resets to  position $ X_r$ with rate $r$.}
    \label{fig1}
  \end{figure}

The position $x(t)$ of the particle at time
$t$ is updated by the following stochastic
rule~\cite{EM1}: in a small time interval $\D t$ 
the position $x(t)$ is updated to
\begin{eqnarray}
 x(t+\D t) & =& X_r \quad {\rm with\,\,probability}\,\, r\, \D t \nonumber \\
& =&  x(t) + \xi(t) (\D t)^{1/2} \quad {\rm with\,\,probability}\,\, (1-r\, \D t)
\label{rule.1}
\end{eqnarray}
where $\xi(t)$ is a Gaussian  random variable with mean  zero and two-time correlator given by
 \begin{eqnarray}
\langle \xi(t)\rangle &=&0 \label{mean}\\
\langle \xi(t)\xi(t')\rangle &=& 2\,D\, \delta(t-t')\;. \label{var}
\end{eqnarray}
The dynamics thus  consists of a stochastic mixture of resetting to
the initial position with rate $r$ (long range move) and ordinary diffusion 
(local  move) with diffusion constant $D$ (see Fig. \ref{fig1}). 

The probability density for the particle to be at position $x$ at time $t$, having started from position $ x_0$ at time $t=0$ with resetting to position $X_r$, should, in principle,  be written as 
$p(x,t|x_0; X_r)$. However in the following, when the context is sufficiently clear, we shall frequently use $p(x,t|x_0)$ (omitting the dependence on $X_r$) or simply $p(x,t)$ (omitting the dependence on both $x_0$ and $X_r$). 

The forward master equation for the probability density for diffusion with resetting rate $r$ to point $X_r$ is 
easily obtained from the update (\ref{rule.1}): averaging over events in  time $t$ to $t+ \D t$ we obtain
\begin{equation}
p(x,t+\D t) =  r \D t \, \delta(x-X_r) + (1-r \D t) \int_{-\infty}^{\infty} {\cal D} \xi\, p(x- \xi (\D t)^{1/2},t) \;,
\end{equation}
where $\int_{-\infty}^{\infty} {\cal D} \xi$ denotes an integral over random variables $\xi$ with a Gaussian distribution.
Expanding in $\D t$ yields
\begin{eqnarray*}
p(x,t+\D t) &=&  r \D t \, \delta(x-X_r) +  (1-r \D t) \int_{-\infty}^{\infty} {\cal D} \xi \nonumber \\
&&\times \left[
p(x,t) -  (\D t)^{1/2}\xi \frac{\partial p(x,t)}{\partial x} +  \D t \frac{\xi^2}{2} \frac{\partial^2 p(x,t)}{\partial x^2} + \ldots \right] .
\end{eqnarray*}
Performing the integrals using (\ref{mean},\ref{var})
and taking the  limit $\D t \to 0$ we obtain 
\begin{equation}
\frac{\partial p(x,t)}{\partial t}
= D \frac{\partial^2 p(x,t)}{\partial x^2} - r p( x,t) + r\delta(x- X_r)\;,
\label{fme}
\end{equation}
with initial condition $p( x,0) = \delta(x- x_0)$. The first term on the right hand side (r.h.s.) of (\ref{fme}) expresses the diffusive spread of probability;
the second term expresses the loss of probability from $x$ due to resetting to $X_r$; the final term corresponds to the gain of probability at $X_r$ due to resetting from all other positions.
We shall refer to (\ref{fme}) as the forward master equation.

In an analogous way one can derive the backward master equation in which the initial position $x_0$ is the variable,
 i.e. averaging over events in the interval $[0,\D t]$ yields
\begin{equation}
p(x,t+\D t|x_0) =  r \D t \, p(x,t | X_r) + (1-r \D t) \int_{-\infty}^{\infty} {\cal D} \xi\, p(x,t| x_0 + \xi (\D t)^{1/2}) \;,
\end{equation}
from which one obtains
\begin{equation}
\frac{\partial p(x,t| x_0)}{\partial t}
= D \frac{\partial^2 p(x,t|x_0)}{\partial x_0^2} - r p( x,t|x_0)+ r p( x,t|X_r)\;.
\label{bme}
\end{equation}
Note that the gain term from resetting (i.e. the final term on r.h.s.) now involves  the probability density
of reaching $x$ at  time $t$ having started from the resetting position~$X_r$.

\subsection{Renewal equation approaches}
\label{sec:renewal}
Instead of beginning from  these master equations one can write down   renewal equations (which  indeed  give the solution to (\ref{fme}), (\ref{bme})) in a simple and intuitive way as follows.

\begin{figure}
\centering
\includegraphics[width=0.6\linewidth]{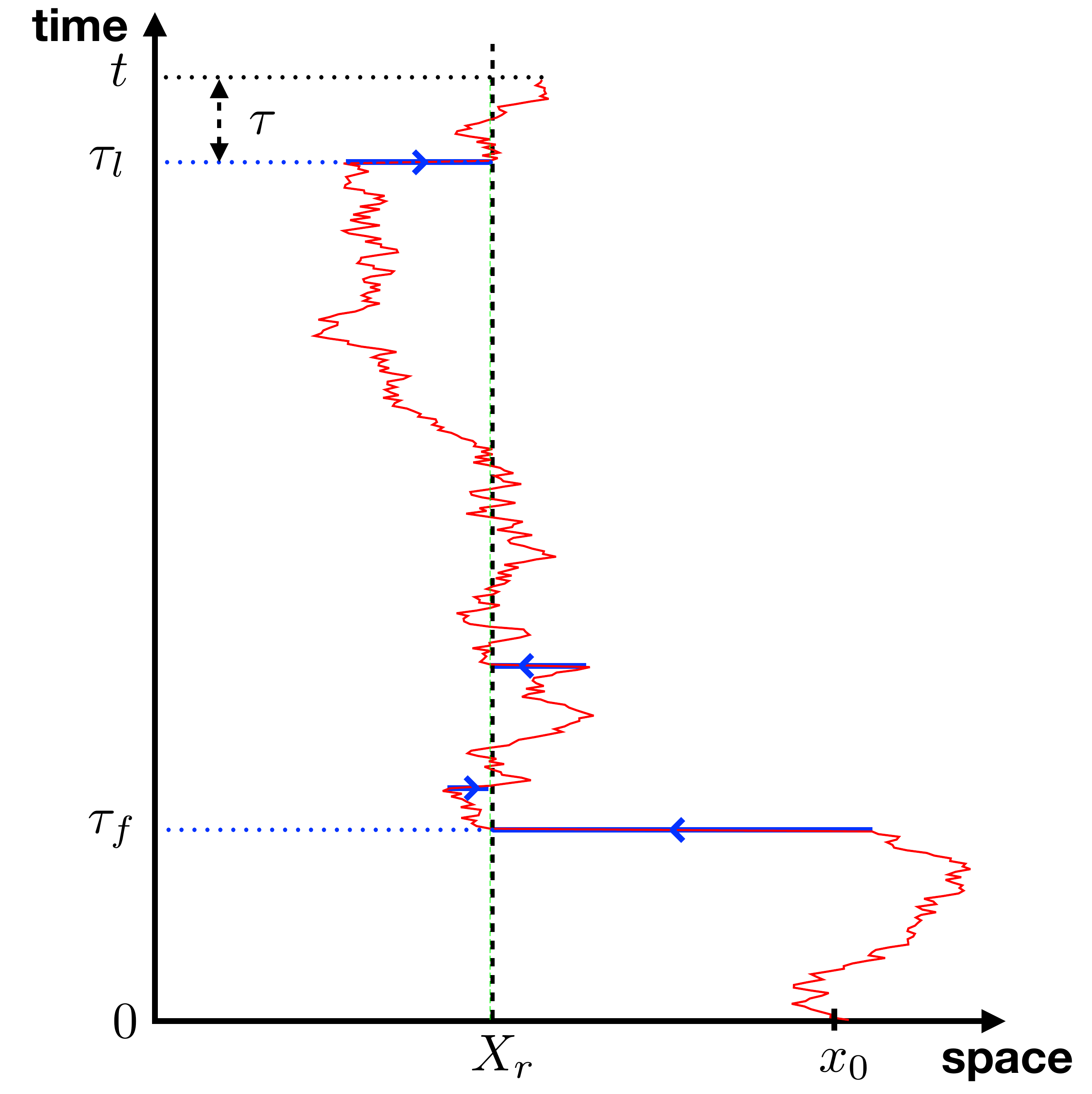}
\caption{Same trajectory as in figure \ref{fig1} of $1d$-diffusion with resetting: here $\tau_l=t-\tau$ and $\tau_f$ denote respectively the time at which the last and first resetting events happen.}\label{Fig_renewal}
\end{figure}

We first note that in the absence of resetting ($r=0$), the  diffusive Green function (also known as the propagator for the diffusion equation) which we denote  $G_0(x,t|x_0)$, satisfies
\begin{equation}
\frac{\partial G_0(x,t|x_0)}{\partial t} = D\frac{\partial^2 G_0(
x,t|x_0)}{\partial x^2}\;,
\end{equation}
with initial condition
$G_0(x,t=0|x_0) = \delta( x - x_0)$, and is given by the familiar Gaussian expression
\begin{equation}
G_0( x,t| x_0) = \frac{ 1}{\left(4 \pi D t\right)^{1/2}} \exp
\left[-\frac{| x-x_0|^2}{4 D t}\right]\;.
\label{Gdef}
\end{equation}
The probability density in the presence of resetting, $p( x,t| x_0)$, is a sum over two
contributions:  one which comes from trajectories where no resetting events
have occurred in time $t$ and a second contribution which comes from  summing
over trajectories where the {\em last} resetting event occurred at time $\tau_l=t-\tau$ (see figure \ref{Fig_renewal}).
For Poissonian resetting (with constant rate $r$), the probability of no resetting events having occurred up to time $t$  is ${\rm e}^{-rt}$
and the probability  density of the last resetting event having occurred
at $\tau_l= t-\tau$ (and no resetting events since) is $r {\rm e}^{-r\tau}$.
Thus the full time-dependent solution to (\ref{fme}) can be written down 
as 
\begin{equation}
p( x,t|x_0) = {\rm e}^{-rt}  G_0( x,t| x_0) + r \int_0^t \D \tau \,
{\rm e}^{-r\tau}  G_0( x,\tau| X_r)\;.
\label{pt}
\end{equation}
We refer to this equation as a  \textit{last renewal} equation as  it involves the time of the last reset
$\tau_l = t-\tau$. Note that this renewal equation holds for  more general stochastic processes, with  propagator  denoted by
$G_0( x,\tau| x_0)$, which can be different from the diffusive case we have considered so far.

We will also consider \textit{first renewal} equations
where instead of the last resetting, we consider the \textit{first
resetting} at time $\tau_f$ having started
from $t=0$ (see figure \ref{Fig_renewal}).
Subsequently, the particle diffuses from $\tau_f$ until time $t$,  under resetting. It is again
straightforward to write down an equation for the probability:
\begin{eqnarray}
p(x,t|x_0)&=&\e^{-rt}G_0(x,t|x_0) \nonumber \\
&+&r\int_{0}^{t}\D\tau_f~\e^{-r\tau_f}~p(x,t-\tau_f|X_r)~,
\label{FR}
\end{eqnarray}
where the 
second term now integrates over trajectories in which  there
has been a first reset to $X_r$ between time $\tau_f$ and $\tau_f+\D\tau_f$
and then there can be multiple resets in the remaining time $t-\tau_f$
which is  why $p(x,t-\tau_f|X_r)$ now  appears inside the integral. 

The equivalence between (\ref{pt}) and
(\ref{FR}) may be shown
by taking  Laplace transforms
of both equations (see Appendix  \ref{app:renewal}). 
For the time being we note that the Laplace transform of the solution to (\ref{pt}) is given by
\begin{equation}
\tilde p( x,s|x_0) =   \tilde G_0( x,r+s| x_0)  + \frac{r}{s}\tilde G_0( x,r+s| X_r) \;,
\label{LRlt}
\end{equation}
where 
\begin{equation}
\tilde{p}(x,s|x_0) = \int_0^\infty \D t\, \e^{-st}\, p(x,t|x_0)
\end{equation}
is the Laplace transform of $p(x,t|x_0)$
and similarly $\tilde{G}_0(x,s|x_0)$ is the Laplace transform of $G_0(x,t|x_0)$.

\subsection{Nonequilibrium stationary state}
The stationary state is attained as  $t\to \infty$ where (\ref{pt}) tends to
the stationary distribution
\begin{eqnarray}
p^*(x) =  r \int_0^\infty \D \tau\,  {\rm e}^{-r\tau }  G_0( x,\tau| X_r)\;,
\label{pss}
\end{eqnarray}
thus the stationary distribution under resetting is related to the Laplace transform
(with Laplace variable $r$) of the propagator in the absence of resetting. This
is actually a generic property, valid for more general processes with a propagator $G_0(x,\tau|x_0)$, when resetting is Poissonian.

In order to evaluate the integral (\ref{pss}) in the case of the diffusive propagator (\ref{Gdef})
we  use the identity (Equation 3.471.9 of \cite{GR})
\begin{equation}
\int_0^\infty \D t\, t^{\nu-1} {\rm e}^{-\frac{\beta}{t} -\gamma t}
= 2\left( \frac{\beta}{\gamma}\right)^{\nu/2}
K_\nu(2 \sqrt{\beta \gamma})\;,
\label{magint}
\end{equation}
where $K_\nu$ is the  modified Bessel function of the second kind  of order $\nu$. The relevant case of this identity
 here is $\nu =1/2$ and using the definition
\begin{equation}
K_{1/2} (y) = \left( \frac{\pi}{2y}  \right)^{1/2} {\rm e}^{-y}\;,
\label{Khalf}
\end{equation}
equation (\ref{magint}) becomes 
\begin{equation}
\int_0^\infty \D t\, t^{-1/2} {\rm e}^{-\frac{\beta}{t} -\gamma t}
= \left( \frac{\pi}{\gamma}\right)^{1/2}
{\rm e}^{-2 (\beta \gamma)^{1/2}}\;.
\label{magint2}
\end{equation}
Then one  obtains from (\ref{Gdef}) (with $x_0 = X_r$) and (\ref{pss}),
\begin{equation}
p^*(x) = \frac{\alpha_0}{2}
{\rm e}^{ -\alpha_0 |x-X_r|}
\label{pssres}
\end{equation}
where
\begin{equation}
\alpha_0 =  \left( \frac{r}{D}\right)^{1/2}\;.
\label{alpha0}
\end{equation}
Of course, we can check directly that (\ref{pssres}) satisfies (\ref{fme}) with the left hand side \textcolor{red}{(l.h.s.)} set to zero  by using the identity
\begin{equation}
\frac{\D^2}{\D x^2}  {\rm e}^{- \alpha_0 |x-X_r|}
= \alpha_0^2   {\rm e}^{- \alpha_0 |x-X_r|}  - 2\alpha_0\delta(x-X_r)\;.
\end{equation}

\begin{figure}[ht]
  \begin{center}
   \includegraphics[width=0.8\textwidth]{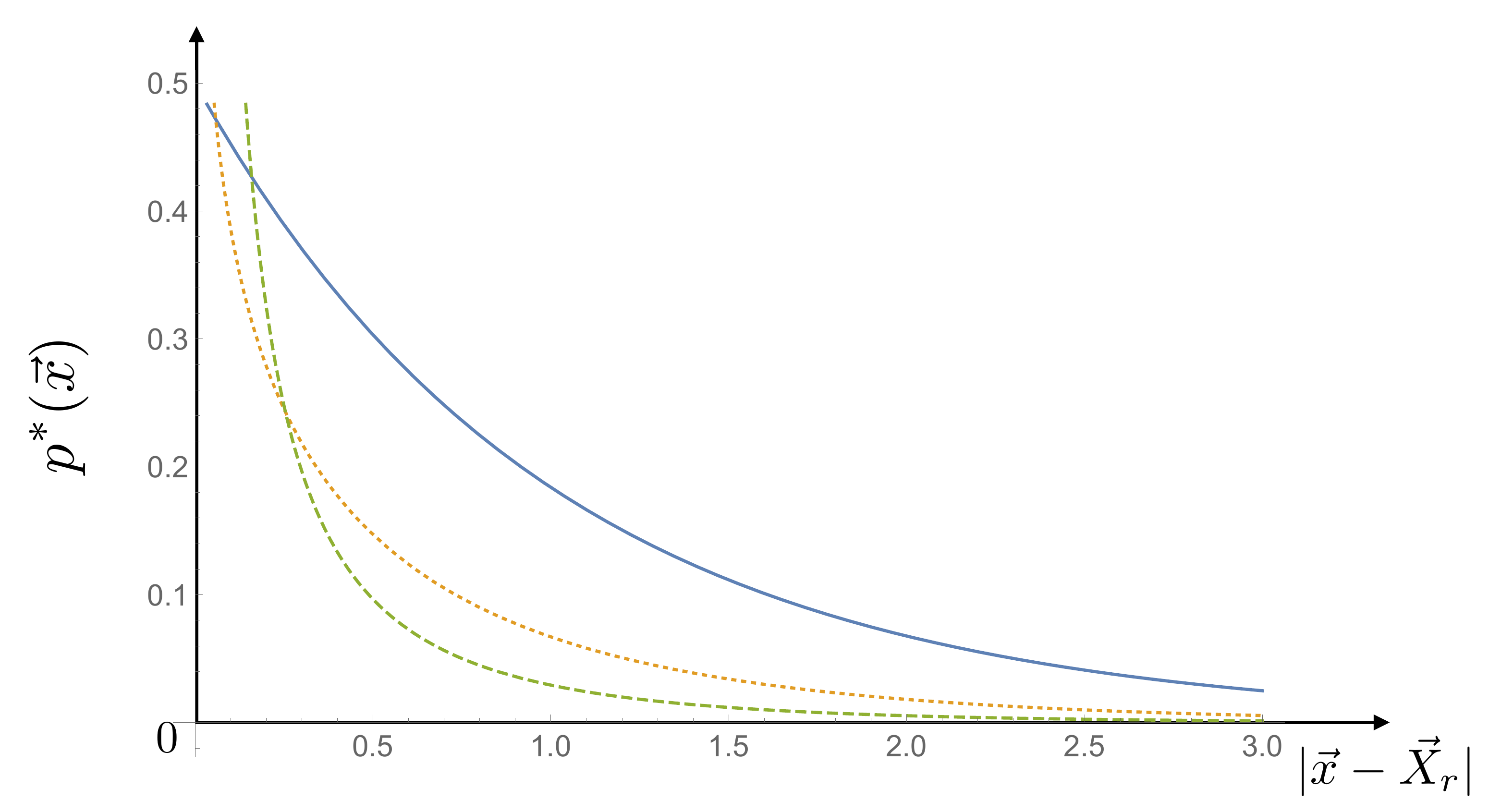}
  \end{center}
  \caption{The stationary probability densities $p^*(\vec x)$ given by (\ref{pssgen}) for the case $\alpha_0 =1$ and $\vec X_r=0$:
$d=1$ full lines; $d=2$ dotted lines; $d=3$ dashed lines.
}
    \label{fig:ss}
  \end{figure}

The first thing to note is that the stationary distribution in the presence of resetting (\ref{pssres}) exhibits exponential decay away from the resetting position $X_r$ in both the $x-X_r>0$ and $x-X_r< 0$ directions. The double exponential decay is known as a Laplace distribution.
Thus the distribution is localised around $X_r$ over a length $1/\alpha_0$ and 
there is  a cusp singularity at $ x=  X_r$ (see figure~\ref{fig:ss}).

Also note that  (\ref{pssres}) is a nonequilibrium stationary state (NESS)
by which it is meant that there is circulation of probability, in contrast to an equilibrium state where detailed balance holds
and probability currents vanish. This is because resetting  
implies a source of probability at $X_r$ while probability 
is lost through resetting from all other values of $ x \neq  X_r$.

\subsection{Diffusion with resetting in potentials}
As an illustration of the utility of (\ref{pss}) one can consider a diffusive particle with a constant drift  $\mu$
in the positive $x$ direction under Poissonian resetting with rate $r$. This corresponds to an unbounded linear potential.
For this case the  Green function in the absence of resetting is
\begin{equation}
G_0( x,t| x_0) = \frac{ 1}{\left(4 \pi D t\right)^{1/2}} \e^{-\frac{| x-x_0- \mu t|^2}{4 D t}}\;,
\label{Gmudef}
\end{equation}
and one finds the stationary state under resetting, using (\ref{pss},\ref{magint2}), 
to be 
\begin{equation}
p^*(x) =  \frac{r}{(4Dr + \mu^2)^{1/2}} 
{\rm e}^{ \frac{(x-X_r) \mu}{2 D}-\frac{(4Dr + \mu^2)^{1/2}}{D}|x-X_r|}\;.
\end{equation}
Here the stationary distribution is  asymmetric about the resetting position $X_r$ with different
exponential decays in the downstream ($x>X_r$) and upstream ($x<X_r$) directions.
This case has  been studied in detail in \cite{RMR19} and  the P\'eclet number ${\rm Pe} = X_r \mu/(2 D)$ identified 
as a key governing dimensionless variable.

We also note that (\ref{pss}) allows one to write down the stationary state under  Poissonian resetting
for various confining potentials (linear or quadratic) or unstable potentials \cite{Pal15}. One just requires the knowledge of $G_0$,
the propagator in each case.

\subsection{Relaxation to stationary state for diffusion with resetting}
\label{sec:relss}

In addition to knowing the stationary state, it is also important to understand how the system relaxes to this  state. 
In order to investigate this relaxation, we start with the exact solution in equation~(\ref{pt}), valid at all time $t$, and analyse it for
large but finite $t$ \cite{MSS15a}. For simplicity, we will set $X_r=x_0$, i.e. we reset the particle to its initial position.
It is further 
convenient to rescale the time $\tau = w\, t$ and rewrite (\ref{pt}) as
\begin{equation}
p( x,t) =  \frac{{\rm e}^{-t \Phi(1,(x-X_r)/t)}}{\sqrt{4 \pi D t}} + \frac{r t^{1/2}}{\sqrt{4 \pi D}}
\int_0^1 \frac{\D w}{w^{1/2}} \,
{\rm e}^{-t \Phi(w, (x-X_r)/t)}
\label{ptw}
\end{equation}
where we have defined
\begin{equation}
\Phi(w, y) = rw + \frac{y^2}{4D w}\;.
\end{equation}
For large $t$ the integral in the second term in (\ref{ptw}) can be analysed by the saddle-point method. We keep 
$y= (x-X_r)/t$ fixed and take the $t \to \infty$ limit. The saddle point of this integral, if it exists, occurs at 
\begin{equation}
w^* = \frac{ |y|}{\sqrt{4Dr}}\;,
\end{equation}
which minimises the function $\Phi(w,y)$, for fixed $y$. If $w^*<1$, the saddle point occurs
within the integration limits $w\in [0,1]$ and one gets,
from \eref{ptw} $p(x,t) \sim \e^{-t\, \Phi(w^*, (x-X_r)/t)}$ for large
$t$, where $\Phi(w^*,y)= \alpha_0\, |y|$ where  $\alpha_0$ is given by (\ref{alpha0}).  In contrast, for $w^*>1$, the
function $\Phi(w,y)$ has its lowest value in $w\in [0,1]$ at
$w=1$. Hence the integrand in the second term is dominated by the
regime at $w=1$ (and is of the same order as the first
term). Physically, this corresponds to trajectories which have
undergone zero (or almost zero) resettings up to time $t$.  One then
gets $p(x,t) \sim \e^{-t\, \Phi(1,(x-X_r)/t)}$, with $\Phi(1,y)=
r+y^2/{(4D)}$. Summarising, we obtain
\begin{subequations}
\label{LD-diffusion}
\begin{equation}
p(x,t)\sim \e^{-t I \left((x-X_r)/t\right)},
\label{LD0-diffusion}
\end{equation}
where the function $I(y)$ is called the rate function or the large deviation function (LDF). In this case, it is given by
\begin{equation}
I(y) = \begin{cases}\\[-2.2ex]\displaystyle
\alpha\, |y| & \text{for}~~|y| < y^*, \\[2mm]
\displaystyle
r + \frac{y^2}{4D} & \text{for}~~ |y| > y^*,\\[3mm]
\end{cases}
\label{I-diffusion}
\end{equation}
\end{subequations}
with $y^*=\sqrt{4Dr}$.

\begin{figure}
\centering
\includegraphics[width = 0.9\linewidth]{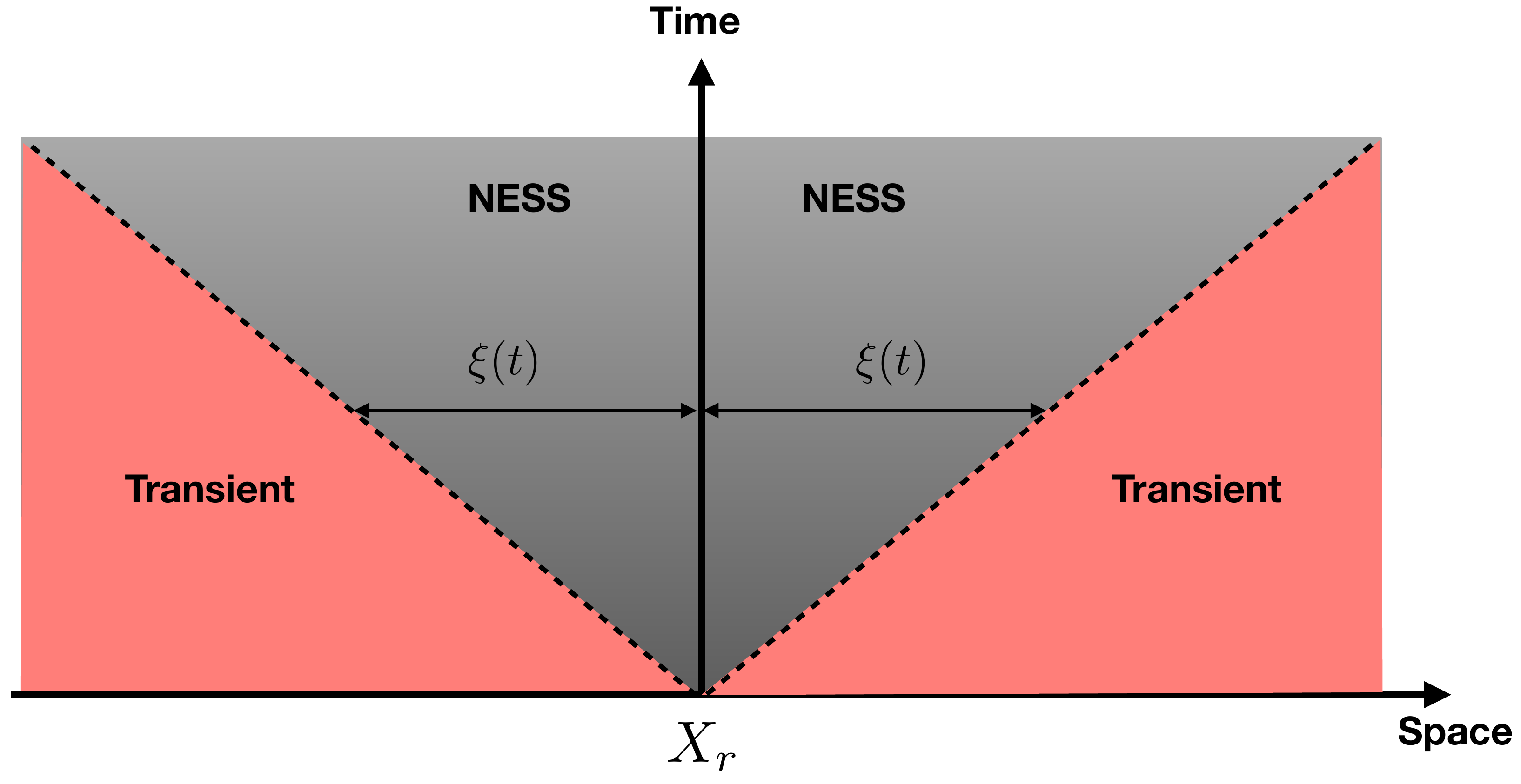}
\caption{A NESS  gets established in a core region around the resetting center $X_r$ whose frontiers $\xi(t)$ grow with time as $\xi(t) \propto t$. Outside the core region, the system is transient.}\label{Fig_relax}
\end{figure}

The appearance of the factor $(x-X_r)/t$ as the argument of the rate function $I$ in 
\eref{LD0-diffusion} indicates that there is a 
growing length scale $\xi(t) \sim t$, much larger than the typical
diffusion length scale $\sim \sqrt{t}$.  The linearity of the LDF for
$|y| < y^*$ implies that, for any large but finite $t$, there is an
interior spatial region $-y^*t < x-X_r < y^*t$, where the NESS has been
achieved, since  $p^*(x,t) \sim \exp(-\alpha |x-X_r|)$ becomes independent of
$t$, in agreement with \eref{pssres}. However, there is still an
exterior region $|x-X_r| > y^*\, t$ that has not yet relaxed to the NESS (see figure \ref{Fig_relax}). 
The boundaries between the two regions move at a constant speed
$y^*$.  From
\eref{I-diffusion}, it is easy to check that while $I(y)$ 
and its first derivative are both continuous at $y=\pm y^*$, its
second derivative has a discontinuity at $y=\pm y^*$.  This signifies a second order
dynamical phase transition.

What is the physical significance of this phase transition? The
probability density $p(x,t)$ can also be interpreted as the density
at time $t$ of a swarm of independent Brownian motions, each subjected
to stochastic resetting with rate $r$, all starting from the origin at
$t=0$. Our calculation shows that at time $t$ the density for
$|x|<y^* \, t$ becomes stationary, while is still time dependent for
$|x|>y^* \, t$. From the analysis above, it is clear that, for
$|x|>y^* \, t$, the density is typically of the form $\sim \e^{-r\,t}
G_0(x,t)$ in \eref{pt}, i.e., it corresponds to particles
that have undergone almost no resetting up to time $t$.  This is of
course a very rare event and these particles in the outer region thus
have very atypical trajectories. In contrast, the particles in the
inner core region correspond to typical trajectories that have
undergone a large number of resettings---leading to a stationary
behaviour in this regime. The LDF $I(y)$ in (\ref{LD0-diffusion})
probes precisely the separation between these two regions, i.e.,
between the typical and the atypical trajectories; the singularity in
the LDF signifies a sharp separation between these two types of
particles. 

In any typical application of resetting, for instance in
the optimisation of search algorithms, we would ideally like to keep,
at any given finite time $t$, only the typical trajectories and not
the atypical ones---since the latter ones do not feel the resetting
at all. The LDF $I(y)$ and its associated singularity, that sharply
separates the two types of trajectories, thus provides a very useful
and practical way to select the typical ones at any given time
$t$. Even though we discuss it here in the context of a single
particle diffusion, it turns out that this physical picture associated with 
the second order dynamical phase transition is quite generic \cite{MSS15a}
and holds for arbitrary stochastic processes undergoing resetting and
even for spatially extended systems, such as fluctuating interfaces \cite{GMS14}
that we discuss later.

\subsection{Diffusion with resetting in arbitrary spatial dimension}
It is straightforward to generalise the formalism of Sections~\ref{sec:1ddiff}--\ref{sec:relss}  to diffusion with resetting in arbitrary spatial dimension \cite{EM14}. The particle  now moves
in $\mathbb{ R}^d$ with
initial position $\vec x_0$ at $t=0$ and resetting to position $\vec X_r$.
In a small time interval $\D t$ each component $x_i$ of
the position vector $\vec x(t)$ becomes 
\begin{eqnarray}
 x_i(t+\D t) & =& ( X_r)_i \quad {\rm with\,\,probability}\,\, r\, \D t \nonumber \\
& =&  x_i(t) + \xi_i(t) (\D t)^{1/2} \quad {\rm with\,\,probability}\,\, (1-r\, \D t)
\label{rule.d}
\end{eqnarray}
where $\xi_i(t)$ is a Gaussian random variable with mean $\langle \xi_i(t)\rangle =0$
and the two-time correlator $\langle \xi_i(t)\xi_j(t')\rangle= 2\,D\,\delta_{ij} \delta(t-t')$.
The forward master equation for the probability density for diffusion with resetting rate $r$ to point $\vec X_r$ now  reads
\begin{equation}
\frac{\partial p(\vec{x},t)}{\partial t}
= D\nabla^2p(\vec x,t) - r p(\vec x,t) + r\delta^d(\vec x-\vec X_r)\;,
\label{fmed}
\end{equation}
with initial condition $p(\vec x,0) = \delta^d(\vec x-\vec x_0)$,
where $\delta^d(\vec x-\vec x_0)$ is the $d$-dimensional Dirac delta function centred on $\vec x_0$.

As before, we  can write down a last renewal equation, which is the solution to (\ref{fmed}), 
as 
\begin{equation}
p(\vec x,t) = {\rm e}^{-rt}  G_0(\vec x,t|\vec x_0) + r \int_0^t \D \tau \,
{\rm e}^{-r\tau}  G_0(\vec x,\tau|\vec X_r)\;,
\label{ptd}
\end{equation}
where the $d$-dimensional diffusive propagator is now
\begin{equation}
G_0(\vec x,t|\vec x_0) = \frac{ 1}{\left(4 \pi D t\right)^{d/2}} \exp
\left[-\frac{|\vec x-\vec x_0|^2}{4 D t}\right]\;.
\label{Gdefd}
\end{equation}
The stationary distribution for the resetting problem is again related to the 
Laplace transform of the  propagator in the absence of resetting
\begin{eqnarray}
p^*(\vec x) =  r \int_0^\infty \D \tau\,  {\rm e}^{-r\tau }  G_0(\vec x,\tau|\vec X_r)\;.
\label{pssd}
\end{eqnarray}
The integral in (\ref{pssd})
may be evaluated using (\ref{magint})
where the relevant case  is now
\begin{equation}
\nu=1-d/2\;,
\end{equation}
and  one obtains from (\ref{Gdefd}) and (\ref{pssd}) 
\begin{equation}
p^*(\vec x) = \left( \frac{\alpha_0^2}{2\pi}\right)^{1-\nu} 
(\alpha_0 |\vec x- \vec X_r| )^\nu 
K_\nu(\alpha_0|\vec x- \vec X_r| )\;,
\label{pssgen}
\end{equation}
where $\alpha_0$ is, as before, given by (\ref{alpha0}).

Expression (\ref{pssgen}) holds for arbitrary $d$ and  one can continue it to noninteger $d$.
Of course the cases of integer $d$ are of special interest (see figure \ref{fig:ss} for a plot). For $d=1$, one recovers the result
given before in (\ref{pssres}), which has a cusp singularity at the resetting point  $x=X_r$. We note that for $d=2$ the singularity
at $X_r$ becomes logarithmic.  For $d=3$ one can use the identity
$K_{-1/2} (y)=K_{1/2} (y) =\left( \frac{\pi}{2y}  \right)^{1/2} {\rm e}^{-y}$ to find a simple form  
\begin{equation}
p^*(\vec x) = \frac{\alpha^2_0}{4 \pi |\vec x-\vec X_r|} \exp( - \alpha_0 |\vec x-\vec X_r|)\;.
\label{pss3d}
\end{equation}
In general, using the asymptotic behaviour $K_\nu(r) \sim r^{-|\nu|}$ as $r\to 0$, one finds that near the resetting position $\vec X_r$, 
the stationary PDF behaves as 
\begin{eqnarray} \label{singular}
p^*(\vec x) \sim 
\begin{cases}
& O(1) \;, \; \hspace*{1.5cm} d < 2 \\
& -\ln (|\vec x-\vec X_r|) \;, \; d = 2 \\
& |\vec x-\vec X_r|^{-(d-2)} \;, \; d > 2 \;.
\end{cases}
\end{eqnarray}
Thus, in $d\geq 2$ the stationary PDF $p^*(\vec x)$ diverges at the resetting position $\vec X_r$ and the divergence gets
stronger as the dimension increases.  Note that, despie the singularity at the resetting point $X_r$,  $p^*(\vec x)$ remains integrable (and normalisable to unity) because in the
integral $\int {\rm d} {\vec x}\, p^*(\vec x)$, after making the change of variable $\vec x' = \vec x - \vec X_r$, the divergence of $p^*(\vec x-  \vec X_r)$
at the origin gets compensated by the volume factor $\propto |\vec x'|^{d-1}$.

\subsection{Resetting distribution and spatially dependent resetting}
\label{sec:resetdist}
We now consider some simple generalisations of the resetting dynamics.
First let us consider  resetting to a distribution of sites rather than to a single preordained site.
We define a {\em resetting
 distribution} ${\cal P}_r (X_r)$ for the resetting process such that
the process is reset to $X_r+ \D X_r$ with probability ${\cal P}(X_r) \D X_r$.
Then the renewal equation for the probability distribution of the process (\ref{fme}) becomes
\begin{equation}
p( x,t|x_0) = {\rm e}^{-rt}  G_0( x,t| x_0) + r \int_0^t \D \tau \,
{\rm e}^{-r\tau}  \int \D X_r {\cal P}_r(X_r) G_0( x,\tau|X_r)\;.
\label{ptrd}
\end{equation}
In the long time limit we find that the stationary distribution is given by 
\begin{equation}
p^*( x) =   \int \D X_r \, {\cal P}_r(X_r)  \,p^*(x|X_r)\;,
\end{equation}
where  here $p^*(x|X_r)$ is the stationary distribution with reset to fixed position $X_r$. This equation is intuitively 
obvious: the stationary state is just that of resetting to a fixed position, averaged over the resetting position
distribution. For the case of a finite number $N$ of resetting positions
$X_{r_i}$ $i=1,\ldots,N$,
 each chosen at a resetting event  with probability ${\cal P}_i$, one has
\begin{equation}
p^*( x) =  \sum_{i=1}^n {\cal P}_i  \,p^*(x|X_{r_i})\;.
\end{equation}

We now turn to a space-dependent resetting rate $r(x)$:  the particle at position $x$ at time $t$ is reset in time $t$ to $t+ \D t$ with probability $r(x)\D t$. In this case, the simplest 
is to use the forward master equation, i.e. a generalization of (\ref{bme}), which
for the case of a one-dimensional  diffusive process reads
\begin{equation}
\frac{\partial p(x,t)}{\partial t}
= D \frac{\partial^2 p(x,t)}{\partial x^2} - r(x) p(x,t) +\int \D x' r(x')p(x',t)\delta(x-X_r)\;.
\label{fme2}
\end{equation}
Although it appears difficult to solve this equation generally,  a specific case of  resetting outside of a  window
($r(x) = 0$ for  $|x|< a$ and $r(x) =r$ for $|x|> a$) has been studied in \cite{EM2}.
Also in \cite{Pinsky19} the case of $r(x)$ decaying with $x$ has been considered and the conditions for which a 
stationary state exists have been derived.
A general path integral approach to the space-dependent resetting problem has been developed in \cite{RG17b}.

\subsection{Non-Poissonian resetting}
\label{sec:nP}
So far we have considered resetting to occur at constant rate $r$ which we refer to as Poissonian resetting.
More generally one can define the resetting process through the waiting time distribution $\psi(t)$ between
resetting events \cite{EM16,PKE16,NG16,Shkilev17}, i.e. after a reset the next reset occurs in time interval $(t,  t + \D t]$ with probability $\psi(t) \D t$. 
The probability, $\Psi(t)$, of no resets up to time $t$ is given by 
\begin{equation}
\Psi(t) = \int_t^\infty \D t' \psi(t') = 1 - \int_0^t \D t' \psi(t')\;.
\label{Psidef}
\end{equation}

For Poissonian resetting (constant  $r$) one obtains as before $\psi(t) = r {\rm e}^{-rt}$ and
$\Psi(t) = {\rm e}^{-rt}$ .
One realisation of non-Poissonian resetting  is to have a time-dependent resetting rate, $r(t),$ then $\psi(t) = r(t) {\rm e}^{-R(t)}$
where $R(t) = \int_0^t r(t') \D t'$ and
$\Psi(t) = {\rm e}^{-R(t)}$ \cite{PKE16}. A time-dependent rate is often referred to as a {\em time-inhomogeneous} Poisson Process. However we stress that here the resetting rate is itself reset, so that $r(t)$ depends on the time $t$  
since the last reset rather than absolute time from the initial condition.
The latter scenario would be strongly non-Markovian in nature as discussed in
\cite{KT19}. However, here we consider the scenario where the whole history of the process is reset.
This means that when a reset happens, the system no longer remembers what happened
before resetting. Thus the process is still Markovian. Non-Poissonian simply means, in this context, that the
waiting time distribution $\psi(t)$ is different from a pure exponential as in Poissonian resetting.

For non-Poissonian resetting it is more difficult to write down a forward master equation analogous to (\ref{fme})
as one must in addition keep track of the time since the last reset. This results in a generalised master equation \cite{EM16}. Here, we use the renewal approach (see e.g. \cite{Pakes97,GMS14,PKE16,PR17,CS18,EM18,LD19,MM19})
which we now review.

In the case of time-dependent resetting, one can again exploit the renewal structure of the process in a simple and
straightforward way. We consider a time interval $[0,t]$ and the particle starts initially at $x_0$.  We want to  compute the probability distribution $p(x,t|x_0)$ in the presence of resetting. In this time interval $[0,t]$ there can be no resetting events, one resetting, two resettings, etc. Consider for example the case of no resetting. The probability for this event is simply $\Psi(t)$ and hence the contribution to the probability distribution  representing no resetting in $[0,t]$ is therefore $\Psi(t) G_0(x,t|x_0)$ where $G_0(x,t|x_0)$ is the bare propagator. If there is one resetting event, say at time $t_1 \in [0,t]$, the contribution to the probability  is given by
\begin{equation}\label{contrib_1}
\int_0^t  \D t_1 \psi(t_1) \Psi(t-t_1) G_0(x,t-t_1|X_r) \;,
\end{equation}
where $\psi(t_1) \D t_1$ is the probability that a reset event  happens in $(t_1, t_1 + dt_1]$, followed by no resetting in
the interval $(t_1,t]$ during which the particle propagates freely. Similarly, if there are two resetting events, the contribution to the probability  is 
\begin{equation}\label{contrib_1}
\int_0^t \D t_1 \int_{0}^{t-t_1}\D t_2\, \psi(t_1) \psi(t_2) \Psi(t-t_1-t_2) G_0(x,t-t_1-t_2|X_r) \;.
\end{equation}
The same pattern holds for $n$ resetting events and we need to sum over all $n \geq 1$. The convolution structure of these terms suggests that it is simpler to work in the Laplace space. 
Taking the Laplace transform and summing over all resetting events, using the geometric series, one immediately obtains
\begin{equation}\label{sum_laplace}
\tilde p(x,s|x_0)  = \int_0^\infty \D t \, \e^{-st} \, \Psi(t) G_0(x,t|x_0) + \frac{\tilde \psi(s)}{1-\tilde \psi(s)}\int_0^\infty \D t \, \e^{-st} \, \Psi(t) G_0(x,t|X_r) \;.\end{equation}
If we now set $X_r=x_0$, a simplification occurs and one gets
\begin{equation}\label{sum_laplace2}
\tilde p(x,s|x_0)  = \frac{1}{s \tilde \Psi(s)}\int_0^\infty \D t \, \e^{-st} \, \Psi(t) G_0(x,t|x_0)  
\;,
\end{equation}
where we used the relation $\tilde \Psi(s) = (1-\tilde \psi(s))/s$. Here we denote by $\tilde \Psi(s)$ and $\tilde \psi(s)$ the Laplace transform of $\Psi(t)$ and $\psi(t)$ respectively.
Note that one can also obtain the result in (\ref{sum_laplace2}) just by renewing the process after the first resetting (we 
referred to this as the first renewal equation in Section~\ref{sec:renewal})
\begin{equation}
p(x,t|x_0)=\Psi(t)~G_0(x,t|x_0) 
+\int_{0}^{t}\D\tau~\psi(\tau)~p(x,t-\tau|X_r)\;.
\label{ptnp}
\end{equation}
Taking the Laplace transform of this equation, upon setting $X_r=x_0$, and using the relation $\tilde \Psi(s) = [1- \tilde \psi (s) ]/s$, one recovers (\ref{sum_laplace2}).

Also one can work from the last renewal equation
\begin{equation}
p(x,t|x_0)= \Psi(t)G_0(x,t|x_0) 
+\int_{0}^{t}\D \tau_l~\Upsilon(\tau_l) \Psi(t-\tau_l)~G_0(x,t-\tau_l|X_r) \;,
\label{LR1}
\end{equation}
where  $\Upsilon(\tau_l)$ is the probability density for a reset to occur
in $(\tau_l , \tau_l + \D\tau_l]$ 
 (without specifying when a previous reset occurred) and  $\Psi(t-\tau_l)$
is the probability that there are no further resets after this.
The distribution $\Upsilon(\tau)$ is implied by $\psi(\tau)$ but is  difficult to write down in closed form. However
in the Laplace  domain it is simply given by
$\displaystyle \tilde \Upsilon(s) = \tilde \psi(s)/(1-\tilde \psi(s))$
and the Laplace transform of (\ref{LR1}) recovers (\ref{sum_laplace}).

A stationary state will only exist as $t\to \infty$ in the case when $\psi(\tau)$ decays to zero quickly enough. 
The stationary state is given by the coefficient of $1/s$ in (\ref{sum_laplace2}) in the limit $s\to 0$ thus
\begin{equation}
\hspace*{-1.cm}p(x,t\to \infty|x_0,X_r=x_0) \to p^{*}(x|x_0,X_r=x_0)=  \frac{\int_{0}^{\infty}\D t\Psi(t) G_0(x,t|x_0)}{\int_0^{\infty} \D t \Psi(t)}
\label{stst}
\end{equation}
provided the limit exists. A sufficient condition for this is
\begin{equation}
\int_0^\infty  \D t \,\Psi(t)  < \infty\;.
\label{sscond}
\end{equation}
This condition implies that the waiting time distribution $\psi(t)$ should decay to zero more quickly
than $1/t^2$. In the case where $\psi(t)$ decays slower than $1/t^2$ then the system does not reach any stationary state \cite{NG16}.

We end up this section by mentioning that the case of a resetting rate that depends on absolute time $t$ elapsed from the initial condition, rather than time since last reset,
was considered in \cite{KT19}.

\subsection{Discrete time random walks and L\'evy flights with resetting}\label{sec:discrete}

Up to now we have considered continuous time stochastic processes with resetting. However, in some cases, it is
relevant to consider discrete time processes. This might be the case, for instance, when studying animal movements which typically
consist of discrete jumps. The simplest example of such processes is the discrete time random walk (RW) subject to resetting.

 We thus consider a random walker on a line, starting from $x_0$ and evolving according to the following rules \cite{KMSS14}
\begin{eqnarray}\label{def_RW}
x_n = 
\begin{cases}
&X_r  \;\;{\rm with \;\; probability \;\;} r\\
&x_{n-1} + \eta_n \;\;{\rm with \;\; probability \;\;} 1-r \;,
\end{cases}
\end{eqnarray}
where $r$ denotes here the probability (and not a probability rate) of a resetting event, and hence $0<r<1$. In (\ref{def_RW}) the jumps
$\eta_n$'s are independent and identically distributed (i.i.d.) random variables each drawn from a probability distribution
function (PDF) $f(\eta)$. Here we will restrict our attention to the case where $f(\eta)$ is continuous and symmetric. 
We may consider ordinary random walks, corresponding to jump distributions 
$f(\eta)$ with a well defined second moment $\sigma^2 = \int_{-\infty}^{+\infty} \eta^2\, f(\eta)\, \D\eta$ (and in that case the RW converges for large $n$ to Brownian motion), as well as L\'evy flights, corresponding to heavy-tailed jump distribution $f(\eta) \sim |\eta|^{-1-\mu}$ with $0<\mu<2$. The tail behaviour of $f(\eta)$ is encoded in the small $k$ behaviour of the Fourier transform $\hat f(k) = \int_{-\infty}^{+\infty} f(\eta)\, \e^{-ik\eta}\, \D\eta$ of the jump distribution,
\bea\label{smallk}
\hat f(k) = 1 - |a\,k|^{\mu} + o(|k|^{\mu}) \;, \; \; 0 <\mu \leq 2 \;,
\eea
where $a$ sets the characteristic scale of the jumps and $\mu$ is called the L\'evy index. The case $\mu = 2$ thus corresponds to ordinary random walk while $0< \mu < 2$ corresponds to L\'evy flights.  

We denote by $p(x,n|x_0, X_r)$ the probability density to find the particle at $x$ at step $n$, starting from $x_0$ with resetting to position $X_r$. As before, we will use the shorthand notations $p(x,n|x_0)$ or even simply $p(x,n)$ where there is no ambiguity. From the evolution (\ref{def_RW}), it is straightforward to derive a forward master equation for $p(x,n|x_0, X_r)$. It reads
\bea \label{forward_p_discrete}
\hspace*{0cm} p(x,n) = (1-r) \int_{-\infty}^{+\infty} p(x-\eta,n-1) f(\eta)\, \D\eta + r \delta(x-X_r) \;, 
\eea  
starting from the initial condition $p(x,0) = \delta(x-x_0)$. This equation (\ref{forward_p_discrete})
is the discrete time counterpart of the continuous time forward equation derived in (\ref{fme}). 

This forward equation (\ref{forward_p_discrete}) can be solved via the use of Fourier transform. If one denotes by $\hat p(q,n) = \int_{-\infty}^{+\infty} \e^{i q x} p(x,n)\, \D x$ the Fourier transform of $p(x,n)$ with respect to $x$, one obtains from (\ref{forward_p_discrete}) that it satisfies the equation
\bea \label{forward_Fourier}
\hat p(q,n) = (1-r) \hat f(q)\,\hat p(q,n-1)  + r\, \e^{i q X_r} \;,
\eea
starting from $\hat p(q,0) = \e^{i q x_0}$. This recurrence equation (\ref{forward_Fourier}) can be easily solved with the result
\bea \label{sol_Fourier}
\hspace*{-1.5cm} \hat p(q,n) = (1-r)^n \left[\hat f(q)\right]^n \left(\e^{iq x_0} - \frac{r\,\e^{i q X_r}}{1-(1-r) \hat f(q)} \right) + \frac{r\, \e^{i q X_r}}{1-(1-r) \hat f(q)} \;.
\eea
Let us focus here on the limit $n \to \infty$ of  expression (\ref{sol_Fourier}). Clearly, since $|\hat f(q)| \leq \int_{-\infty}^{+\infty} f(\eta) \, \D\eta = 1$,  one has $|(1-r) \hat f(q)| < 1$ and therefore in the limit $n \to \infty$ the only term that remains in (\ref{sol_Fourier}) is the last one. Hence, one finds that $p(x,n)$ reaches a stationary distribution which is given by the inverse Fourier transform of the last term in  (\ref{sol_Fourier}), independently of $x_0$,
\bea\label{p*}
p(x,n) \underset{n \to \infty}{\longrightarrow} && p^*(x) = \int_{-\infty}^{+\infty} \frac{\D q}{2\pi} \e^{-iq(x-X_r)} \hat p^*(q) \;, \\
\mbox{where} \quad&& \hat p^*(q)=\frac{r}{1-(1-r)\hat f(q)} \;.
\eea
For an arbitrary jump distribution $f(\eta)$, it is very hard to compute explicitly $p^*(x)$ from this formula (\ref{p*}) for all $x$ (one exception being the double exponential jump distribution, see below). However, from (\ref{p*}) one can rather easily extract the large $x$ behaviour of the stationary distribution $p^*(x)$, which turns out to be very different for the two cases $\mu=2$ and  $0<\mu < 2$.  \\

\noindent{\bf The case ${\mu = 2}$}. It is instructive to study the case of a double exponential jump distribution $f(\eta) = 1/(2a) \e^{-|x|/a}$, for which 
the integral over $q$ in (\ref{p*}) can be performed explicitly and one finds, 
\bea \label{pst_exp}
p^*(x) = r \delta(x-X_r) + (1-r) \frac{\sqrt{r}}{2a} \e^{-\sqrt{r}\,\frac{|x-X_r|}{a}} \;.
\eea
 In this case $\hat f(q) = 1/(1+(aq)^2) \approx 1 - (a\,q)^2$ for small $q$ (and hence indeed $\mu = 2$ from (\ref{smallk})) 
which is very similar to the stationary state found for continuous time diffusion (\ref{pssres}), apart from the term $r\,\delta(x-X_r)$ which exists only in the case of discrete
time RW. (We note that the delta-peak at the resetting position is generic feature in discrete time resetting problems.)
In particular, in the limit of large $|x|$ the stationary distribution has an exponential tail 
\bea\label{exp_tail}
p^*(x) \approx \e^{-|x|/\xi(r)} \;, \; \;\;|x| \to \infty\;,
\eea
with $\xi(r) = a\,/\sqrt{r}$. In fact, such an exponential tail is quite generic for $\mu = 2$.  The reason is that for $\mu = 2$, $\hat p^*(q)$ in (\ref{sol_Fourier}) is an analytic function in the complex $q$-plane and the large $x$ behaviour of the integral over $q$ in (\ref{sol_Fourier}) will be dominated, say for $x \to +\infty$, by the pole of smallest modulus of the integrand (in the lower half complex $q$-plane, i.e. for ${\rm Im}(q)<0$). Therefore  $p^*(x)$ decays exponentially as in (\ref{exp_tail}) where $\xi(r)$ is the largest solution of $1-(1-r)\hat f(i/z) = 0$ for $z>0$.  \\

\noindent{\bf The case $0<{\mu <2}$}. In this case, the situation is quite different since $\hat p^*(q)$ is non analytic near $q = 0$, where it behaves as $\hat p^*(q) \approx (1 - |a q|^\mu(1-r)/r)$. Hence, for large $|x|$ the integral over $q$ in (\ref{sol_Fourier}) is dominated by this non-analyticity which implies that $p^*(x)$ decays as a power law for $|x| \to \infty$
\bea \label{largex_powerlaw}
p^*(x) \sim A_\mu(r) \, |x|^{-1-\mu} \;,\; A_\mu(r) =  a^\mu \, \frac{1-r}{r} \sin{\left(\frac{\pi \mu}{2} \right)} \frac{\Gamma(\mu+1)}{\pi}  \;,
\eea
which is markedly different from the exponential decay (\ref{exp_tail}) for $\mu = 2$.

\section{Survival in the presence of an absorbing target}
\label{sec:surviv}
\setcounter{equation}{0}

We now consider stochastic processes under resetting with a target to be achieved by the process.
In the case of diffusive processes we consider a spatial target which absorbs the diffusive particle
(the searcher for the target) and arrests the process.

\begin{figure}[ht]
  \begin{center}
   \includegraphics[width=0.6\textwidth]{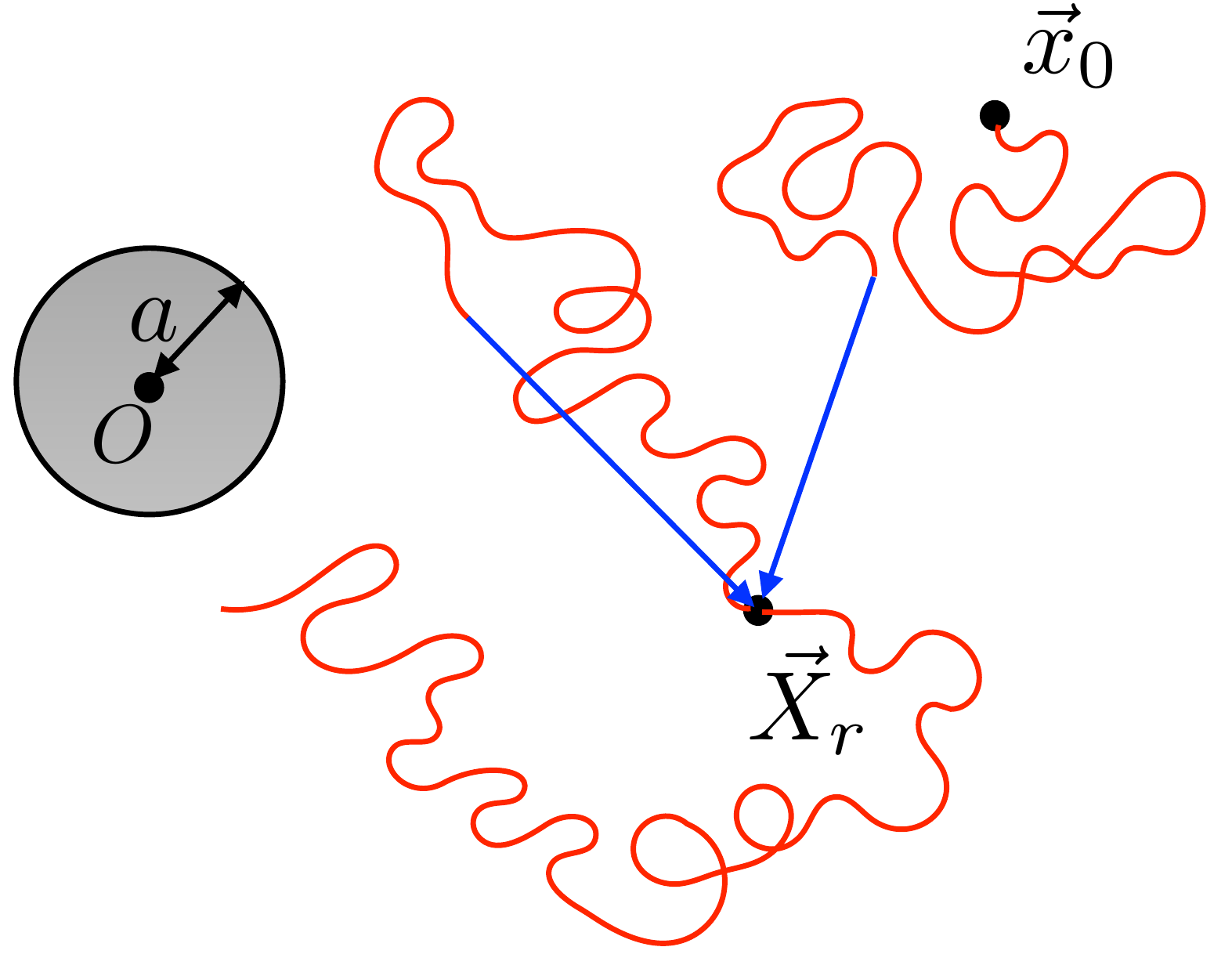}
  \end{center}
  \caption{Illustration in $d=2$ of  the diffusion of a particle with initial
position $\vec x_0$ and  resetting 
to $\vec X_r$, in the presence of an absorbing trap of radius $a$ with centre at the origin $O$.}
   \label{fig:2d}
  \end{figure}

We begin by considering the one-dimensional diffusive case of Section \ref{sec:1ddiff}.
The particle (or searcher) starts at the 
initial position $x_0$ and undergoes diffusion with diffusion constant $D$ and stochastic resetting
to  $X_r$ with a constant rate $r$. When it  reaches
the target, the particle is absorbed  (see figure \ref{fig:2d}).
We wish to compute the survival probability,  $Q_r( x_0,t| X_r)$,  of a diffusive particle
at time $t$, having {\em started} at $x_0$ at $t=0$ with resetting to $X_r$. The subscript $r$ emphasises that this quantity pertains to the process with resetting.  As we have already seen, the results are simplified when the initial position coincides with the resetting position $X_r=x_0$. 
In the following we will  have recourse to $Q_0( x_0,t|X_r)$ which denotes the survival probability in the absence of resetting.

There are several approaches to compute the survival probability:
for example, one can use the forward master equation, the backward master
equation or a renewal equation approach. Here we will present
the renewal equation approach. We refer the reader to the literature 
\cite{EM1,EM2,EM14}
for the backward master equation approach.

\subsection{Renewal equation approach for Poissonian resetting}
\label{sec:renewal_first}
For Poissonian resetting,  for a generic process, it is possible to relate in a simple way the survival probability with resetting, $Q_r$, to that without resetting, $Q_0$. 
A convenient way to establish this relation is to use a last renewal equation which reads
\begin{equation}
Q_r( x_0,t) = {\rm e}^{-rt}  Q_0(  x_0, t)+ r \int_0^t \D \tau \,
{\rm e}^{-r\tau}  Q_0(  X_r, \tau) Q_r(  x_0, t- \tau)\;,
\label{Qt}
\end{equation}
where, to lighten the notation, we have used the shorthand $Q_r(x_0,t|X_r) = Q_r(x_0,t)$ and similarly for $Q_0$. The first term in (\ref{Qt}) represents trajectories in which there has been no resetting. The second term 
represents trajectories in which resetting has occurred. The integral is over $\tau$, the time elapsed since  the last
 reset and we have a convolution of survival probabilities: survival starting from $x_0$ with resetting up to time  $t-\tau$ (the time of the last reset) and survival starting from $X_r$ in the absence of resetting for duration $\tau$
(see figure~\ref{Fig_renewal}).

We now define the Laplace transform
 \begin{equation}
\tilde Q_r( x_0,s) = \int_0^\infty \D t\,  {\rm e}^{-rt}  Q_r(  x_0, t)\;.
\end{equation}
Then  Laplace transforming (\ref{Qt}) yields
\begin{equation}
\tilde Q_r( x_0,s) = \tilde Q_0(  x_0, r+s) +r \tilde Q_0(  X_r, r+s ) \tilde Q_r(  x_0, s)
\end{equation}
from which we readily obtain
\begin{equation}
\tilde Q_r( x_0,s) = \frac{ \tilde  Q_0(  x_0, r+s)}{1- r \tilde Q_0(  X_r, r+s)} \;.
\label{Qrt}
\end{equation}
This is a very general result for Poissonian resetting, relating
the Laplace transform of the survival probability in  the presence of resetting
to that in the absence of resetting. We shall use it repeatedly in this section.

In the specific case where the initial position and resetting position coincide, i.e. $x_0 =X_r$, (\ref{Qrt}) simplifies to
\begin{equation}
\tilde Q_r( X_r,s) = \frac{ \tilde  Q_0(  X_r, r+s)}{1- r \tilde Q_0(  X_r, r+s)} \;.
\label{Qrt2}
\end{equation}
From these expressions (\ref{Qrt}) and (\ref{Qrt2}) we obtain the survival probability with Poissonian resetting 
from that without resetting, as claimed above. Various first-passage observables in the presence of resetting can 
then be computed. 

For example, the mean time to absorption (MTA), with coincident initial and resetting positions
$X_r = x_0$, can be computed from the survival probability $Q_r(X_r,t)$.
First note that the first-passage time density is given by
$\displaystyle -\frac{\partial Q_r(X_r,t)}{\partial t}$. Averaging the
first-passage time over this density, then integrating by parts yields
\begin{eqnarray}
\langle T(X_r) \rangle  &=& - \int_0^\infty \D t\, t \frac{\partial Q_r(X_r,t)}{\partial t}\nonumber\\
&=& \tilde Q_r(X_r,s=0) = \frac{\tilde Q_0(X_r,r)}{1- r\, \tilde Q_0(X_r,r)}\;.\label{MTA}
\end{eqnarray}
The last equality follows from (\ref{Qrt2}) and it relates the MTA with resetting to the Laplace transform of the survival probability without resetting, for any arbitrary stochastic process with Poissonian resetting.
Let us now consider an application of these results to our prototypical case of one-dimensional diffusion.

\subsection{Survival probability for diffusion with resetting}

The expression for $Q_0(  x_0, t)$, the survival probability of a diffusive particle starting from $x_0$ and its Laplace transform
$\tilde Q_0(  x_0, s)$ are standard results in the literature (see e.g. \cite{Redner}). For completeness we derive 
$\tilde Q_0(  x_0, s)$ here,  first using  a general renewal 
equation approach for a first-passage process  and then using the backward master equation for the diffusive case.

We can write a general equation for the propagator  from $x_0$ to $x$ as an integral over 
the {\em first} time to reach $x$
\begin{equation}
G_0( x,t|x_0) =  \int_0^t \D \tau \phi_0(  x, \tau|x_0) G_0(x, t-\tau|x)
\end{equation}
where $\phi_0(  x, \tau|x_0) $ is the probability density of reaching $x$ for the first time at $t$.
Taking the Laplace transform yields
\begin{equation}
\tilde \phi_0(x, s|x_0)=  \frac{ \tilde G_0(x, s|x_0)}{\tilde G_0(x, s|x)}\;.
\label{FPdist}
\end{equation}
This is a general result for the first-passage distribution for  Markovian  processes, which expresses its  Laplace transform  in
terms of the Laplace transform of the propagator $G_0$ for the process.

Now $\phi_0(x, s|x_0)$ is equivalent to  the rate of absorption at an absorbing target at $x$, thus for our case of an absorbing target at the origin
\begin{equation}
\phi_0(0,t|x_0) = - \frac{\partial}{\partial t} Q_0(x_0, t)\;.
\end{equation} 
Taking the Laplace transform yields
\begin{equation}
\tilde \phi_0(0,s|x_0) = 1- s \tilde Q_0(x_0, s)\;.
\end{equation} 
Thus  we obtain from (\ref{FPdist})
\begin{equation}
\tilde Q_0(x_0, s) =\frac{1}{s} - \frac{1}{s}   \frac{ \tilde G_0(0, s|x_0)}{\tilde G_0(0, s|x)}\;.
\end{equation}
This is a general result relating survival probability and hence first passage distribution to the
propagator of the process.

Finally using the form of the 
Laplace transform of the diffusive propagator $\tilde G_0(x, s|x_0)$ 
\begin{equation}
\tilde G_0(x, s|x_0) = \frac{1}{2(Ds)^{1/2}} {\rm e}^{-(s/D)^{1/2} |x-x_0|}
\end{equation}
we obtain
\begin{equation}
\tilde Q_0( x_0,s) = \frac{1- {\rm e}^{-(s/D)^{1/2} x_0}}{s}\;.
\label{Q0s} 
\end{equation}
We note for future reference that the Laplace transform in (\ref{Q0s}) can be simply inverted (see e.g. \cite{Redner}), yielding 
\begin{equation}
 Q_0( x_0,t) = \mbox{erf}\left( \frac{x_0}{2 (Dt)^{1/2}}\right) \;.
\label{Q0t}
\end{equation}

Using (\ref{Qrt}) we deduce
\begin{equation}
\tilde Q_r(x_0,s) = \frac{1 - \exp(-\alpha x_0)}{s+r \exp(-\alpha X_r)}
\label{Qrs}
\end{equation}
where
\begin{equation}
\alpha(s) = \left( \frac{r+s}{D}\right)^{1/2}\;.
\label{alphadef}
\end{equation}
We note that  $\alpha(0) = \alpha_0$ given by (\ref{alpha0}).
In the case where the resetting position $X_r$ coincides with the initial position $x_0$ we have
\begin{equation}
\tilde Q_r(X_r,s) = \frac{1 - \exp(-\alpha X_r)}{s+r \exp(-\alpha X_r)} \;.
\label{Qrs2}
\end{equation}

Having obtained   this  expression for the survival probability in the presence of resetting  one would ideally wish to invert the Laplace transform in order to obtain 
$Q_r( X_r,t)$.
However, it is a difficult task to invert the
Laplace transform (\ref{Qrs}) explicitly for all parameters.  We will discuss the late time asymptotics
in Section~\ref{sec:qasymp}.

For completeness let us also derive  (\ref{Q0s}) from the backward Fokker-Planck equation for the survival probability which reads
\begin{equation}
\frac{\partial Q_0(x_0,t)}{\partial t}
=D\frac{\partial^2 Q_0(x_0,t)}{\partial x_0^2}
\label{QFP}
\end{equation}
with boundary condition $Q_0(0,t)=0$ and initial condition  $Q_0(x_0,t=0)=1$
for $x_0 \neq 0$.
The Laplace transform obeys 
\begin{equation}
D\frac{\partial^2 \tilde Q_0(x_0,s)}{\partial x_0^2}
- s \tilde Q_0(x_0,s) = -1 \;,
\end{equation}
whose general solution, satisfying  the additional boundary condition that
$ \lim_{x  \to \pm \infty}\tilde Q_0(x_0,s) < \infty$, is given by
\begin{equation}
\tilde Q_0(x_0,s) =  A \e^{-(s/D)^{1/2}|x_0|}  + \frac{1}{s} \;.
\end{equation}
The constant $A$ is then fixed by the boundary condition of (\ref{QFP}) at $x_0 =0$ which translates to   $\tilde Q_0(0,s)=0$
and we obtain 
\begin{equation}
\tilde Q_0(x_0,s) =   \frac{1}{s} \left[ 1-  \e^{-(s/D)^{1/2}|x_0|}\right]  \end{equation}
which recovers (\ref{Q0s}).

\subsection{Mean time to absorption for diffusion with resetting}

\begin{figure}[ht]
  \begin{center}
   \includegraphics[width=0.8\textwidth]{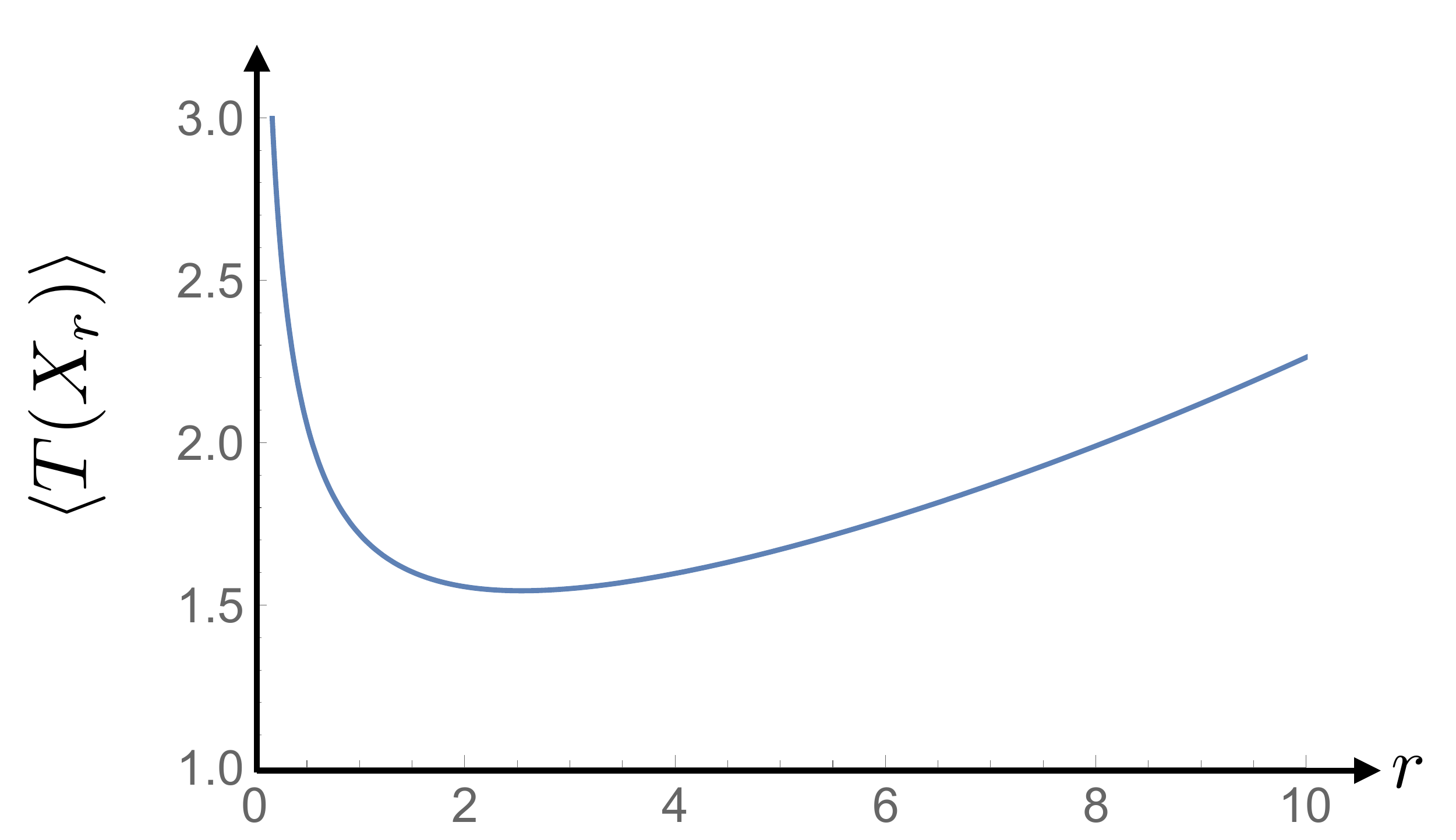}
  \end{center}
  \caption{Mean time to absorption as a function of $r$ (for $D=1$ and $X_r = x_0 = 1$).}
\label{fig:Tr}
  \end{figure}

The mean time to absorption in the case of one-dimensional diffusion  with resetting 
is obtained from (\ref{MTA}) by setting $s=0$ in (\ref{Qrs})
\begin{equation}
\langle T(X_r)\rangle = \frac{1}{r} \left( \exp(\alpha_0 X_r ) -1 \right)\;,
\end{equation}
where we recall that
\begin{equation}
\alpha_0  = \left( \frac{r}{D}\right)^{1/2}\;.
\end{equation}
We note that   $\langle T(X_r)\rangle$ diverges as $r\to 0$ as $\langle T(X_r)\rangle \sim  r^{1/2}$, which recovers the well-known result that the mean time for a diffusive particle to reach the origin (in the absence of resetting) is infinite. Also $\langle T(X_r)\rangle$ diverges as $r \to \infty$, the explanation being that as the reset rate increases the diffusing particle has less time between resets to reach the origin. In between these two divergences there is a single  minimum of $\langle T(X_r)\rangle$ (see figure
\ref{fig:Tr}) which we now study.

For convenience we introduce the dimensionless quantity
\begin{eqnarray}
\gamma &=& \alpha_0  X_r \label{gammadef} 
\end{eqnarray}
which is the ratio of two length scales: $X_r$ is the distance from the resetting position to the target
and $1/\alpha_0$ is the typical diffusion length between resets.

We now seek to minimise the MTA with respect to $r$.
The equation 
\begin{equation}
\frac{\D \langle T(X_r) \rangle}{\D r} =0
\end{equation}
reduces in terms of $\gamma$ to the transcendental equation 
\begin{equation}
\frac{\gamma}{2} = 1- {\rm e}^{-\gamma}
\label{gamma}
\end{equation}
which has a unique non-zero solution $\gamma^*= 1.5936\ldots$.
Thus the minimal mean time to locate the target is achieved when the ratio of
the distance $X_r$  to the target to the typical distance diffused between resets is $\gamma^*$.

\subsection{Optimal resetting: general considerations for diffusive problems}

As we have seen the resetting rate $r$  to position $X_r$ may be chosen to minimise the mean time to absorption for a target
at the origin. This is our first instance of optimal resetting i.e. choosing the resetting rate or the distribution of resetting sites so as to optimise some measure of efficiency such as the mean time to absorption by a target.
Of course in more realistic search problems we may have only partial information about the  target. For example,  we may merely know that the target position $x_T$ is drawn from some  distribution ${\cal P}_T(x_T)$.
In \cite{EM2} various optimisation problems concerning  optimal resetting
with a target distribution were considered.
Optimal resetting implies choosing  (most generally) the space-dependent resetting rate $r(x)$, or the  resetting distribution ${\cal P}_r(X_r)$  (see  Section~\ref{sec:resetdist}),  to
minimise the mean time to absorption for a given target distribution ${\cal P}_T(x_T)$.

In the case of a precisely located target (where ${\cal P}_T$ is a delta function distribution)  it was shown in \cite{EM2} how  a non-resetting window around the coincident initial and resetting  positions $x_0= X_r$ can reduce the mean time to absorption, provided that $x_0$ is sufficiently far from the target. The non-resetting window
is given by a space-dependent resetting rate 
\begin{equation}
r(x) = \begin{cases} 0\quad\mbox{for}\quad |x-x_0|<a \\
 r\quad\mbox{for}\quad |x-x_0|\geq a\;. 
\end{cases}
\end{equation}

For an exponentially decaying target distribution centred at the origin
${\cal P}_T(x_T) = (\beta/2) {\rm e^{-\beta |x_T|}}$,
it was shown  in \cite{EM2}  that a transition in the optimal resetting distribution occurs as $\beta$ decreases i.e. the target distribution broadens. The critical  value of $\beta$ is 
$\beta_c = 2 \alpha_0$ where $\alpha_0 = \sqrt{r/D}$.
For a narrow target distribution $\beta >\beta_c$ the optimal resetting distribution is simply a delta function at the origin.
However, for a broader target distribution $\beta < \beta_c$ the optimal resetting distribution becomes  a delta function at the origin  plus an exponentially decaying piece:
\begin{equation}
{\cal P}^*_r(X_r) =
\begin{cases}
\displaystyle \delta(X_r) \quad \mbox{for}\quad \beta>\beta_c \\[2ex]
\displaystyle  \frac{\beta}{4}\left[1- \frac{\beta^2}{4 \alpha_0^2}\right]{\rm e^{-\beta |X_r|/2}} +
\frac{\beta^2}{4 \alpha_0^2}\delta(X_r) \quad\mbox{for}\quad \beta \leq \beta_c \;.
\end{cases}
\end{equation}

A  related optimisation question is: when does diffusion with resetting perform better than diffusion in a confining potential?
In the former scenario the resetting process confines the particle and creates a nonequilibrium stationary  state, whereas in the latter scenario  the confining potential  creates an equilibrium stationary  state. The question is: which class of dynamics  gives the lower mean time to absorption? In \cite{EMM13} it was shown that
the optimal  mean time to absorption under resetting with optimised constant rate $r^*$, is less than that for an effective equilibrium Langevin process with a potential that generates  the same stationary distribution.
In \cite{KBGN17}  the  optimal potential for a Langevin process (without resetting) for  a given target distribution  was computed  exactly. The mean time to absorption was then compared  to that of resetting with a constant rate $r^*$ optimised for the target distribution. Whether the Langevin dynamics in a potential or the resetting dynamics performs better depends on the  particular choice of target distribution.

As mentioned in the introduction other search  strategies may be considered e.g.
Gelenbe \cite{Gelenbe} considered searchers that have some probabilistic lifetime after which another searcher will be sent out, and
computed  mean times to absorption. In \cite{BBR16}, several searchers under resetting to a single home base were considered and the optimisation of the search time and associated search cost (i.e., the number of searchers times the search time) was studied. Also, in the mathematical literature the mean first passage time for random walkers that have the option of restarting at the initial position has been considered \cite{JP12}.

Returning to our original problem of optimising the  MTA by tuning the resetting rate, we have seen that for diffusive processes with Poissonian resetting in one dimension, there exists an optimal resetting rate $r^*$ that minimises the  MTA to the target. However, it turns out that this optimisation paradigm holds for a wide class of stochastic processes with both Poissonian and non-Poissonian resetting. These generalisations to arbitrary stochastic processes will be discussed in Section \ref{sec:genresetfp}.

\subsection{Late time asymptotics of the survival probability and connection with extreme value statistics}
\label{sec:qasymp}
We now consider the inversion of the Laplace transform of the survival probability (\ref{Qrs2})
\begin{equation}
Q_r(X_r|t)  = \int_{-i \infty + c}^{+i\infty +c}\frac{\D s\,  {\rm e}^{st} }{2 \pi i}
\frac{1 - \exp(-\alpha X_r)}{s+r \exp(-\alpha X_r)}
\end{equation}
where $c$ is a real number chosen so that the integration contour is to the right of any singularities
in the complex $s$ plane.

The singularity structure of the integrand is as follows. There is a simple pole at $s_0$ given by
the solution of
\begin{equation}
s_0 +r \exp\left[- \left( \frac{r+s_0}{D}\right)^{1/2} X_r\right] =0\;.
\label{s0def}
\end{equation}
There is also a branch point singularity at $s=-r$.
One can check that $0> s_0 > -r,$ which implies that for large $t$ the dominant contribution to the inversion will come from the pole and therefore
\begin{equation}
Q_r(X_r|t)  \simeq A {\rm e}^{s_0t} 
\label{Qres}
\end{equation}
where the constant $A$ is determined by the residue of the pole as
\begin{equation}
A= \frac{1+s_0/r}{1+s_0X_r/(2(r+s_0)^{1/2}D^{1/2})}\;.
\end{equation}
Now let us consider the limit $\gamma = \sqrt{r/D}\, X_r \gg 1$ in which case
\begin{equation}
s_0 \simeq -r {\rm e}^{-\gamma}
\end{equation}
is very small and we find from (\ref{Qres})
\begin{equation}
Q_r(X_r|t)  \simeq \exp( -rt\, {\rm e}^{-\gamma} )\;.
\label{QGumb}
\end{equation}
Interestingly expression (\ref{QGumb}) has the form of a Gumbel distribution which occurs in the theory of extreme value statistics of i.i.d. random variables \cite{Gumbel}.

To understand better the reason for this we can make an heuristic derivation of the survival probability.  After a long time $t$ we expect $N=rt$ resets (with corrections of order $t^{1/2}$) to have occurred.  After each reset the diffusive particle will perform an excursion from the reset position $X_r$, which is independent of the previous excursions. For survival until $t$, each excursion must not reach the origin. We have already seen that the survival probability for a diffusive particle (in the absence of resetting) is given by (\ref{Q0t}).  The duration of each excursion is distributed exponentially, thus the survival probability for an excursion $Q_0(X_r,\tau)$, averaged over the duration of the excursion $\tau$ is given by \begin{equation}
 \overline{Q_0(X_r)} =
\int_0^\infty \D \tau\, r {\rm e}^{-r\tau} \mbox{erf}\left( \frac{X_r}{2 (D\tau)^{1/2}}\right)
= 1- {\rm e}^{- (r/D)^{1/2} X_r} \;, \label{Q0bar}
\end{equation}
where we have used the result for the Laplace transform of an error function (\ref{Q0s}).
Thus we  deduce
\begin{equation}
Q_r(X_r,t) \approx \left[ 1- {\rm e}^{-\gamma}\right]^{rt}
\end{equation}
and if $\gamma$ is large this recovers (\ref{QGumb}). We note that the only approximation in this argument
is that  we fix the number of resets to be $N=rt$ and allow fluctuations in the times between resets,
rather than fixing the total  duration of the resets to be $t$.

The connection with extreme value statistics is now clear. The renewal picture implies that we have 
a large number $N \simeq rt$ of resets and we  require the probability that amongst these the largest excursion to the left is less than $X_r$. This coincides with the Gumbel distribution which  is a cumulative probability that the largest of $N$ i.i.d. random variables is less than some value. The Gumbel distribution indeed applies when the  distribution
of each of the random variables has a  tail which decays  exponentially  or faster, which is the case here [see (\ref{Q0bar})]. When the distribution of the random variables has a power-law tail, the corresponding distribution of the maximum belongs to the so-called Fr\'echet class (for a recent pedagogical review on extreme value statistics see \cite{MPS20}). In the context of resetting, if the time between resets is drawn from a power law distribution (as e.g. in non-Poissonian resetting discussed in Section \ref{sec:nP}), one can show that the distribution of the maximum of the reset process is given by the Fr\'echet law, appropriately centred and scaled. This was in fact demonstrated for a ballistic process with reset in one dimensoin ~\cite{VM18}. Another classical extreme value distribution is the Weibull distribution, which occurs when the i.i.d. random variables are each drawn from a bounded distribution \cite{MPS20}. Hence, one would expect the Weibull distribution to appear in the resetting problem by appropriately choosing the time interval between resets (see e.g. \cite{VM18}).

\subsection{Quasi-stationary state for diffusion with resetting and absorption}
It is of  interest to consider the distribution of the particle when we condition on survival ---at long times this converges to a quasi-stationary state \cite{Doorn91,FKMP95}:
\begin{equation}
p(x,t|x_0) \to Q_r(x_0,t) p_{qs}(x)
\label{qs}
\end{equation}
where $Q_r(x_0,t)$ is the survival probability and  $p_{qs}(x)$ is the quasi-stationary state.

For one-dimensional diffusion with resetting, the forward master equation reads
\begin{equation}
\frac{\partial p(x,t)}{\partial t}
= D \frac{\partial^2 p(x,t)}{\partial x^2} - r p( x,t) + r Q_r(x_0,t)\delta(x- X_r)\;,
\label{fmeqs}
\end{equation}
with initial condition $p( x,0) = \delta(x- x_0)$ and boundary condition
$p(0,t)=0$ due to the  absorbing target at the origin. As before, $Q_r(x_0,t)$
is the survival probability at time $t$ having started from $x_0$.

Substituting (\ref{qs}) into the forward master equation and dividing by $Q_r(x_0,t)$ yields
\begin{equation}
 D \frac{\partial^2 p_{qs}(x)}{\partial x^2} - \left(r + \frac{1}{Q_r(x_0,t)} \frac{\partial Q_r(x_0,t)}{\partial t}\right) p_{qs}( x) 
= - r \delta(x- X_r)\;,
\end{equation}
As we have seen [see Equation (\ref{Qres})], for $t \gg 1$, $Q_r(x_0,t) \sim  {\rm e}^{s_0 t}$
and using this we  obtain
\begin{equation}
 D \frac{\partial^2 p_{qs}(x)}{\partial x^2} - (r + s_0) p_{qs}( x) 
= - r \delta(x- X_r)\;,
\end{equation}
with boundary condition $p_{qs}(0) =0$.

\begin{figure}[ht]
  \begin{center}
  \includegraphics[width=0.8\textwidth]{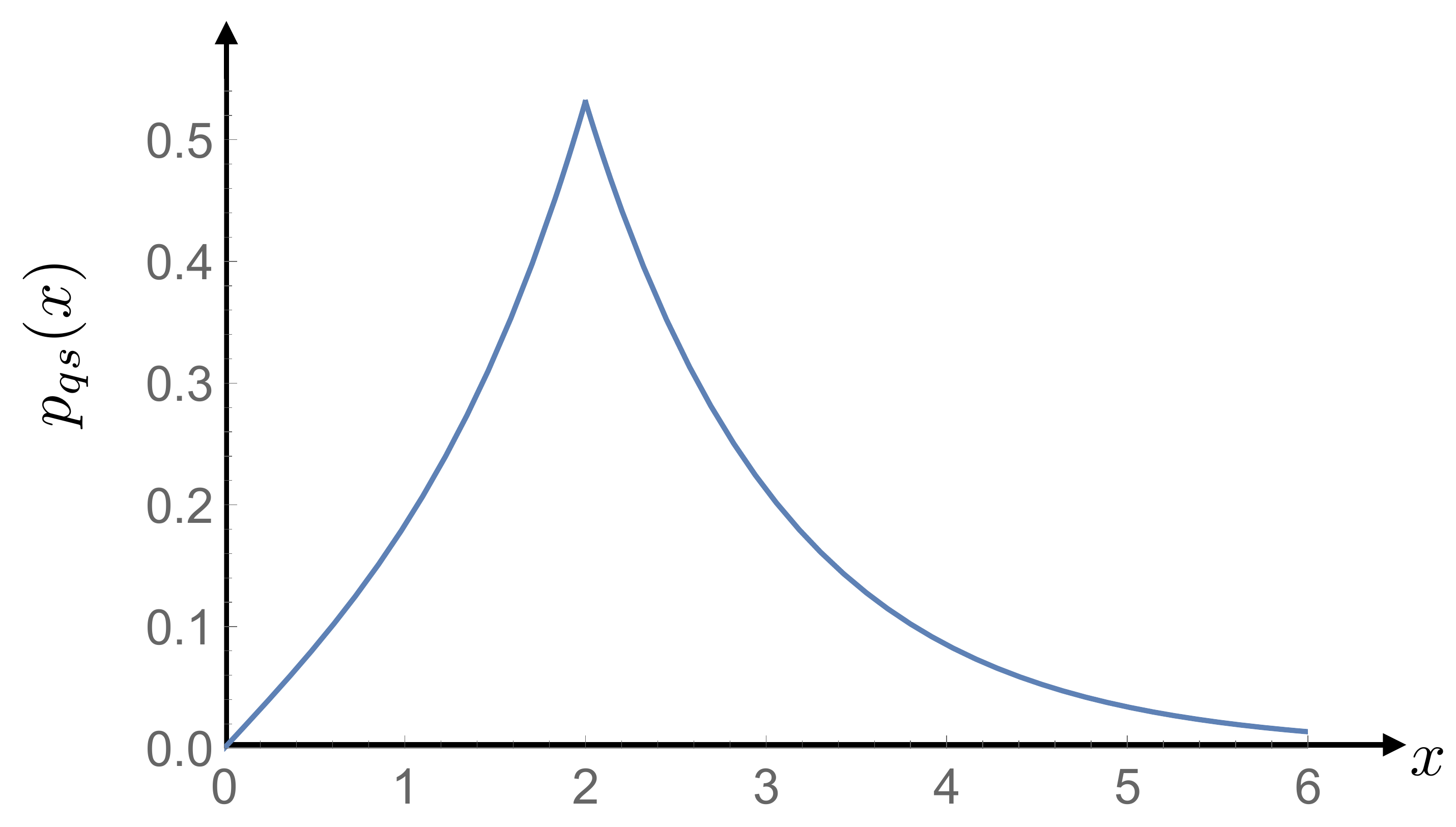}
  \end{center}
  \caption{Plot of the quasi-stationary state distribution given by Eqs. (\ref{qss1}) and (\ref{qss2}). Here $X_r = 2$ and $r=D =1$, such that $s_0 \simeq -0.16$ from (\ref{s0def}).}
\label{fig:qss} 
 \end{figure}

The solution of this equation is obtained by standard means as
\begin{eqnarray}
\hspace*{-0.5cm}p_{qs}(x) &=& \frac{\alpha(s_0)}{{\rm e}^{\alpha(s_0) X_r} -1} \sinh ( \alpha(s_0) x) \;\;\mbox{for}\;\; x< X_r \label{qss1}\\
 &=& \frac{\alpha(s_0)}{{\rm e}^{\alpha(s_0) X_r} -1} \sinh (\alpha(s_0) X_r)  {\rm e}^{-\alpha(s_0)(x-X_r)}\;\;\mbox{for}\;\; x> X_r  \;,
\label{qss2}
\end{eqnarray}
where
\begin{equation}
\alpha(s_0) = \left( \frac{r+s_0}{D} \right)^{1/2}\;.
\end{equation}
The distribution is shown in  figure \ref{fig:qss} which  illustrates the asymmetric distribution decaying more steeply to zero at $x=0$ and with a cusp at the resetting site $X_r$.

\subsection{Arbitrary spatial dimension}
As for the stationary state we can easily generalise the calculation of the survival probability to arbitrary spatial
dimension $d$.
However we have to generalise the point target at the origin in the one-dimensional case 
to an absorbing  $d$-dimensional sphere of radius $a$  (see figure \ref{fig:2d} for
a two-dimensional illustration)
centred at  $\vec x=0$.
The particle starts at the 
initial position $\vec x_0$ (with $|{\vec x_0}|>a$) and undergoes diffusion with diffusion constant $D$ and stochastic resetting
to  ${\vec X_r}$ with a constant rate $r$. When it reaches
the surface of the target sphere, the particle is absorbed.

There is now an additional length scale
in the system, $a$ the radius of the trap.
We generalise the dimensionless variable $\gamma$ (\ref{gammadef})
\begin{eqnarray}
\gamma = \alpha_0  R_r \label{gammadef2} 
\end{eqnarray}
where $R_r = |\vec X_r|$ is the distance from the resetting position to the target
and  we  define an additional  dimensionless reduced variable
\begin{equation}
\epsilon =\frac{a}{R_r}\label{epsdef}
\end{equation}
which  is simply the ratio of the radius  of the absorbing sphere to 
 the distance $R_r$ of the reset point to the target at the  origin.

As before, we  write down a last renewal  equation satisfied by the survival probability
\begin{equation}
Q_r( \vec x_0,t) = {\rm e}^{-rt}  Q_0( \vec x_0, t)+ r \int_0^t \D \tau \,
{\rm e}^{-r\tau}  Q_0( \vec  X_r, \tau) Q_r( \vec  x_0, t- \tau)
\label{Qtd}
\end{equation}
where the first term represents trajectories in which there has been no resetting
and the integral in the second term  is over $\tau$, the time elapsed since  the last reset (see figure~\ref{Fig_renewal}).

We  define the Laplace transform
 \begin{equation}
\tilde Q_r( \vec x_0,s) = \int_0^\infty \D t\,  {\rm e}^{-rt}  Q_r(  \vec x_0, t)
\end{equation}
and we obtain from (\ref{Qtd}) on setting $\vec x_0 = \vec X_r$,
\begin{equation}
\tilde Q_r( \vec X_r,s) = \frac{ \tilde  Q_0(  \vec X_r, r+s)}{1- r \tilde Q_0( \vec  X_r, r+s)}\;.
\label{Qrthighd}
\end{equation}

The expression for $\tilde Q_0(  \vec x_0, s)$, the Laplace transform of the survival probability of a diffusive particle starting from $\vec x_0$ with an absorbing
sphere at the origin, is given by (see e.g. \cite{Redner})
\begin{equation}
\tilde Q_0(\vec x_0, s) =\frac{1}{s} - \frac{1}{s}  \frac{R_r^\nu}{a^\nu} \frac{ K_\nu (\alpha_0 R_r)}{ K_\nu (\alpha_0a)}\;,
\end{equation}
where $K_\nu (z)$ is the modified Bessel function of the second kind, with index $\nu = 1-d/2$, and we obtain
\begin{equation}
\tilde Q_r(\vec X_r,s )  
= 
\frac{
 a^\nu  K_\nu( \alpha a) - R_r ^\nu 
 K_\nu( \alpha R_r)
 }
{r  R_r^\nu 
 K_\nu( \alpha R_r)+ sa^\nu 
 K_\nu( \alpha a)
}\;,
\label{qhdfin}
\end{equation}
where  $R_r =|\vec X_r |$. 
Expression (\ref{qhdfin}) is the exact expression for the Laplace transform of the survival probability
with resetting  in arbitrary dimension $d$.

\subsection{Partial absorption by a target}
\label{sec:pabsorb}
A natural generalisation of the absorbing target, is a target which has some reduced probability of absorbing
the process,  i.e. a partially absorbing target. The formula (\ref{Qrt2}) may still be applied for Poissonian resetting. Thus the problem reduces to finding the survival probability for diffusion with a partially absorbing target at the origin.

This problem may be formulated by introducing an `absorption velocity' $b$ (note that we use $b$ here rather than
$a$ as in \cite{WEM13} to avoid a clash of notation with other sections). The limit $b\to \infty$ corresponds to the absorbing target and the limit $b\to 0$ corresponds to a reflecting boundary at the target (no absorption).
In \cite{WEM13} it is shown that $b$ may be implemented by either applying a boundary condition (sometimes referred to as a radiation boundary condition \cite{SLW84,BRW93}) to the survival probability (without resetting)
\begin{equation} 
\left. \frac{\partial Q_0(x_0,t)}{\partial x_0}\right|_{x_0= 0} = \frac{b}{D} Q_0(0,t)
\end{equation}
or adding a sink term at the origin to a master equation for the survival probability. For example the backward master equation becomes
\begin{equation}
\frac{\partial Q_0(x_0,t)}{\partial t}
= D \frac{\partial^2 Q_0(x_0,t)}{\partial x_0^2} -bQ_0(0,t) \delta(x_0)\;.
\label{bmepabs}
\end{equation}
We begin from (\ref{bmepabs}), the
Laplace transform  of which   becomes
\begin{equation}
D \frac{\partial^2 \tilde Q_0(x_0,s)}{\partial x_0^2}
 -s \tilde Q_0(x_0,s) + 1 =  b \tilde Q_0(0,s) \delta(x_0)\;.
\label{bmepabslt}
\end{equation}
The  general solution of the homogeneous equation (where the right hand side has been set to zero), satisfying  the additional boundary condition that
$ \lim_{x  \to \pm \infty}\tilde Q_0(x_0,s) < \infty$, and the condition that
$\tilde Q_0(x_0,s)$ is continuous at $x_0=0$, is
\begin{equation}
\tilde Q_0(x_0,s) =  A \e^{-(s/D)^{1/2}|x_0|}  + \frac{1}{s} \;.
\end{equation}
The constant $A$ is then fixed by the discontinuity condition
on the first derivative at $x_0 =0$ (which comes from integrating (\ref{bmepabslt}) over $x_0=0$)
\begin{eqnarray}
\lim_{\epsilon \to 0} \left[\frac{\partial \tilde Q_0(x_0,s)}{\partial x_0}\right]^{x_0 = +\epsilon}_{x_0= - \epsilon}  &=&   \frac{b}{D} \tilde Q_0(0,s)\;,
\end{eqnarray}
which yields
\begin{equation}
\tilde Q_0(x_0,s) =   - \frac{b}{2 \sqrt{sD}} \tilde Q_0(0,s)  \e^{-(s/D)^{1/2}|x_0|}  + \frac{1}{s} \;.
\end{equation}
Setting $x_0 =0$ fixes  $\tilde Q_0(0,s)$  self-consistently as
\begin{equation}
\tilde Q_0(0,s) = \frac{1}{s} \frac{2 \phi}{1+2 \phi}  
\end{equation}
where $\phi = \sqrt{sD}/b$. Finally we obtain
\begin{equation}
\tilde  Q_0(x_0,s) = \frac{1}{s} \left[ 1- \frac{1}{1+2 \phi}\e^{-(s/D)^{1/2} |x_0|}\right]\;.
\label{Qpdiff}
\end{equation}
This expression may then be inserted into (\ref{Qrt2}) to obtain the Laplace transform of the  survival probability in the presence of resetting \cite{WEM13}. Note that the limit $b \to \infty$ of (\ref{Qpdiff}) recovers our previous result for a fully absorbing target (\ref{Q0s}). Also note that in the limit $b \to 0$, $\tilde Q_0(x_0,s) \to 1/s$, which implies $Q_0(x_0,t) = 1$, consistent with no absorption at the target.

\subsection{Survival probability with resetting on a finite domain}

So far we have seen that the  introduction of resetting can render finite
an expected time to complete a task that would otherwise diverge. Our archetypal example is diffusion with resetting rate $r$. In the absence of resetting the mean for a diffusive process to locate a target, or equivalently the mean time to absorption (MTA),  diverges in any dimension. Introducing resetting results in a finite MTA.
However we have so far assumed that the domain of the diffusive process is infinite. If, instead, the diffusion is on a finite domain, for example a finite interval in one dimension with reflecting boundaries, then the mean time to locate a target is always finite. In this case resetting to an initial condition may or may not reduce the mean time to absorption.

Several  recent works have studied diffusion with resetting on a finite domain \cite{CS15,DLLJ19}.
Christou and Schadschneider \cite{CS15} considered the problem just alluded to, that of  diffusion with resetting in the interval $[0,L]$ with reflecting boundaries. They considered an arbitrary number $N$ of possible resetting positions (see section \ref{sec:resetdist}) $X_{r_i}$ with $i=1,\dots N$, each chosen at a resetting event  with probability  ${\cal P}_{X_{i}}$. The  equation for the stationary distribution $p^*(x)$ reads
\begin{equation} 
D \frac{\partial^2 p^*(x)}{\partial x^2} = r p^*(x)
-\frac{r}{N} \sum {\cal P}_{X_{r_i}} \delta(x-X_{r_i})
\label{manyresetss}
\end{equation}
with boundary conditions 
\begin{equation}
\left. \frac{\partial p^*}{\partial x}\right|_{x=0,L} =0\;.
\label{reflectbc}
\end{equation}
The problem may be  solved by using a decomposition in terms of eigenfunctions 
$\phi_n(x)$ with eigenvalue $\epsilon_n = -n^2 \pi^2/L^2$ of the Laplacian with boundary conditions (\ref{reflectbc})
\begin{equation}
\phi_n(x) = \sqrt{2/L} \cos( n \pi x/L)
\end{equation}
for $n$ integer.
This results in the solution of  (\ref{manyresetss})
\begin{equation}
p^*(x) = p_0 + \sum_{n=1}^\infty \frac{2r}{NL(r-D \epsilon_n)}
\cos( n \pi x/L) \sum_{i=1}^N \cos( n \pi X_{r_i}/L)
\end{equation}
where $p_0$ is chosen  to normalise the probability.

In \cite{CS15} the survival probability $Q_r(x_0,t)$ for a single resetting site $X_r$,
on a finite domain $0 \leq x \leq L$ with reflecting boundaries and a partially absorbing site with absorption velocity $b$ was considered. The solution was worked out from the forward master equation for the survival probability $Q_r(x_0,t)$ and an eigenfunction expansion.  We note that it is also straightforward to obtain a closed form  solution using the general result (\ref{Qrt}) for Poissonian resetting.
Then one just needs to obtain $Q_0(x_0,t)$ the survival probability for a diffusive particle
on a finite domain with a partial absorption site. 

As the expression for MTA in \cite{CS15}  involves an infinite sum, rather than minimising the MTA the authors optimised the resetting rate by minimising $s_0$ which appears in the survival probability as $ Q_r \sim \e^{s_0 t}$ and,  as discussed in Section \ref{sec:qasymp}, is the dominant pole in the Laplace transform.
Then the value optimal value of the resetting rate, $r^*$, is that which minimises the survival probability at late times.
It was found that the  $r^* >0$ only in a region where the resetting site $X_r$
is sufficiently close to the target site $X_B$.
Typically only if $|X_B-X_r| \leq L/2$ is resetting advantageous.

In \cite{CCS18} the case of a searcher diffusing on a disc
with a target on the perimeter was considered.
When the searcher reaches the perimeter of the disc the diffusion becomes one-dimensional i.e. the searcher then sticks to the perimeter.  Resetting occurs whereby the searcher is returned to the initial condition, somewhere
in the disc away from the perimeter. Thus the search process  comprises two steps: two-dimensional diffusion from the initial position to the perimeter then one-dimensional diffusion around the perimeter (with periodic boundary conditions) and an absorbing site. The diffusion around the circular perimeter with an absorbing site  is equivalent to a finite domain with  absorbing boundaries at both ends.

The latter problem  has also been studied in the case where absorption at 
the different ends is distinguished \cite{CS18,Belan18,PP19a} so that
absorption at one end is interpreted as  successful completion of a task but absorption at the other end is  a failure of the task. Splitting probabilities and conditional 
MTAs are computed  and conditions for the resetting  to expedite successful completion of the task are analysed in \cite{CS18,Belan18,PP19a}.

\subsection{Survival probability and mean time to absorption for non-Poissonian resetting}
\label{sec:survnp} 
We now derive expressions for the survival probability for non-Poissonian resetting, as defined in Section \ref{sec:nP}, in terms of transforms of the survival probability in the absence of resetting.

In the non-Poissonian case it  is convenient to use  a first renewal equation for the survival probability
which reads
\begin{equation}
Q_r( x_0,t) = \Psi(t)  Q_0(  x_0, t)+  \int_0^t \D \tau_f \,
\psi(\tau_f)  Q_0(  x_0, \tau_f) Q_r( X_r, t- \tau_f)\;,
\label{Qtf}
\end{equation}
where $\psi(t)$ is the distribution of the time period between two successive resets and $\Psi(t) = \int_t^\infty \psi(t')\,\D t'$. 
The first term in (\ref{Qtf}) represents trajectories in which there has been no resetting. The second term 
represents trajectories in which resetting has occurred. The integral is over $\tau_f$, the time of the first reset and 
we have a convolution of survival probabilities: survival starting from $x_0$ without resetting up to time  $\tau_f$  and survival starting from $X_r$ in the presence of resetting for duration $t-\tau_f$ (see figure~\ref{Fig_renewal}).

We now take the Laplace transform of (\ref{Qtf}) and get
\begin{eqnarray}
\tilde Q_r( x_0,s) &=& \int_0^\infty \D t\,\e^{-st}\Psi(t)  Q_0(  x_0, t)\\
&+&\tilde Q_r( X_r, s) \int_0^\infty \D \tau_f \,
\e^{-st\tau_f} \psi(\tau_f)  Q_0(  x_0, \tau_f) \;.
\label{Qtflt}
\end{eqnarray} 
 Setting $x_0 = X_r$, we   obtain
\begin{equation}
\tilde Q_r( X_r,s) = 
\frac{\int_0^\infty \D t\,\e^{-st}\Psi(t)  Q_0(  X_r, t) }{1-
\int_0^\infty \D t\,\e^{-st}\psi(t)  Q_0(  X_r, t) }\;.
\label{Qtflt2}
\end{equation} 
The formula for Poissonian resetting (\ref{Qrt2}) is recovered when we take $\Psi(t) = \e^{-rt}$ and
$\psi(t) = r\e^{-rt}$,
in which  case the integrals on the r.h.s. reduce to Laplace transforms.

Integration by parts in the denominator allows  formula (\ref{Qtflt2}) to be written as
\begin{equation}
\tilde Q_r( X_r,s) = 
\frac{\int_0^\infty \D t\,\e^{-st}\Psi(t)  Q_0(  X_r, t) }{s \tilde \Psi(s)
- \int_0^\infty \D t\,\e^{-st}\Psi(t) {\displaystyle \frac{\partial Q_0(  X_r, t)}{\partial t}} }\;.
\label{Qtflt3}
\end{equation} 
The mean time to absorption becomes
\begin{equation}
\langle T(X_r)\rangle = -
\frac{\int_0^\infty \D t\, \Psi(t)  Q_0(  X_r, t) }{
\int_0^\infty \D t\,\Psi(t) {\displaystyle \frac{\partial Q_0(  X_r, t)}{\partial t}} }\;.
\label{Qtflt3}
\end{equation} 
In the case of a diffusive particle, it was shown in \cite{PKE16}
that the mean time to absorption is optimised  for  deterministic  resetting with suitably chosen period $t^*$ i.e.
\begin{equation}
\psi(t) = \delta(t-t^*)\;.
\end{equation}
We shall see later that this is a general feature  which is valid beyond the case of simple diffusion--see section~\ref{sec:genresetfp}.

\subsection{Mean time to absorption for discrete time random walks and L\'evy flights with resetting}\label{sec:survival_discrete}

In this section, we study the survival probability for the discrete random walk model with resetting discussed in Section \ref{sec:discrete}. We have in mind a searcher that moves in discrete time, on a line, according to the resetting dynamics specified in (\ref{def_RW}), starting from $x_0 >0$. Here we consider a broad class of continuous jump distributions characterized by a L\'evy index $\mu$ [see (\ref{smallk})] with $0<\mu \leq 2$. We recall that the case $\mu = 2$ corresponds to ordinary random walks, while $\mu < 2$ describes L\'evy flights, where the jumps are broadly distributed.

The target is located at the origin and, for simplicity, we restrict our discussion to the case where the resetting position $X_r$ coincides with the initial position $X_r = x_0$. To characterise the efficiency of the search process, it is useful to compute the MTA $\langle T(X_r) \rangle$ which, for a fixed resetting position $X_r$, depends here on the probability of a resetting event $r$ as well as on the L\'evy index $\mu$. To compute the MTA we introduce the cumulative distribution function 
\bea \label{survival_discrete}
Q_r(X_r,n) = {\rm Prob}[T(X_r) \geq n)] \;,
\eea
which is precisely the survival probability, i.e. the probability that the walker, starting at $X_r$, does not cross the origin up to step $n$. Of course, the MTA can then be computed from the relation (analogous to (\ref{MTA}) in the continuous time case)
\bea \label{MTAd}
\langle T(X_r) \rangle = \sum_{n \geq 0} Q_r(X_r,n) \;.
\eea

\begin{figure}[ht]
  \begin{center}
\includegraphics[width=0.8\textwidth]{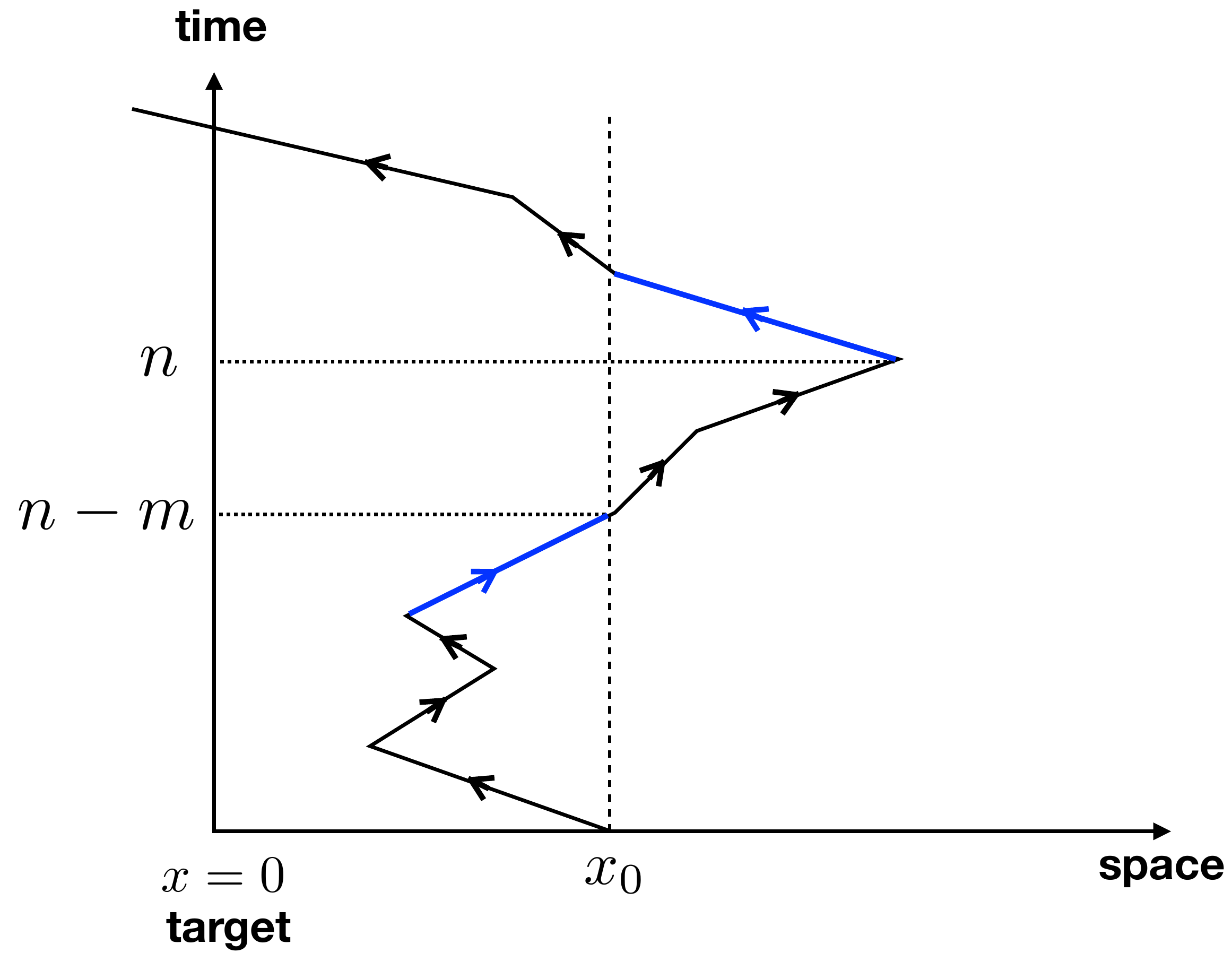}
  \end{center}
  \caption{Illustration of random walk in one dimension with resetting to  the initial position $x_0$  and first passage to
the target at the origin. The integers $n$ and $m$ here illustrate the notation in the renewal equation (\ref{renewal_Qr}).}
\label{fig_discrenewal}
\end{figure}

As done in the case of continuous time processes (\ref{Qt}), one can write a last renewal equation for $Q_r(X_r,n)$  \cite{KMSS14}
\begin{equation} \label{renewal_Qr}
\hspace*{-0cm} Q_r(X_r,n) = \sum_{m=0}^{n-1} r (1-r)^m Q_r(X_r, n-m-1) Q_0(X_r,m) + (1-r)^n Q_0(X_r,n) \;,
\end{equation}
where, as before $Q_0(X_r,n)$ is the survival probability in the absence of resetting (i.e. $r=0$). The first term on the right hand side of (\ref{renewal_Qr}) accounts for the event where the last resetting before step $n$ takes place at step $n-m$ (see figure \ref{fig_discrenewal}) with $0\leq m \leq n-1$. The evolution from step $n-m$ to step $n$ occurs without resetting and the survival probability during this period is $Q_{0}(X_r,m)$, 
while $Q_{r}(X_r,n-m-1)$ accounts for the survival probability from step $1$ to step $n-m-1$ in the presence of resetting. 
The last term in (\ref{renewal_Qr}) corresponds to the case where there is no resetting event at all up to step $n$, which occurs with probability $(1-r)^n$. 

Equation (\ref{renewal_Qr}) can be solved by introducing the generating function $\tilde{Q}_{r}(X_r,z)=\sum_{n\geq0} Q_{r}(X_r,n)
z^n$. Multiplying both sides of (\ref{renewal_Qr}) by $z^n$ and
summing over $n$, we arrive at the result
\begin{equation}
\tilde{Q}_{r}(X_r,z)=\frac{\tilde{Q}_{0}(X_r,(1-r)z)}{1-r z \tilde{Q}_{0}(X_r,(1-r)z)} \;.
\label{Eq:Qr_result}
\end{equation} 
This formula (\ref{Eq:Qr_result}) relates the survival probability in the presence of resetting ($r\geq 0$) to that
without resetting ($r=0$). It is reminiscent of the relation (\ref{Qrt2}) obtained for continuous time processes. Interestingly the Laplace transform of the survival probability in the absence of resetting $\tilde{Q}_{0}(X_r,z)$, with respect to $X_r$, can be computed using the so-called Pollaczek-Spitzer formula \cite{P52,S56,P75,M10}
\begin{subequations}
\label{PS}
\begin{align}
&\int_0^\infty \tilde Q_{0}(X_r,z)\,\e^{-\lambda X_r} \, {\mathrm d}X_r = \frac{1}{\lambda\sqrt{1-z}} \varphi(z,\lambda) \;,  \\
&\varphi(z,\lambda) =  \exp{\left[-\frac{\lambda}{\pi} \int_0^{\infty} \frac{\mathrm d q}{\lambda^2+q^2}\ln{\left(1-z\hat{f}(q)\right)}\right]} \;,
\end{align}
\end{subequations}
which is valid for any continuous and symmetric jump distribution $f(\eta)$, including L\'evy flights (we recall that $\hat f(q) = \int_{-\infty}^{+\infty} \e^{i q\eta} f(\eta)\,\D\eta$). Therefore (\ref{Eq:Qr_result}) together with (\ref{PS}) allow one to compute the cumulative distribution of $T(X_r)$ (\ref{survival_discrete}). In fact by noting the identity $\langle T(X_r)\rangle = \tilde Q_r(X_r,1)$, which follows directly from the definition (\ref{MTAd}), one obtains from (\ref{Eq:Qr_result})
\bea \label{MTA_final}
\langle T(X_r)\rangle = \tilde Q_r(X_r,1) = \frac{\tilde Q_0(X_r,1-r)}{1-r\,\tilde Q_0(X_r,1-r)} \;,
\eea
where $\tilde Q_0(X_r,1-r)$ can, in principle, be computed from (\ref{PS}).  

In Ref. \cite{KMSS14} a detailed analysis of this formula for the MTA (\ref{MTA_final}) was performed for the class of L\'evy stable jump distributions, characterized by $\hat f(k) = \e^{-|ak|^\mu}$, with L\'evy index $0 < \mu \leq 2$ (and we set $a=1$ in what follows). We summarise here the main results obtained there and refer the reader to \cite{KMSS14} for more details. For a fixed $X_r$, it is natural to 
minimise the MTA $\langle T(X_r) \rangle$ with respect to the two parameters $\mu$ and $r$
and find the optimal parameters $\mu^*(X_r)$ and $r^*(X_r)$ as a function of $X_r$. 

It turns out that these optimal values $\mu^*(X_r)$ and $r^*(X_r)$
exhibit a rather rich and surprising behaviour, as functions of
$X_r$. Indeed there exists a critical value $X_r^* \simeq 0.58$ (determined numerically) such that the optimal strategy depends
crucially on whether $X_r > X_r^*$ or $X_r < X_r^*$. When $X_r>X_r^*$,
the optimal parameters are independent of $X_r$, and are given by
\begin{subequations}
\begin{align}
&\mu^*(X_r>X_r^*) = 0 \;,\qquad r^*(X_r>X_r^*) = r_>^* \;, \label{rstar_mu0_a}
\\
\label{rstar_mu0}
& {\rm where}\; r_>^* = \frac{\sqrt{\e-1}}{2} (\sqrt{\e}-\sqrt{\e-1})= 0.22145\ldots \;. 
\end{align}
\end{subequations}
In (\ref{rstar_mu0_a}), $\mu^*=0$ actually means the limit $\mu^* \to 0$. 
On the other hand, for $X_r< X_r^*$, the optimal values $\mu^*(X_r)$
and $r^*(X_r)$ depend continuously on $X_r$, both of them being
monotonously decreasing functions of $X_r$. Interestingly, it was found in \cite{KMSS14} that the optimal parameters $\mu^*(X_r)$ and $r^*(X_r)$ exhibit a discontinuity as $x_0$ crosses the value $X_r^*$. This behaviour is a typical characteristic of a first order transition at $X_r^*$. 
Note that related phase transitions  have also been reported for the case of random walks with exponentially distributed flights under resetting \cite{CM15}.

Here, we have studied L\'evy flights by considering discrete time random walks with heavy-tailed jump distributions $f(\eta) \propto |\eta|^{-1-\mu}$ as $|\eta| \to \infty$, in the limit of a large number of steps $n \gg 1$. It is also possible to study L\'evy flights in the context of continuous time random walks and we refer the reader to \cite{KGN15,KGN19,Santos19} for a study of the MTA for continuous time L\'evy flights with resetting.

\section{Multiparticle diffusive systems}
\label{sec:multi}
\setcounter{equation}{0}

So far we have considered the problem of single particle dynamics under resetting. We now turn to
the problem  of multiple non-interacting particles and show  how the time to find a target is affected \cite{EM1}.

We consider  $N$ independent particles labelled by $j =1,\ldots, N$, each of which is
reset to its own resetting position $X_j $. We will consider here the Poissonian case  where each particle is reset with the same rate $r$. One may think of a team of searchers seeking a target and the whole process stops when any one of the searchers finds the target (see e.g. \cite{MOS11,BBR16}). We are interested in the survival probability $P_s(t)$ of the target, i.e. the probability that none of the particles have reached the target up to time $t$. To simplify matters, we
also take the initial position of each particle to be identical to
its resetting position.

\subsection{Average and typical behaviour: annealed and quenched averages}

As the $N$ particles move independently, the survival probability of the target is given by
\begin{equation}
P_s(t) = \prod_{j=1}^N Q_r( X_j,t)
\label{Psdef}
\end{equation}
where $Q_r( X_j,t)$ is the survival probability, in the presence of resetting, in the single particle problem
considered in Section \ref{sec:surviv}. 

We consider the $N$ resetting positions to be random variables
distributed uniformly with density $\rho$ and consequently, $P_s(t)$
will itself have a distribution.  Its average is simply
$P_s^{\rm av}(t)=\langle P_s(t)\rangle_{X}$ where
$\langle \cdot \rangle_{ X}$ denotes averages over $X_j$'s.
However, as we shall see, the {\em typical} value of the survival
probability $P_s(t)$ is not captured by the average.  This is because
the average may be dominated by rare samples of the resetting
positions of the searchers for which the survival probability is much
larger than the typical value of the survival probability. We will discuss this effect in more detail in section~\ref{sec:evs}.

To  compute the average behaviour of (\ref{Psdef}) 
we write
\begin{eqnarray}
P_s^{\rm av}(t) &=&
 \left[ \langle  Q_r(X,t)\rangle_{ X}\right]^N \\
&=&  \exp\left\{ N \ln\left[1- \langle 1- Q_r(X,t) \rangle_{ X}\right] \right\}\;.
\end{eqnarray}

We begin by considering  each $X_j$ to be
distributed uniformly over a finite interval $[ -L/2,L/2]$ (and later take the limit $L\to \infty$). We obtain
\begin{equation}
\langle 1-  Q_r(X,t)\rangle_{X}
= \frac{1}{L} \int_{-L/2}^{L/2} \D X
\left( 1- Q_r(X,t)\right) \;.
\end{equation}
Letting $N,L \to \infty$
but keeping the density of walkers $\rho=N/L$ fixed, 
we obtain
\begin{equation}
P_s^{\rm av}(t) \to  \exp\left[ - 2\rho I_1(t)\right]\;,
\label{Psann}
\end{equation}
where we define
\begin{equation}
I_1(t) \equiv \int_0^\infty \D X (1- Q_r(X,t)) \;,
\label{I1def}
\end{equation}
and we have assumed that $Q_r(X,t)$ is an even function of $X$, i.e. $Q_r(X,t) = Q_r(-X,t)$.

On the other hand the typical behaviour of  $P_s(t)$ can be found
by first averaging the logarithm of $P_s(t)$ followed by 
exponentiating 
\begin{equation}
P_s^{\rm typ}(t)= \exp\left[\langle \ln P_s(t) \rangle_{X} \right] \;.
\end{equation} 
One can draw an analogy to a disordered system with $P_s(t)$
playing the role of a partition function $Z$ and $X_\mu$'s corresponding to
disorder variables.
Thus the average and typical behaviour correspond respectively 
to the {\em annealed} average (where one averages the partition function $Z$)
and the {\em quenched} average (where one averages the free energy $\ln Z$)
in disordered systems. In the limit  $N,L \to \infty$,
with  density of walkers $\rho=N/L$ fixed, 
 we can express  $P_s^{\rm typ}(t)$ as
\begin{equation}
P_s^{\rm typ}(t)
= \exp \left\{ N \langle \ln\left[ Q_r( X,t)\right] \rangle_{ X}\right\}
\to  \exp\left[ - 2\rho I_2(t)\right]\;,
\end{equation}
where 
\begin{equation}
I_2(t) \equiv -\int_0^\infty \D X  \ln Q_r(X,t)\;,
\label{I2def}
\end{equation}
assuming again that $Q_r(X,t)$ is an even function of $X$. 

Thus the determination of the average and typical behaviour reduces to the  evaluation of two integrals $I_1(t)$ in (\ref{I1def}),
$I_2(t)$ in (\ref{I2def}).

\subsection{Average behaviour for one-dimensional diffusion with resetting}
\label{sec:Pav}
The Laplace transform of (\ref{I1def}) ${\tilde I_1}(s)=\int_0^{\infty} I_1(t) \e^{-st} \D t$
can be determined as follows:
\begin{eqnarray}
{\tilde I_1}(s) &=&   \int_0^\infty\D X\left[\frac{1}{s}-\tilde Q_r(X,s)\right]
\nonumber \\
&=&  \frac{(r+s)}{s} \int_0^\infty\D X\,\left[\frac{  {\rm e}^{-\alpha X}}{s+r  {\rm e}^{-\alpha X}} \right]\\
&=&
\frac{(r+s)}{sr \alpha }  \ln \left( \frac{s+r}{s}\right)\;,
\label{I1s}
\end{eqnarray}
where, in the second line, we have used the expression of $\tilde Q_r(X,s)$ given in (\ref{Qrs}). The Laplace transform can be inverted to yield \cite{EM1}
\begin{equation}
I_1(t) = \left( \frac{D}{r}\right)^{1/2}
\int_0^{rt} \D v \frac{1-{\rm e}^{-v}}{v}\left[ \mbox{erf}[(rt-v)^{1/2}] + \frac{{\rm e}^{-(rt-v)}}{\sqrt{\pi(rt-v)}}\right]\;.
\label{I1int}
\end{equation}
The asymptotic behaviours  of $I_1(t)$ are given by
\begin{eqnarray}
I_1(t) \simeq
\begin{cases}
& \dfrac{2}{\sqrt \pi}  (Dt)^{1/2}\quad \mbox{for}\quad  rt \ll 1\;,\\ 
& \\
& \left( \dfrac{D}{r}\right)^{1/2}\left[ \ln(rt) + \gamma_e\right]\quad \mbox{for}\quad  rt \gg 1 \label{I1_larget} \;,
\end{cases}
\end{eqnarray}
where 
$\gamma_e$ is Euler's constant which may be defined as
\begin{equation} \label{gammaE}
\gamma_e = 
\int_0^\infty \D u\, {\rm e}^{-u}\ln u = 0.577216\ldots\;.
\end{equation}
The first limit $rt \ll 1$ can be obtained from the small $v$ expansion of  the integral (\ref{I1int}).
The $rt \gg1 $ behaviour can be most easily obtained from the  $s\to 0$ of the Laplace transform (\ref{I1s})
\begin{equation}
{\tilde I_1}(s) = \frac{2}{s\alpha_0}
 \ln s +O(1/s)\;,
\end{equation}
together with the identity
\begin{equation}
\int_0^\infty \D t \,{\rm e}^{-st} \ln t = \frac{1}{s} \int_0^\infty \D u\, {\rm e}^{-u}\ln \frac{u}{s} = - \frac{\gamma_e}{s} - \frac{\ln s}{s} \;.
\end{equation}

By substituting the result for $I_1(t)$ given in (\ref{I1_larget}) in (\ref{Psann}), one finds that 
the long-time behaviour ($rt \gg 1$) of the average survival probability is a power law with an exponent that varies
 continuously with the density 
\begin{equation}
P_s^{\rm av}(t)  \sim t^{-2\rho (D/r)^{1/2}}\;.
\label{Pav}
\end{equation}
Such behaviour is somewhat unexpected, since in the absence of resetting the average survival probability decays like $P_s(t) \sim e^{-b \sqrt{t}}$ \cite{BMS13, BB02,BB03}.

\subsection{Typical behaviour for one-dimensional diffusion with resetting}
For  the typical behaviour (the quenched case)  we recall that
in the long time limit the inversion of the Laplace transform 
$\tilde Q_r(X,s)$ is dominated by a pole at $s_0$ (\ref{s0def}) in the complex $s$ plane  endowing
$Q_r(X,t)$ with exponential dependence on time.  Therefore at large times,  $I_2(t)$ in (\ref{I2def}) becomes
\begin{equation}
I_2(t) \simeq 
- t\int_0^\infty \D X\, s_0(X)\;.
\label{I2int}
\end{equation}
The integral may be computed exactly as follows.
We first define $u_0$ through
\begin{equation}
s_0 = r(u_0 -1) \;.
\end{equation}
Then we may express
$X$ as a function of $u_0$ from (\ref{s0def})
\begin{equation}
u_0 = 1- {\rm e}^{-\gamma u_0^{1/2}}\;,
\end{equation} 
where $\gamma$ is defined in (\ref{gammadef}).
 In particular we have
\begin{equation}
\alpha_0 X =  -\frac{\ln(1-u_0)}{u_0^{1/2}}\;,
\end{equation}
allowing one to transform the integration from $X$ to   $u$ with range $ 0 < u <1$.
One finds
\begin{eqnarray}
-\int_0^\infty \D X\, s_0(X)&=&
\frac{r}{\alpha_0} \int_0^1 \D u \left[
\frac{(1-u)}{u^{3/2}}\ln(1-u)+ \frac{1}{u^{1/2}}\right] \\
&=&
(D r)^{1/2} 4(1-\ln 2)\;.
\end{eqnarray}
Thus,  the  asymptotic decay of the typical  total survival probability is exponential with explicit  decay constant \cite{EM1}
\begin{equation}
P^{\rm typ}_s(t) \sim \exp \left[ - t \rho (D r)^{1/2} 8(1- \ln 2) \right] \;,
\label{Ptyp}
\end{equation}
which behaves quite differently from the average survival probability (\ref{Pav}).

\subsection{Explanation in terms of extreme value statistics}\label{sec:evs}
It is important to note   that the average and typical survival probabilities 
have distinct asymptotic behaviours---the average behaviour (\ref{Pav}) decays far more slowly than the typical behaviour (\ref{Ptyp}).
In fact the average behaviour has a different functional form, power law rather than exponential decay with time, which is a surprising result.

In order to understand the  asymptotic form of the average survival probability (\ref{Pav}),
we consider the following simple picture.  At long times we assume that  the absorption  probability of the target will be 
dominated by the searcher which started nearest to the target (taken to be at the origin).
Denoting  the position of this searcher by $y$,  the average survival probability
for the many searcher problem  should then be recovered by averaging the  single searcher's survival probability (\ref{Qres}) over the distribution of the  position $y$.
What we will now show is that this average is dominated by  rare configurations of the searcher initial positions where $y$ is large.

In order to obtain the  distribution of the distance $y$ of the nearest
searcher to the origin, we consider first the probability that a single searcher, distributed uniformly in a box of size $L$, starts at distance $X>y$ from the origin
\begin{equation}
\mbox{Prob}(X > y) =\frac{2}{L} \int_y^{L/2} \D X
= \left[ 1 - \frac{2y}{L} \right]\;.
\end{equation}
Then it follows that the probability that all $N$ searchers start at distance $X>y$ from the origin is given by, in the limit of large $N,L$ with $\rho$ fixed,
\begin{equation}
[\mbox{Prob}(X > y)]^N  \to \exp(-2\rho y)
\end{equation}
and we obtain  the distribution of the distance $y$ of the nearest
searcher from the   target as
\begin{equation}
P(y) = - \frac{\D }{\D y }
\exp\left[ -2 \rho y \right] =2 \rho \exp \left(
-2 \rho y
\right)\;.
\label{Py}
\end{equation}

As we have seen for  a single searcher starting at $y$ the long time behaviour of the survival probability, in the presence of resetting, is given by (\ref{Qres})
\begin{equation}
Q_r \simeq   \exp\left[ - | s_0(y)| t\right]\;,
\label{Psy}
\end{equation}
where the function $s_0(y)$  is given by (\ref{s0def}). Within the approximation that the survival  probability of the target  is
dominated by the  searcher initially nearest  to the target, at distance $y$, 
we obtain $P^{\rm av}_s(t)$ as the  average of (\ref{Psy}) with respect to (\ref{Py})
\begin{equation}
P^{\rm av}_s(t) 
 \simeq \int_0^\infty \D y\,  \exp\left[- 2 \rho  y  + t s_0(y) \right]\;.
\label{PSint}
\end{equation}
For large $t$, we expect
the integral to be dominated by the  value $y^*$  that maximises the integrand with respect to $y$. Thus
\begin{equation}
- 2 \rho   +t s_0'(y^*)=0\;.
\end{equation}
For large $t$ we expect $y^*$ to be  large and $s_0(y^*) \simeq -r \exp(-\alpha_0 y^*)$ small.
The  maximum $y^*$ is then given by
\begin{equation}
 - 2 \rho   + r\alpha_0t {\rm e}^{-\alpha_0 y^*} =0
\end{equation}  
which implies that asymptotically
\begin{equation}
y^* 
\sim \frac{\ln t}{\alpha_0}\;.
\end{equation}
The dominant behaviour of the  integral (\ref{PSint}) is then 
\begin{equation}
P^{\rm av}_s(t)
\sim \exp\left[-2\rho y^*\right]
\end{equation}
which recovers the asymptotic result (\ref{Pav}) 
of Subsection \ref{sec:Pav}.

Thus  we have deduced that at long times $t$ the average survival probability is dominated by initial arrangements of searchers  in which  the nearest searcher is at distance $y^* \sim \ln t/\alpha_0$.
These initial configurations of searchers are atypical as may be seen by comparing with the distribution
of nearest searcher distance (\ref{Py}). As time progresses rarer and rarer  initial configurations of the  searchers with nearest searcher at distance
$y \sim \ln t$ from the target dominate the average.
This  reflects  a strong  dependence on the initial conditions whose memory is retained through resetting.

\section{General resetting and first passage processes}
\label{sec:genresetfp}
\setcounter{equation}{0}

\subsection{Arbitrary stochastic process with Poissonian resetting}

In the previous sections, we mainly discussed diffusive processes with Poissonian and non-Poissonian
resetting (see figure~\ref{fig1}). The resetting can be generalized to arbitrary stochastic processes, going
beyond simple diffusion as follows:
\begin{itemize}
\item[$\bullet$]{Consider any process $x(t)$ evolving freely under its own dynamics (it can be deterministic or stochastic) during
a certain interval of time.}

\item[$\bullet$]{At the end of this random period, the process is reset to a new starting point $X_r$ (which can in particular be the initial position $X_r = x_0$) and
then its dynamics restarts afresh.}

\item[$\bullet$]{The interval of free evolution between resets is drawn independently from a distribution $\psi(\tau)$ (hence naturally it is a renewal process). For Poissonian resetting, $\psi(\tau) = r\, \e^{-r\,\tau}$.}

\end{itemize}

In this subsection, we first focus on Poissonian resetting. Exploiting the renewal structure, we can then relate observables in the presence of resetting to the same observables in the absence of resetting, for arbitrary stochastic processes, as was done for diffusive processes before (see e.g.   (\ref{pt})). For example,  
$p_r(x,t|x_0)$, defined as the probability density to reach $x$ at time $t$ in the presence of resetting is related to the propagator without resetting, $G_0(x,t|x_0)$,
\begin{eqnarray}
p_r(x,t|x_0) = \e^{-r\,t}G_0(x,t|x_0)  + r\,\int_0^t \D\tau \, \e^{-r \tau} G_0(x,\tau|X_r) \; \label{renewal_p_1} \;.
\end{eqnarray}
This is the analogue of (\ref{pt}) for the diffusive process. The derivation of this relation (\ref{renewal_p_1}) is straightforward, as in the diffusive case. The first term refers to no resetting in $[0,t]$. In the second term, $\tau$ denotes the time between $t$ and the last resetting before $t$. In this term, $r \D\tau e^{-r\tau}$ denotes the probability that there is no resetting during $\tau$ followed by a resetting during $\tau$ and $\tau + \D\tau$. During this interval $\tau$, the particle evolves freely with the propagator $p_0$, since there is no resetting event in the interval $(t-\tau,t]$ (as in figure~\ref{Fig_renewal}). One then takes the product of these two terms and integrate over all $\tau$ in $[0,t]$. Therefore, if we know the free propagator $G_0$, in principle one can compute $p_r$ in the presence of resetting. Finally, the stationary state, if it exists, can be obtained by taking the limit $t \to \infty$ in (\ref{renewal_p_1}). This gives 
\begin{eqnarray}
p^*_r(x) = r\,\int_0^\infty \D\tau \, \e^{-r \tau} G_0(x,\tau|X_r) \; \label{renewal_p_2} \;.
\end{eqnarray}  
The stationary state is thus given by the Laplace transform of the free propagator (up to the constant factor $r$), provided the integral in (\ref{renewal_p_2}) is finite. This is then a very general relation for any stochastic process with resetting. 

One can also relate other observables between processes with and without resetting, going beyond the one-point function discussed above. For instance, let us define the two-point correlation function for any process as 
\begin{eqnarray}\label{def_correl_r}
C(t_1, t_2) = \langle x(t_1) x(t_2)\rangle -  \langle x(t_1) \rangle \langle x(t_2)\rangle \;.
\end{eqnarray}
For Poissonian resetting, one can relate the correlator $C_r(t_1, t_2)$ for the process with reset to $C_0(t_1, t_2)$ referring to the correlator
in the absence of resetting. For resetting to the initial condition $X_r=x_0$, this exact relation has been derived recently in \cite{MO18} and it reads (for $t_1\leq t_2$)
\begin{equation}\label{renewal_correl}
C_r(t_1, t_2) = \e^{-r(t_2 - t_1)} \left[\e^{-r t_1} C_0(t_1, t_2) + r \int_0^{t_1} d\tau \,\e^{-r\tau} C_0(\tau, t_2-t_1+\tau) \right] \;.
\end{equation}
The derivation of this relation exploits the renewal structure of the reset process. This relation was then used to obtain the power spectrum of various stochastic processes with reset, such as the fractional Brownian motion (fBm) \cite{MO18}.

Similarly, one can relate the survival probability for arbitrary stochastic processes with Poissonian resetting to that in the absence of resetting via the renewal equation 
\begin{equation}
Q_r( x_0,t) = {\rm e}^{-rt}  Q_0(  x_0, t)+ r \int_0^t \D \tau \,
{\rm e}^{-r\tau}  Q_0(  X_r, \tau) Q_r(  x_0, t- \tau)\;,
\label{Qt_gen}
\end{equation}
which was already presented in Section~\ref{sec:renewal_first} and was exploited to obtain explicit results for diffusive processes. Consequently, the MTA for the process with resetting can be related to the Laplace transform of the survival probability without resetting, for any arbitrary stochastic process with Poissonian resetting [see (\ref{MTA})].

As an application of these general results valid for arbitrary processes with Poissonian resetting, going beyond the simple diffusion, we just give the example of a run and tumble particle (RTP) subject to stochastic resetting \cite{EM18,SS17}. The position of a RTP in one dimension, in the absence of resetting, evolves via the stochastic equation of motion
\begin{eqnarray}\label{eq:RTP}
\frac{\D x}{\D t} = v_0 \,\sigma(t)
\end{eqnarray}
where $\sigma(t)$ is a dichotomous noise that switches between two states $\sigma(t) = \pm 1$ with rate $\gamma$. The correlation function of the noise
then decays as $\langle \sigma(t_1) \sigma(t_2)\rangle = \e^{-2 \gamma|t_1-t_2|}$. For finite $\gamma$, the noise has thus a memory. This motion is 
sometimes referred to as a persistent random walk and has been the subject of renewed recent interest in the context of active particles. Consider this RTP being subjected to Poissonian resetting, and using the general results above, one can compute various observables in the presence of resetting in terms of those without resetting. For example, the propagator $G_0(x,t|x_0=0)$ for an RTP starting at the origin with equal probability for the initial velocity to be $\pm 1$ can be computed exactly. Its Laplace transform reads simply
\begin{equation}\label{Laplace_RTP}
\tilde p_0(x,s|0) = \int_0^\infty \D t \, \e^{-st} \,G_0(x,t|x_0=0) = \frac{\lambda(s)}{2s}\, \e^{-\lambda(s) |x|} \;, \;{\rm where} \; \lambda(s) =\sqrt{ \frac{{s(s+2\gamma)}}{v_0} } \;.
\end{equation} 
Under Poissonian resetting of the position to the initial position $x_0=0$ with rate $r$, and randomisation of the velocity after each resetting, the stationary state $p^*_r(x)$ is then given by (\ref{renewal_p_2}) and one gets \cite{EM18}
\begin{equation} \label{Pstat_RTP}
p^*_r(x) = r\, \tilde p_0(x,s=r|0) = \frac{\lambda(r)}{2} \e^{-\lambda(r)\, |x|} \quad, {\rm where} \quad \lambda(r) =
\sqrt{ \frac{{r(r+2\gamma)}}{v_0} } \;.
\end{equation} 
It turns out that this stationary state is robust, i.e., it does not depend on the precise velocity randomisation protocol following each reset of the position \cite{EM18}. In the limit $v_0 \to \infty$, $\gamma \to \infty$ keeping the ratio $v_0^2/(2 \gamma)=D$ fixed, the RTP is known to reduce to ordinary diffusion. By taking this limit in (\ref{Pstat_RTP}), one indeed recovers the diffusive stationary state given in Eqs. (\ref{pssres}) and (\ref{alpha0}). Similarly, one can also derive the survival probability and the MTA of the RTP with reset \cite{EM18}, from (\ref{Qt_gen}) using the known result for the survival probability of the RTP without reset \cite{malakar,rtp_noncrossing}.    We also mention that the telegrapher's equation, which 
naturally arises in the context of RTP, has been studied under resetting \cite{Masoliver19}. 

\subsection{General resetting and completion times}

So far we have discussed Poissonian resetting for arbitrary stochastic processes. One can easily generalise these ideas to non-Poissonian
resettings, as we have already seen for the diffusive process. The stationary state for non-Poissonian resetting and arbitrary stochastic process 
can be read off from  (\ref{stst}),  with $G_0(x,t|x_0)$  the
bare propagator  of the stochastic process without reset. For non-Poissonian resetting, the first-passage probability with reset can also be generalised to arbitrary processes, as we discuss below.

\begin{figure}[ht]
  \begin{center}
\includegraphics[width=0.8\textwidth]{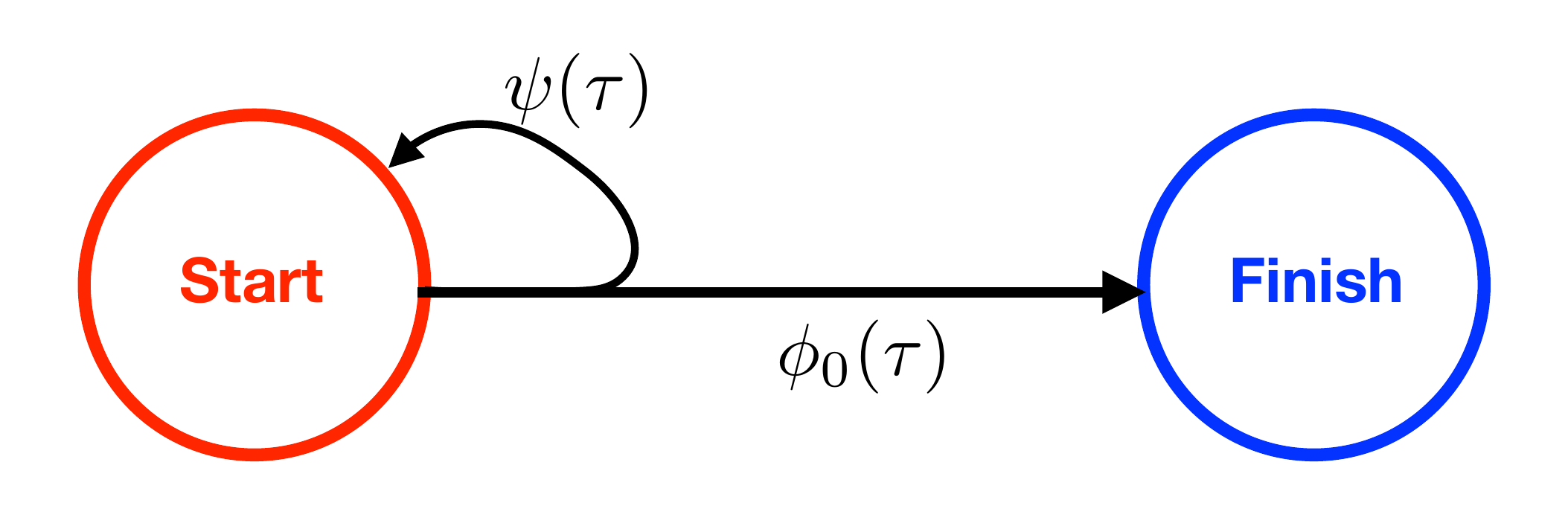}
  \end{center}
  \caption{Schematic illustration  of task to be completed  with
bare completion distribution $\phi_0(\tau)$ and a restart process waiting time $\psi(\tau)$ \cite{Reuveni16}.}\label{Fig:completion}
\end{figure}

As already noted in Section \ref{sec:nP} one can also choose non-Poissonian resetting wherein the distribution of the  waiting time $\tau$ to the next reset,  $\psi(\tau)$, is specified. The Poissonian case corresponds to $\psi(\tau) = r\exp(-r \tau)$. For this, one can of course study the standard first-passage probability. This first-passage probability to find a target can be thought of, in a more general context, as the distribution of the time to complete a task; 
let us call this distribution $\phi_0(\tau)$ (see figure \ref{Fig:completion}). The diffusive first-passage problem then  is an example of
a distribution $\phi_0(\tau)$ which decays asymptotically as $\phi_0(\tau) \sim \tau^{-3/2}$ i.e. it has infinite mean and variance.
As we have seen, in this case resetting dramatically improves the mean time to completion.

More broadly one can consider  a general  completion time distribution $\phi_0(\tau)$ for the 
stochastic process, with for example  finite mean, and ask whether resetting will improve the
completion rate of the task \cite{Reuveni16}.

The  first renewal equation of  Section~\ref{sec:survnp} is exactly  applicable in this case and one obtains the result (\ref{Qtflt2})
which now reads
\begin{equation}
\tilde Q_r(s) = 
\frac{\int_0^\infty \D t\,\e^{-st}\Psi(t)  Q_0( t) }{1-
\int_0^\infty \D t\,\e^{-st}\psi(t)  Q_0(  t) }\;.
\label{Qrgen}
\end{equation} 
where the survival probabilities,  $Q_*(t)$ with $*=0,r$, are now the   probabilities that the task is not completed in time $t$
($Q_r(t)$  is the survival probability with resetting while $Q_0(t)$ is the survival probability without resetting).

If we define the completion time distribution of the reset process as $\phi_r(\tau)$,  we then have
\begin{equation}
\phi_*(\tau) = - \frac{\partial Q_*(\tau)}{\partial \tau}\;,
\end{equation}
so that the Laplace transforms are related through
\begin{equation}
\tilde \phi_*(s) = 1- s\tilde Q_*(s) \;,
\label{phiQeq}
\end{equation}
again with $* = 0,r$.
Substituting (\ref{phiQeq})  in (\ref{Qrgen})  and using  integration by parts one finds
\begin{equation}
s\int_0^\infty \D t\,\e^{-st}\Psi(t)  Q_0( t) = 1- 
\int_0^\infty \D t\,\e^{-st}\left[\psi(t)  Q_0( t) + \Psi(t)  \phi_0( t)\right],
\end{equation} 
which yields
\begin{equation}
\tilde \phi_r(s) = 
\frac{\int_0^\infty \D t\,\e^{-st} \Psi(t) \phi_0( t) }{1-
\int_0^\infty \D t\,\e^{-st}\psi(t)  Q_0(  t) }\;.
\label{phirgen}
\end{equation} 
Equations (\ref{Qrgen}) and (\ref{phirgen}) are the general results for resetting time distribution $\psi(\tau)$ and completion time distribution $\phi_0(\tau)$,  for the underlying stochastic process. As noted they are derived in \cite{PKE16} for the diffusive process and generalised to an arbitrary stochastic process in \cite{CS18}.

\subsection{Poissonian reset and general completion time distribution}

In the case of Poissonian reset the formulas (\ref{Qrgen}) and  (\ref{phirgen}) 
simplify to
\begin{equation}
\tilde Q_r(s) = 
\frac{\tilde  Q_0( r+s) }{1-
r \tilde Q_0( r+s) }\;,
\label{QrPgen}
\end{equation} 
and
\begin{equation}
\tilde \phi_r(s) = 
\frac{(r+s) \tilde \phi_0(r+s)}{s + r \tilde \phi_0(r+s)}\;,
\label{phirPgen}
\end{equation} 
where, as usual, $\tilde f(s)$ is the Laplace transform with Laplace variable $s$ of the function $f(t)$ in the time domain. 
Formula (\ref{phirPgen}) was derived by Reuveni \cite{Reuveni16} using an alternative recursion relation for the completion time under reset
$T_r$ :
\begin{equation}
T_R = \begin{cases}\\[-2.2ex]\displaystyle
T & \text{for}\quad T<R \\[2mm]
\displaystyle
R + T_R'& \text{for}\quad T \geq R
\end{cases}
\label{rvrecur}
\end{equation}
where $T_R$ and $T_R'$ are two  completion times drawn from $\phi_r(t)$, $R$ is a reset time drawn from $\psi(t)$
and $T$ is a completion time drawn from $\phi_0(t)$. This recursion involving the i.i.d.random variables $T_R$, $T_R'$ 
contains essentially the same information as the first renewal equation for the probability distribution of $T_R$.

From (\ref{phirPgen}) all moments of the completion time $T_R$ can easily be computed, in particular
\begin{eqnarray}
\langle T_r \rangle &=& \frac{1}{r}\frac{1- \tilde \phi_0(r)}{\tilde \phi_0(r)}\\
\langle T^2_r \rangle &=& \frac{2}{r^2}\frac{r \frac{\rm d \tilde \phi_0(r)}{\rm d r} - \tilde \phi_0(r) +1}{\tilde \phi_0(r)^2} \;.
\end{eqnarray}
Then the optimal resetting rate is given by extremising $\langle T_r \rangle$ which yields
\begin{equation} 
\frac{{\rm d} \tilde \phi_0(r)}{{\rm d} r} = \frac{\tilde \phi_0(r)(1- \tilde \phi_0(r))}{r}
\label{ropt}
\end{equation}
at $r= r^*$.

Reuveni made the interesting observation that at $r=r^*$ the coefficient of variation (defined as standard deviation over mean)  is unity,
\begin{equation}
\frac{\sigma(T_{r^*})}{\langle T_{r^*} \rangle} = 1,
\label{CoV}
\end{equation}
implying universality as the result  does not depend on $\phi_0(\tau)$. 
Furthermore, (\ref{CoV}) trivially implies that
\begin{equation}
\langle T_{r^*} \rangle  = \frac{1}{2} \frac{\langle T_{r^*}^2\rangle}{\langle T_{r^*} \rangle} 
\end{equation}
which may be interpreted as follows. At the optimal resetting rate the mean time to completion (from the initial condition)
is equal to the {\em mean residual life time} \cite{gallagher} which is the mean time to completion of the process without resetting
from a randomly chosen time during an (incomplete) run.

\subsection{General resetting time distribution and general completion time distribution}
Pal and Reuveni \cite{PR17} considered general resetting time and  general completion time distributions
and sought to answer the question posed in \cite{PKE16,BBR16} of whether a sharp restart distribution $\psi(\tau)
= \delta(\tau-\tau^*)$, with $\tau^*$ suitably chosen
is always optimal, i.e. minimises 
$\langle T_r \rangle$. They showed through a probabilistic argument that this is indeed the case and moreover
at the optimal resetting the coefficient of variation obeys
\begin{equation}
\frac{\sigma(T_r)}{\langle T_r \rangle} \leq  1\;.
\end{equation}
Here we  present a subsequent alternative  proof, given by Chechkin and Sokolov \cite{CS18}, that a sharp reset distribution is optimal, which follows easily from (\ref{phiQeq}).

The mean time to completion $\langle T_r \rangle$ is given, as usual, by setting $s=0$ in the expression of the Laplace transform of the survival probability (\ref{Qrgen})
\begin{equation}
\langle T_r \rangle  = 
\frac{\int_0^\infty \D t\,\Psi(t)  Q_0( t) }{1-
\int_0^\infty \D t\, \psi(t)  Q_0(  t) }\;.
\label{Tgen}
\end{equation} 
Defining $F_0$ as the probability of completion before time $t$ 
\begin{equation}
F_0(t) =1 - Q_0(t)\;,
\end{equation}
the denominator of  (\ref{Tgen}) becomes 
$\int_0^\infty F_0(t) \psi(t)$  (assuming $\int_0^\infty {\rm d} t\, \psi(t) =1$).
The numerator of (\ref{Tgen}) may be rewritten using the definition (\ref{Psidef}) as
\begin{equation}
\int_0^\infty \D t\,\Psi(t)  Q_0( t)  = \int_0^\infty \D t' \psi(t') \int_0^{t'} \D t\, Q_0(t)
= \int_0^\infty \D t' \psi(t') \left[ t' - \int_0^{t'}  \D t F_0(t) \right]\;.
\end{equation}
Thus, after a trivial relabelling of integration variables,  (\ref{Tgen}) becomes
\begin{equation}
\langle T_r \rangle  = 
\frac{\int_0^\infty \D t\,\psi(t) F_0(t)  H( t) }{
\int_0^\infty \D t\, \psi(t)  F_0(  t) }\;,
\label{Tgen2}
\end{equation} 
where
\begin{equation}
H(t) = \frac{ t - \int_0^{t} \D t' F_0(t') }{F_0(t)}\;.
\label{Hdef}
\end{equation}
(We note that we have used the notation $H(t)$ rather than $G(t)$  of \cite{CS18}  to avoid a clash of notation
with Green function.) Since $\psi(t)  F_0( t) $ is a positive quantity we immediately get a lower bound from (\ref{Tgen2})
 \begin{equation}
\langle T_r \rangle  \geq 
\min_{0\leq t < \infty} H(t)\;.
\label{Tlb}
\end{equation} 
The lower bound is achieved by choosing $\psi(t) = \delta(t-t_r)$ where $t_r$ gives the global minimum of $H(t)$, as can be easily checked from (\ref{Tgen2}).
Thus the minimum mean time to completion is realised by a sharp resetting distribution.

The  optimal value of the  resetting time, $t_r$, can be obtained by extremising $H(t)$ (assuming the minimum is an extremum rather than a boundary value)
to obtain
\begin{equation}
\frac{1-F_0(t)}{F_0(t)} - \frac{ t - \int_0^{t} \D t' F_0(t') }{F_0^2(t)} F_0'(t) =0\;.
\end{equation}
Note that this equation can be written in a slightly more compact form in terms of $Q_0(t) = 1 -F_0(t)$ as \cite{VM18}
\begin{eqnarray} \label{optimal_Q0}
Q_0(t) - Q^2_0(t) + Q'_0(t) \, \int_0^t dt' Q_0(t') = 0 \;. 
\end{eqnarray}
If this equation has no solution the minimum will be at $t_r \to \infty$, i.e. in the limit of  no resetting. In addition, if $Q_0(t) = e^{-\alpha t}$ with $\alpha >0$, one finds that this equation (\ref{optimal_Q0}) is automatically satisfied for all time $t$, which means that there is no optimal $t_r$ in this case.

\subsection{Michaelis-Menten reaction scheme (MMRS)}
An interesting context in which to frame  general questions of restarting a complex stochastic process  has been proposed by  Reuveni, Urbakh and Klafter
and entails a (generalised) Michaelis-Menten Reaction Scheme (MMRS) \cite{RUK14,RRU15}.
These authors  envisage a molecular interaction  involving an enzyme molecule which, when bound to a substrate, triggers
the start of a complex  process leading to the production of a product. This process has a bare (in the absence of resetting) completion time distribution $\phi_0(\tau)$.
The  enzyme unbinds  and binds  reversibly, thus stopping and restarting the process, and the unbinding rate corresponds to the resetting rate. Also note that in this scenario
the unbound enzyme is in a quiescent state where the stochastic process has been switched off.
Schematically the reaction scheme is
$$
E + S \rightleftharpoons ES \rightarrow E+P\;.
$$
Most generally there are {\em three} waiting times with associated  distributions here:   the duration of the quiescent period when the enzyme is unbound, the duration of  the period during which the enzyme is bound  and the production process is active, and the time to completion of the production process (given that the enzyme is bound for sufficiently long time).
Reuveni et al initially considered an exponential distribution for the unbinding times which corresponds to Poissonian resetting.
In this case they obtained using a recursion similar to (\ref{rvrecur}), namely \cite{RUK14}
\begin{eqnarray}
\langle T_r \rangle &=& \frac{\langle \tau \rangle}{\tilde \phi_0(r)}+ \frac{1}{r}\frac{1- \tilde \phi_0(r)}{\tilde \phi_0(r)}
\label{Trrefac}
\end{eqnarray}
where $\langle \tau \rangle$ is the mean duration of  the quiescent period when the enzyme is unbound, which we shall refer
to as the refractory period.  Note that (\ref{Trrefac}) is Equation (4) of \cite{RUK14} transcribed into our current notation.
The equation illustrates that the refractory period gives an additive constant to  the mean time to completion.
The coefficient of variation of the completion time at the optimal restart  rate is now \cite{Reuveni16}
\begin{equation}
\frac{\sigma(T_{r^*})}{\langle T_{r^*} \rangle} = \sqrt{1 + \frac{ \langle \tau^2 \rangle -2 \langle \tau \rangle^2}{\tilde \phi_0(r) \langle T_{r^*}\rangle^2}},
\label{CoV2}
\end{equation}
which is now no longer universal as it depends on $\tilde \phi_0(r)$.

However, one can  usefully define a critical  value of the coefficient of variation
\begin{equation}
C_v^* = \sqrt{1 + \frac{2 \langle \tau \rangle}{\langle T_r \rangle}}
\end{equation}
which allows one to categorise the behaviour of
$\langle T_{r} \rangle$ at small $r$, i.e. how the introduction of resetting changes the  mean completion time:
If $C_v < C_v^* $  there is an inhibitory effect, i.e. the  mean time to completion increases;
if $C_v >  C_v^* $  there is an excitatory effect, i.e. the  mean time to completion decreases (linearly with $r$);
if $C_v \to \infty$  there is a superexcitatory effect i.e. the  mean time to completion decreases nonanalytically with $r$;
if $\langle T_0 \rangle  \to \infty$  there is a  restorative effect i.e. an infinite mean time to completion is rendered finite.
The example of diffusion with resetting falls into  the restorative category, which simply means that $\langle T_r \rangle$ decreases as $r$ increases for small $r$. However, there can be situations where the opposite happens, i.e., $\langle T_r \rangle$ increases with $r$ for small $r$. In fact, by changing system parameters, such as reaction rates, it is possible to induce a transition between the two scenarios -- such ``restart transitions'' in a generic setting have been discussed in several recent papers~\cite{RRU15,PP19b,ANBND19,RMR19}.

It was also shown in \cite{RRU15} how the equation for the optimal resetting rate (\ref{ropt}) generalises in the case of a refractory period with mean $\langle \tau \rangle$ to
\begin{equation} 
 \frac{\tilde \phi_0(r)(1- \tilde \phi_0(r))}{r^2  \tilde \phi'_0(r)} -\frac{1}{r} =\langle \tau \rangle
\label{ropt2}
\end{equation}
at $r= r^*$.

\subsection{Effects of refractory period}
The effects of a refractory period have been further studied in \cite{HK17} and \cite{EM19} where a spatial stochastic process with
propagator $G_0(x,t)$ was considered in the presence of resetting with a refractory period. (Note that in that work
the convention is taken that the refractory period occurs after a reset, so that the initial condition is slightly different
from \cite{RUK14,RRU15}.)   A first renewal equation was used to derive the   probability distribution in the absence of an absorbing target and the Laplace transform of the survival probability in the presence of an absorbing target.

For the case of Poissonian resetting with rate $r$ to the origin ($X_r=0$)
the nonequilibrium stationary state has the interesting feature of a delta peak at the  resetting position, due to the refractory period \cite{EM19}
\begin{equation}
p^*( x)     = \frac{r}{1 + r \langle \tau \rangle} \left[  \tilde G_0(x,r) + \langle \tau \rangle \delta(x)\right]\;.
\end{equation}
The relative weight of the peak is  equal to the ratio of the mean refractory period to the mean resetting period.
If the mean refractory period diverges then $p^*(x) \to \delta(x)$. The emergence of the peak has been analysed and it was shown
how  slow relaxation can emerge when the refractory period distribution $W(\tau)$ has a power law tail.

In addition, the case of a {\em correlated} resetting time and refractory period was considered,
 a simple example being Poissonian resetting with rate $r$ but now with a correlated refractory period
\begin{equation}
H(t,\tau) = r{\rm e}^{-rt} W(\tau|t)
\end{equation}
where $W(\tau|t)$ is the refractory period distribution conditioned on a preceding resetting time $t$. 
Finally, the joint active time and first passage time distribution was calculated.

An extension of the idea of a refractory period following a reset is to have a  different dynamics, which commences on resetting and returns the
process to the origin over some finite time. Such  two-phase, reset-return  processes have been studied in \cite{PKS19a,PKS19b,BS19}.

\section{Extended systems with resetting}
\setcounter{equation}{0}

Even though we have so far considered resetting of a single particle stochastic process (with the exception
of Section~\ref{sec:multi} which is still a system of non-interacting particles with random initial conditions). Resetting dynamics
can be easily generalised to any extended system with interacting degrees of freedom as we discuss in this section. 

\subsection{Resetting dynamics for general extended systems}

Consider any extended system, e.g. a fluctuating interface where heights of the interface at different space points are the relevant degrees of freedom that 
fluctuate in time according to some prescribed stochastic dynamics. Similarly, one can consider for instance an Ising model 
where the spins are the degrees of freedom that evolve, under say the Glauber dynamics at some temperature $T$. One can also think
of a polymer chain consisting of $N$ monomers (the degrees of freedom) that evolve via say the Rouse dynamics. Let $p_0({\cal C},t|{\cal C}_{\rm in})$ denote
the probability that the system is in a given configuration ${\cal C}$ at time $t$, starting from the initial configuration ${\cal C}_{\rm in}$. For example, in the case of the Ising model, a configuration ${\cal C}$
corresponds to a spin configuration. For a fluctuating interface, ${\cal C}$ is specified by a height profile $h(x,t)$ in $1+1$ dimensions. 
All these systems have their own microscopic dynamics by which the configuration ${\cal C}$ evolves in time,
but at this point, we do not need to specify  the dynamics.

Now imagine that we introduce the
resetting process  whereby the configuration ${\cal C}$ at time $t$ is reset to a specific reset {\em configuration} ${\cal C}_r$ with a constant rate $r$. This is a generalisation
of the single particle case discussed earlier where the configuration ${\cal C}$ is specified by the position of the particle $x$ and $X_r$ denotes the resetting position. Here we will discuss only the Poissonian resetting for simplicity, though it can be easily generalised to non-Poissonian resetting as well. 

More precisely, in time $\D t$, the configuration ${\cal C}$ is reset to ${\cal C}_r$ with probability $r \D t$ and, with the complementary probability $1 - r \D t$, the system continues to evolve by its own dynamics. Let $p_r({\cal C},t|{\cal C}_{\rm in})$ denote the probability that the system is in configuration ${\cal C}$ in the presence of
resetting with rate $r$, starting from the initial configuration ${\cal C}_{\rm in}$. Then, as in the single particle case (see
Section \ref{sec:renewal} ), one can express $p_{r}({\cal C},t|{\cal C}_{\rm in})$ in terms of $p_0({\cal C},t|{\cal C}_{\rm in})$ using a renewal approach, which takes into account the event of last
resetting before time $t$. This reads
\bea\label{p_ren_ext}
p_r({\cal C},t|{\cal C}_{\rm in}) = \int_0^t r\, \e^{-r\tau} p_0({\cal C},\tau|{\cal C}_{r})\, \D \tau + \e^{-r t} \, p_0({\cal C},t|{\cal C}_{\rm in}) \;.
\eea
The second term represents the case where there is no resetting in the interval $[0,t]$, which happens with probability $\e^{-r\,t}$ -- in this case the system
evolves by its own dynamics (without reset) from time $0$ till time $t$, explaining the occurrence of $p_0({\cal C},t|{\cal C}_{\rm in})$ in the second term. The first term can also be explained easily. Let the last resetting event before time $t$ occur at time $\tau_l = t-\tau$. Looking backwards in time from the instant $t$, this means that there is no resetting in the interval $[0,\tau]$ followed by a resetting event between $\tau$ and $\tau + \D \tau$ -- the probability for this event is $r\, \e^{-r \tau}\, \D \tau$. During this time interval $\tau$ followed by the last resetting, the system evolves freely (without resetting) from configuration ${\cal C}_r$ to ${\cal C}$ by its own dynamics, which happens with probability $p_0({\cal C},\tau|{\cal C}_{r})$.

Even though the system's own dynamics may not lead to a stationary state, the resetting drives the systems into a non-equilibrium stationary state (as in the single particle case). The corresponding stationary state is obtained by taking the $t \to \infty$ limit in (\ref{p_ren_ext}), leading to
\bea \label{stat_ext}
p_{r}^{\rm stat}({\cal C}) = \int_0^\infty r\, \e^{-r\tau} p_0({\cal C},\tau|{\cal C}_{r})\, \D \tau  \;.
\eea
Note that even though the stationary state is independent of the initial configuration ${\cal C}_{\rm in}$, it does depend on the resetting
configuration ${\cal C}_r$.

Various models of extended systems subject to resetting have been studied recently. This includes
fluctuating interfaces \cite{GMS14,GN16}, reaction diffusion systems \cite{DHP14},  exclusion processes \cite{BKP19}, Ising model \cite{MMS20} etc. In the following, we discuss in detail  a specific example of an extended system under resetting, namely a fluctuating $1+1$-dimensional interface.

\subsection{Fluctuating interfaces: non-equilibrium steady states}

We consider a $1+1$ dimensional interface
characterized by a height field  
$H(x,t)$ at position $x$ and time $t$. Starting from an initially flat profile: $H(x,0) = 0
~\forall~ x$, the heights evolve according to the Kardar-Parisi-Zhang (KPZ) equation~\cite{KPZ86, HHZ95,Krug97}:
\bea
\frac{\partial H}{\partial t}=\nu \frac{\partial^2 H}{\partial
x^2}+\frac{\lambda}{2}\Big(\frac{\partial H}{\partial x}\Big)^2+\eta(x,t)
\;,
\label{kpz-eom}
\eea
where $\nu$ is the diffusivity, $\lambda$ accounts for the
nonlinear term, while $\eta(x,t)$ is a Gaussian noise of zero mean and correlations $\langle
\eta(x,t)\eta(x',t')\rangle=2D\delta(x-x')\delta(t-t')$. In the case where $\lambda=0$, the non-linear term
disappears and the height field $H(x,t)$ becomes Gaussian -- in this case the KPZ
equation (\ref{kpz-eom}) reduces to the Edwards-Wilkinson (EW) equation \cite{EW82}.

For an interface of length $L$ evolving according to (\ref{kpz-eom}),
the spatially averaged height $\overline{H(x, t)}=\int_0^L {\rm d}x ~H(x,
t)/L$ grows with time with velocity $v_\infty=(\lambda/2)\int_0^L {\rm
d}x ~\langle (\partial H/\partial x)^2 \rangle$. Let us define the height fluctuation as
\bea\label{h_fluct}
h(x,t) = H(x,t)-\overline{H(x, t)} \;.
\eea
For a given sample of the interface, we define the (empirical) variance of the height fluctuation
as
\bea\label{def_sigma}
\sigma^2(L,t) = \frac{1}{L} \int_0^L h^2(x,t) \, \D x \;.
\eea
Note that $\sigma^2(L,t)$ is still a random variable, fluctuating from sample to sample. The interface width $W\equiv W(L,t)$ is then defined as 
\bea \label{def_w}
W(L,t) = \sqrt{\langle \sigma^2(L,t) \rangle} \;.
\eea
where the average $\langle \cdot \rangle$ is an ensemble average over different realisations  of the noise
$\eta(x,t)$.
In the thermodynamic limit $L \to \infty$, we expect that $\sigma^2(L,t)$ approaches its expectation value. As time grows beyond a non-universal microscopic time scale $T_{\rm micro}\sim O(1)$, the width initially grows as a power law $W(L,t) \sim t^\beta$ as long as $T_{\rm micro} \ll t \ll T^* \sim L^z$, where $\beta$ and $z$ are known as the growth and the dynamical exponents respectively. For times $t \gg T^*$, the width saturates to an $L$-dependent value $\sim L^{\alpha}$. In other words
\bea \label{width_scaling}
W(L,t) \sim
\begin{cases}
&t^\beta \;, \; T_{\rm micro} \ll t \ll T^* \sim L^z\\
&L^\alpha \;, \; t \gg T^* \;.
\end{cases}
\eea
The former is called the ``growing'' regime while the latter is called the ``stationary'' regime. The width in these two regimes is connected
via the Family-Vicsek scaling form \cite{FV85}: $W(L,t) \sim L^\alpha {\cal W}(t/T^*)$ where the scaling function ${\cal W}(s)$  behaves as a constant as $s\to \infty$,
and as $s^\beta$ as $s \to 0$ where $\beta = \alpha/z$. For the one-dimensional KPZ equation (\ref{kpz-eom}) with $\lambda \neq 0$, the
dynamical exponent is $z=3/2$ while the width exponent $\alpha = 1/2$, and hence the growth exponent $\beta = 1/3$. In contrast, for $\lambda = 0$ (i.e. the
EW equation), the exponents are $z=2$, $\alpha = 1/2$ and $\beta = 1/4$. Indeed, generically at long times $t \gg T^*$, the full probability distribution of the height fluctuations  (\ref{h_fluct})
reaches a stationary state in a {\em finite} system at long times $t \gg T^*$. In fact, both for the KPZ and the EW cases in $(1+1)$ dimensions, 
the stationary height distribution $p_0^{\rm stat}(h,L)$ turns out to be a simple Gaussian~\cite{BS95}. If, however, we take the $L \to \infty$
limit first, such that $T^*$ diverges, the system never reaches a stationary state and it is always
in a ``growing'' regime where the height fluctuations typically grow as a power law in time and where the distribution of the height fluctuations is time-dependent.  

In the following, we will restrict ourselves to this growing regime where $t \ll T^* \sim L^z$ and
switch on the resetting that interrupts the growth and restarts the
system from its initial flat configuration. This resetting move drives the system to a non-equilibrium
stationary state (NESS) as discussed in  (\ref{stat_ext}) in the general context. Below we characterise precisely
the height distribution in this reset driven NESS. We characterise a configuration by ${\cal C} = \{h(x,t) \}_{0\leq x \leq L}$   
that specifies the height fluctuation at each space point. Furthermore, we integrate out the heights at all points except one, say at the
origin at $x=0$ and denote by $p_r(h,t)$ as the height distribution at $x=0$ at time $t$, in the presence of the resetting at constant rate $r$.
Focusing thus on this marginal distribution $p_r(h,t)$ at $x=0$, the general (\ref{stat_ext}) then reads
\bea \label{p_h_gen}
p_r(h,t) = \int_0^t r\, \e^{-r\tau} p_0(h,\tau) \, \D \tau + \e^{-rt} p_0(h,t) \;,
\eea 
where $p_0(h,\tau)$ is the height distribution at time $\tau$ in the growing regime, starting from a flat configuration in the absence of resetting. This equation is valid for arbitrary time $t$. Taking the $t \to \infty$ limit, the stationary height distribution at $x=0$ is given by
\bea \label{stat_h}
p^{\rm stat}_{r}(h) = \int_0^\infty \D \tau r \,\e^{-r\tau} p_0(h,\tau) \;.
\eea

We start with the simpler EW case where $\lambda = 0$ in (\ref{kpz-eom}). The resulting linear equation can be trivially solved using Fourier transform and it gives a Gaussian distribution for $p_0(h,\tau)$
\bea \label{p0_EW}
p_0(h,\tau) = \frac{1}{\sqrt{2 \pi W^2}}  \e^{-\frac{h^2}{2 W^2}} \;, \;  
\eea
where $W \equiv W(L\to \infty,\tau)$ is the time-dependent width of the interface in the thermodynamic limit $L \to \infty$ and is given by
\bea \label{W_EW}
W(\tau) = D \sqrt{\frac{2 \tau}{\nu\,\pi}} \;.
\eea 
Plugging this result (\ref{p0_EW}) in (\ref{stat_h}), the reset induced stationary height distribution can be expressed in the scaling form
\bea
p^{\rm stat}_{\rm r}(h) \sim \sqrt{\gamma} \, r^{1/4} \,G^{\rm
EW}(h\sqrt{\gamma}r^{1/4}) \;,
\label{ph-scaling}
\eea
where $\gamma=\sqrt{\pi \nu}/(D2^{3/2})$ and $G^{\rm EW}(x)$ is given by
\bea
G^{\rm EW}(x)=\frac{1}{\sqrt{\pi}}\int_0^\infty \frac{{\rm
d}y}{y^{1/4}}\exp\Big(-y-\frac{x^2}{\sqrt{y}}\Big) \;,
\label{GEW-1}
\eea
which is symmetric in $x$, $G^{\rm EW}(-x) = G^{\rm EW}(x)$, yielding
zero mean, and variance $\int_{-\infty}^{+\infty} x^2 G^{\rm EW}(x) dx = \sqrt{\pi/4}$. From the scaling form in (\ref{ph-scaling}),
one obtains the scaling of the stationary width with $r$ as $W^{\rm EW}_{\rm r} \sim r^{-1/4}$. One can show that $G^{\rm
EW}(x)$ behaves asymptotically as \cite{GMS14}
\begin{eqnarray}\label{asympt_GEW}
G^{\rm EW}(x) \sim 
\begin{cases}
&\frac{1}{\sqrt{\pi}}\Big[\Gamma \left(\frac{3}{4}\right)-x^2 \Gamma
\left(\frac{1}{4}\right)+\frac{8}{3} \sqrt{\pi } |x|^3\Big] \;,\; x \to 0 \,, \\
& \\
& c|x|\exp[-3/2^{2/3} \, |x|^{4/3}] \;, \; x \to \pm \infty \;,
\end{cases}
\end{eqnarray}
where $\Gamma(x)$ is the Gamma function and $c$ is a computable
constant. Interestingly, due to the
$|x|^3$ term in (\ref{asympt_GEW}), $G^{\rm EW}(x)$ is non-analytic close to $x=0$. In the limit $x \to \pm \infty$, the stretched
exponential behaviour (\ref{asympt_GEW}) is significantly different from a Gaussian tail. These analytical results have been also verified
numerically in \cite{GMS14}.

We now turn to the KPZ case. Here, it is known that for times $T_{\rm
micro} \ll t \ll T^*$, and for a flat initial profile, the interface height $H(x,t)$ has a deterministic
linear growth with stochastic $t^{1/3}$ fluctuations \cite{SS10a,SS10b,CDR10,CD11,Dotsenko10,ACQ11,TW94,TW96,BBF08}:
\bea
H(x,t)=v_\infty t+(\Gamma t)^{1/3}\chi(x) \;.
\label{H-form}
\eea
Here, $\Gamma\equiv\Gamma(\nu,\lambda,D)$ is a constant, while $\chi$ is a time-independent random variable distributed according to the celebrated
Tracy-Widom distribution corresponding to Gaussian Orthogonal Ensemble (GOE), $f_1(\chi) = F_1'(\chi)$, which can be written explicitly in terms of the Hastings-McLeod solution of the Painlev\'e II equation~\cite{TW96}. 
In particular, $f_{1}(\chi)$ has
asymmetric non-Gaussian tails \cite{TW96,BBF08}: 
\bea \label{TW}
f_1(\chi) \sim
\begin{cases}
& \exp(-|\chi|^3 /24) \;, \; \chi \to - \infty \\
& \exp(-2\chi^{3/2} /3) \;, \; \chi \to +\infty\;.
\end{cases}
\eea
Equation (\ref{H-form}) gives \bea
h=(\Gamma
t)^{1/3}\Big[\chi-(1/L)\int_0^L {\rm d}x~ \chi(x)\Big]\;.
\eea
Knowing that $f_1(\chi)$ has a finite mean $\langle \chi \rangle < 0$, it follows
from the law of large numbers that in the limit $L \to \infty$, the
second term on the r.h.s. converges to $\langle \chi \rangle$,
so that $\langle h \rangle=0$. In this case, in the limit $\tau \to \infty$, $h \to \infty$, keeping $h/\tau^{1/3}$ fixed, $p_0(h,\tau)$ takes the scaling form
\bea
p_0(h,\tau) \sim
\frac{1}{(\Gamma \tau)^{1/3}}\widehat{f}_1\Big(\frac{h}{(\Gamma \tau)^{1/3}}\Big) \;,
\label{kpz-ht-scaling}
\eea
where $\widehat{f}_1(x)\equiv f_1(x+\langle \chi \rangle)$. Note that this scaling form is valid only in the large $\tau$ limit. In contrast, in (\ref{stat_h}), the integral is over all $\tau$. Therefore, unfortunately, we can not replace $p_0(h,\tau)$ by its scaling form (\ref{kpz-ht-scaling}) which is only valid in the 
large $\tau$ limit. However, this is possible in the $r \to 0$ limit. To see this, we make a change of variable $\tau' = r \tau$ in (\ref{stat_h}) and get  
\bea \label{stat_h_2}
p^{\rm stat}_{r}(h) = \int_0^\infty \D \tau'  \,\e^{-\tau'} p_0(h,\tau'/r) \;.
\eea
One now sees that, in the limit $r \to 0$, the effective time $\tau'/r$ inside $p_0(h,\tau'/r)$ becomes large and, hence, we can replace it by its scaling form (\ref{kpz-ht-scaling}). Hence, for $r \to 0$, $h \to
\infty$, with $h \, r^{1/3}$ fixed, we get
\bea
p_r^{\rm stat}(h) \sim (r\Gamma^{-1}) ^{1/3}G^{\rm
KPZ}\left[(r\Gamma^{-1})^{1/3}h\right] \;,
\label{ph-scaling-kpz}
\eea
where the scaling function $G^{\rm KPZ}(x)$ is given by
\bea
G^{\rm KPZ}(x)=\int_0^\infty {\rm d}y~\frac{\e^{-y}}{y^{1/3}}\widehat{f}_1\left(\frac{x}{y^{1/3}}\right) \;.
\label{ph-analytic-kpz}
\eea
In contrast to $G^{\rm EW}(x)$, $G^{\rm KPZ}(x)$ is not
symmetric in $x$. Since $\hat{f}_1$ has zero mean, it follows that 
$G^{\rm KPZ}$ has also vanishing mean, but is still asymmetric, 
with a variance $\int_{-\infty}^{+\infty} x^2 G^{\rm KPZ}(x) \D x \approx 1.44$.  
From (\ref{ph-scaling-kpz}), the stationary
width scales as $W^{\rm
KPZ}_{\rm r} \sim r^{-1/3}$. Its asymptotic
behaviours for $x \to \pm \infty$, obtained from the corresponding
behaviours of $\hat f_1(x)$ combined with a saddle point analysis, are
\begin{eqnarray}\label{asympt_KPZ}
G^{\rm KPZ}(x) \approx
\begin{cases}
& \exp(-|x|^{3/2}/\sqrt{6}) \;, \; x \to -\infty \\
& \\
& \exp(-3^{1/3}x) \;, \; x \to +\infty \;.
\end{cases}
\end{eqnarray}
Equation (\ref{ph-analytic-kpz}) implies that $G^{\rm KPZ}(x)$ has a
non-analytic behaviour as $x \to 0$: $G^{\rm KPZ}(x) \sim A + B x +
C x^2 \ln x$, with $A,B,C$ being constants. Non-analyticity at the
resetting value was also
observed for the EW interfaces, (\ref{asympt_GEW}), 
and, hence, is a
generic feature of stochastic resetting. A quick comparison between Eqs. (\ref{asympt_GEW}) and (\ref{asympt_KPZ}) shows
that the resetting induced steady state height distribution is rather different in the two cases. This is in contrast to the stationary 
state in a finite system of size $L$ without resetting where both have the identical Gaussian distribution. Thus resetting is able to 
distinguish between the two cases.  

Non-Poissonian resetting (see section \ref{sec:nP})  of the interface  with a power-law waiting time distribution between resets
$\psi(\tau)  \sim \tau^{-(1+\alpha)}$ with $\alpha  > 0$ has been considered in \cite{GN16}. For
$\alpha >1$ the height attains a stationary distribution, which exhibits heavy tails, whereas for $\alpha <1$ the distribution remains time dependent for large times. 

\subsection{Fluctuating interfaces: relaxation to the stationary state}

To study the relaxation to the reset induced non-equilibrium stationary state in an extended system such as
fluctuating interfaces, our starting point is the finite $t$ renewal equation (\ref{p_h_gen}) for the height distribution
at fixed point in space, say at $x=0$. We now want to analyse this equation for finite but large time $t$. For this purpose, we
first note from Eqs. (\ref{p0_EW}) and (\ref{kpz-ht-scaling}) that there is a scaling regime for large $t$, large $h$, keeping the ratio
$h/t^\beta$ fixed such that the single-site height distribution in the absence of resetting $p_0(h,t)$ 
exhibits a scaling form, both for the EW and the KPZ equation,
\bea \label{scaling_p0}
p_0(h,t) \approx \frac{1}{(\Gamma \,t)^\beta} \, g\left(\frac{h}{(\Gamma t)^\beta} \right) \;.
\eea  
In the above equation, $\Gamma$ is a microscopic constant, the growth exponent $\beta = 1/4$ (for EW) and $\beta=1/3$ (for KPZ) and the scaling
function $g(x)$ is also different in the two cases. For EW, $g(x)$ is a simple Gaussian while, for the KPZ, $g(x)$ is the shifted Tracy-Widom GOE as discussed below (\ref{kpz-ht-scaling}). Unlike the EW case where $g(x) \sim \e^{-x^2}$ for large $x$ both on the positive and the negative side, for the KPZ the scaling function $g(x)$ has asymmetric tails [see (\ref{TW})] with $g(x)\sim \e^{- |x|^3/24}$ for $x \to -\infty$ while $g(x) \sim \e^{-(2/3)x^{3/2}}$ for $x \to +\infty$. Hence to investigate the approach to the NESS, we consider the generic
case when $g(x)\sim \exp(-a_\pm |x|^{\gamma_\pm})$ as
$x\to \pm \infty$. For example, for the KPZ with flat initial condition,
$\gamma_+=3/2$, $a_+=2/3$ and $\gamma_{-}=3$, $a_{-}=1/{24}$.

In order to know $p_r(h,t)$ from (\ref{p_h_gen}) we need to know $p_0(h,\tau)$ for all $\tau \in [0,t]$. However, except for the EW interface, where
$p_0(h,\tau)$ is an exact Gaussian at all times $\tau$,  we  typically have information on $p_0(h,\tau)$ only in the scaling limit when $\tau$ and $h$ both are large, while the ratio $h/\tau^\beta$ is held fixed, as discussed in (\ref{scaling_p0}). Thus, to use this scaling form in the equation for $p_r(h,t)$ in (\ref{p_h_gen}) we focus in the regime where $h$ is large (i.e., in the scaling regime). We substitute this scaling form (\ref{scaling_p0}) in (\ref{p_h_gen}) and rescale, as before, the time $\tau = w t$. This gives 
\begin{align}
p_r(h,t)&\approx (\Gamma t)^{-\beta} \e^{-r t} g \bigl((\Gamma
t)^{-\beta}h\bigr) \notag\\ +& rt (\Gamma t)^{-\beta} \int_0^1 \D w\,
w^{-\beta} \e^{-rt w} g \bigl((\Gamma t)^{-\beta} h w^{-\beta}\bigr).
\label{master-interface}
\end{align}
This solution has been analysed in detail in Ref. \cite{MSS15a}. The main result is the following. We consider a scaling regime where $h$ and $t$ large but the ratio $h/t^\nu_{\pm}$ is fixed (where $\pm$ refers to positive or negative $h$) and the exponent 
\bea \label{nu_pm}
\nu_{\pm} = \frac{\gamma_{\pm}}{1 + \beta \gamma_{\pm}} \;,
\eea
where we recall that $\beta$ is the growth exponent and $\gamma_{\pm}$ specify the behaviour at the tails of the scaling function $g(x)$ on the positive and negative side respectively (see the previous paragraph). In this scaling regime, it has been shown that $p_r(h,t)$ admits a large deviation form 
\begin{subequations}
\begin{equation}
p_r(h,t) \sim \e^{-t\, I(h\, t^{-1/\nu_\pm})},
\label{prob_height} 
\end{equation}
\end{subequations}
where the rate function is given by
\begin{subequations}
\begin{equation}
I(y) = \begin{cases}\displaystyle
\frac{r\,|y|^{\nu_\pm}}{\beta\nu_\pm (y_\pm^*)^{\nu_\pm}}\,
 & \text{for}~~|y| < y_\pm^*, \\[3mm] 
\displaystyle
r + b_\pm |y|^{\gamma_\pm} & \text{for}~~ |y| > y_\pm^* \;,
\end{cases}
\label{I-interface}
\end{equation}
\end{subequations}
\begin{figure}
\centering
\includegraphics[width = 0.9\linewidth]{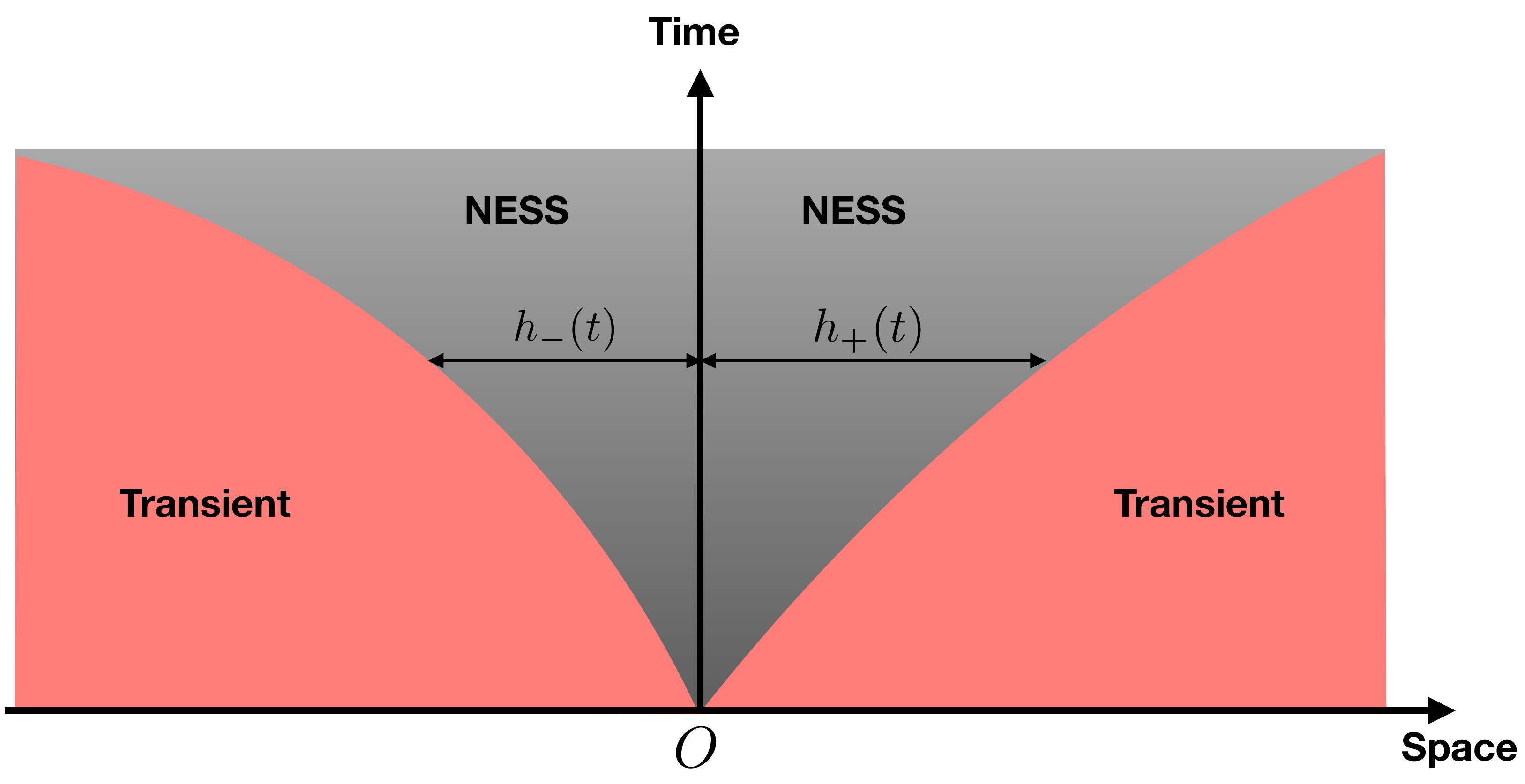}
\caption{A NESS gets established in a core region around the resetting center $O$ (corresponding here to a flat profile $h=0$) whose right and left frontiers $h_\pm(t)$ grow with time with (a priori) different exponents $h_{\pm}(t) \sim t^{1/\nu_{\pm}}$. Outside the core region, the system is transient.}\label{Fig_relax_height}
\end{figure}
where the singular points of the rate function $y^*_{\pm}$ on both sides have been computed explicitly \cite{MSS15a}. The second derivative
of $I(y)$ is discontinuous at both $y^*_{\pm}$, indicating a second order dynamical phase transition. Essentially, in the height space, there are 
two growing length scales $h_{\pm}(t) \sim t^{1/\nu_{\pm}}$ growing in the opposite direction. For $h_-(t) < h < h_+(t)$, the distribution $p_r(h,t)$ becomes independent of time and reaches a NESS, while for $h$ outside this range, the height distribution still depends on $t$ and is transient (see figure \ref{Fig_relax_height}).

\section{Resetting with memory of history}
\setcounter{equation}{0}

So far we have mainly considered resetting to a fixed reset point which may be chosen to coincide with the initial condition. We saw in Section~\ref{sec:resetdist} that this may be easily generalised to a resetting distribution from which the
reset point is sampled at each reset event.

In this section we review works where the process is reset to its value at some selected time in its history.
This is most naturally illustrated in the case of a discrete time random walk in which with some reset probability the 
walker is returned to  its previous position at a randomly selected time from the past. These dynamics
are examples of a more general class of models referred to as reinforced random walks \cite{Davis90,FGP09}.

\subsection{Preferential visit model}

Boyer and Solis-Salas \cite{BSS14} considered a minimal model for animal mobility proposed in the ecological
literature \cite{GM05,GM06}, which they called the {\em Preferential Visit Model} (PVM). The idea is that an animal can either explore territory locally (by
random walk dynamics) or relocate to places visited in the past (via a stochastic resetting move). For simplicity, we start by presenting the model on a lattice, the generalization to the continuum space is discussed later. At each discrete time step, $t \rightarrow t+1$, the walker moves with probability $1-r$ to a randomly chosen nearest neighbour site on the lattice and, with probability $r$, the walker relocates to a site it has visited in the past. In the simplest case
this relocation is implemented by selecting a previously visited site with a probability proportional to the number of past visits to that site. One of the key observations that leads to the solvability of some aspects of this model is the fact that this relocation protocol is exactly equivalent to choosing a past time at random \cite{BSS14}. Note that this model is different from the so-called ``elephant random walk'' \cite{ST04} which is also a non-Markovian process but there the stochastic rules are different from the PVM discussed above. In the ``elephant random walk'', one again chooses a past time uniformly at random but one actually resets the increment of the jump rather than the position of the walker.

For simplicity we consider a one-dimensional lattice and let $P(n,t)$ denote the probability that the walker is at site $n$ at time $t$. One
can write down the master equation for the evolution of $P(n,t)$ and it reads
\begin{equation}\label{pvm_1}
P(n,t+1) = \frac{1-r}{2} P(n-1,t) + \frac{1-r}{2}P(n+1,t) +\frac{r}{t+1} \sum_{n'=-\infty}^\infty \sum_{t'=0}^t P(n',t; n,t') \;,
\end{equation}
where $P(n',t; n,t')$ denotes the joint probability that the walker is at site $n'$ at time $t$ and at site $n$ at time $t'\leq t$. The first two terms on the r.h.s. of (\ref{pvm}) represent the standard random walk dynamics. The last term can be explained as follows. Suppose that the particle is at $n'$ at time $t$ and makes a transition to site $n$ at time $t+1$ via the relocation move. For this transition to occur, the walker must have been at site $n$ at some previous time $t'$. The probability for this event of being at $n'$ at time $t$ and at $n$ at time $t'$ is simply the joint probability $P(n',t; n,t')$. The prefactor $r/(t+1)$ in the third term in (\ref{pvm_1}) is just the probability of the relocation via this event. Finally, the transition can occur from any site $n'$ at time $t$ to site $n$ at time $t+1$ -- hence one has to sum over all possible $n'$. In addition, one has to sum over all possible $t'$. This explains the third term in (\ref{pvm_1}). Fortunately, when one sums the two-point probability distribution over all $n'$, one gets back a one-point distribution
\begin{eqnarray} \label{identity}
\sum_{n' = -\infty}^\infty P(n',t;n,t') = P(n,t') \;.
\end{eqnarray} 
Consequently, (\ref{pvm_1}) becomes a closed equation for $P(n,t)$  \cite{BSS14, BRC14}
\begin{equation}
P(n,t+1) = \frac{1-r}{2} P(n-1,t) + \frac{1-r}{2}P(n+1,t) + \frac{r}{t+1} \sum_{t'=0}^t P(n,t')\;.
\label{pvm}
\end{equation}
Even though this equation is linear, it is nonlocal in time and hence the solution is nontrivial as we will see below.

As a first step, consider the mean squared displacement 
\begin{equation}
M_2(t) = \sum_{n=-\infty}^{\infty} n^2P(n,t) \;.
\end{equation}
It obeys an equation obtained by summing (\ref{pvm})
\begin{equation}
M_2(t+1) = (1-r) + (1-r)M_2(t) + \frac{r}{t+1} \sum_{t'=0}^t M_2(t').
\label{M2PVM}
\end{equation}
The solution to this equation, with initial condition $M_2(0) =0$
is given by
\begin{equation}
M_2(t) = \frac{1-r}{r} \sum_{k=1}^t \frac{ 1-(1-r)^k}{k}
\end{equation}
as may be checked by substitution into (\ref{M2PVM}).
For large  $t$
\begin{equation} \label{M2_larget}
M_2(t) \simeq \frac{1-r}{r}\left[ \ln( rt) + \gamma_e\right]
\end{equation}
where $\gamma_e$ is Euler's constant [see (\ref{gammaE})] and we have used the large $t$ asymptotics
\begin{equation}
\sum_{k=1}^t \frac{1}{k} \simeq \ln t + \gamma_e
\end{equation}
and
\begin{equation}
\sum_{k=1}^t \frac{(1-r)^k}{k} \simeq -\ln r\;.
\end{equation}
Thus the width of the distribution grows as $(\ln t)^{1/2}$. Similarly, equations for higher moments of the
displacement may be written down and eventually solved for large $t$. Indeed, it can be shown that the full distribution converges
at late times to a Gaussian form \cite{BSS14} 
\begin{eqnarray}\label{pvm_gaussien}
P(n,t) \simeq \frac{1}{\sqrt{2 \pi M_2(t)}} \, \e^{-\frac{n^2}{2 M_2(t)}} \;,
\end{eqnarray}
where the variance $M_2(t)$ grows extremely slowly, i.e., logarithmically at late times, as given in (\ref{M2_larget}). Thus the PVM provides a simple mechanism for anomalously slow sub-diffusive growth. Similar slow subdiffusion is known to arise in diffusion in disordered medium, such as the Sinai model. There the slowdown occurs due to the disorder that blocks the particle motion. In contrast, in the PVM, the slow subdiffusion arises even in the absence of disorder, simply by the dynamics of memory-driven resetting.

In \cite{BRC14} the PVM model was generalised to include biased sampling of the past history during relocations.
Specifically at a relocation event at time $t$ the past time $t'$ is selected with a probability
\begin{equation}
F(t-t') =B (t-t'+1)^{-\beta}
\end{equation}
where $\beta \geq 0$ and $B$ is a normalisation constant. The case $\beta=0$ recovers the previous PVM model.
The master equation (\ref{pvm}) is modified to
\begin{equation}
P(n,t+1) = \frac{1-r}{2} P(n-1,t) + \frac{1-r}{2}P(n+1,t) + r \sum_{t'=0}^t F(t-t') P(n,t')\;.
\label{pvm2}
\end{equation}
In this case the large time behaviour of the mean squared displacement depends on the value of $\beta$ \cite{BRC14}
\begin{eqnarray}
\beta >2\quad & M_2(t) \simeq& \left( \frac{1-r}{1+r \langle \tau \rangle}\right) t \;,\\
1< \beta <2\quad  &M_2(t)  \propto & t^{\beta -1} \;,\\
\beta <1\quad  &M_2(t)  \propto & \ln t \;.
\end{eqnarray}
Moreover, in the large time limit, the probability distribution of the position takes the scaling form
\begin{equation}
P(n,t) \simeq \frac{1}{\sqrt{M_2(t)}}g\left(\frac{n}{\sqrt{M_2(t)}}\right)
\end{equation}
where the scaling function $g(y)$ was found to be Gaussian in the cases $\beta >2$ and $\beta <1$ but to have a nontrivial form
for $1<\beta <2$.

Thus for $\beta >2$ (fast decaying memory) the probability distribution is that of the simple random walk
but with a modified diffusion constant. For $\beta <1$ (slowly decaying memory) one obtains a narrow Gaussian distribution with width $\sim (\ln t)^{1/2}$ and for  $1<\beta <2$ one obtains an anomalous width $\sim t^{(\beta-1)/2}$
and nontrivial scaling distribution \cite{BRC14,BP16}.

\subsection{Interpolation between the preferential visit model and resetting to the initial condition}
A continuous time and space version of  resetting with memory was considered in \cite{BEM17}.
The master equation now reads
\begin{equation}
\frac{\partial p(x,t)}{\partial t}
= D\frac{\partial^2 p( x,t)}{\partial x^2} - r p( x,t) + r\int \D \tau  K(\tau,t) p(x,\tau)\;.
\end{equation}
The third term represents the
gain of probability into $x$ by choosing a time $\tau$ in the past with probability density $K(\tau,t)$ and relocating to $x$ with probability density $p(x,\tau)$. The memory kernel $K(\tau,t)$ is normalised so that
\begin{equation}
\int_{0}^t \D \tau K(\tau,t)  =1\;.
\end{equation}
In \cite{BEM17} a memory kernel was chosen that allows one to recover the case of resetting to the initial condition
and to interpolate to the PVM:
\begin{equation}
K(\tau,t) = \frac{ \phi(\tau)}{\int_0^t \D \tau \phi(\tau)}\;.
\label{Kphi}
\end{equation}
Thus $K(\tau, t)$ only depends on the present time $t$ through the normalisation denominator in (\ref{Kphi}). By choosing $\phi(\tau)  = \delta(\tau)$ one recovers resetting to the initial condition, which has a stationary state,
and by choosing $\phi =1$ one recovers a continuous time and space version of PVM, which does not have a stationary state but exhibits a time-dependent distribution whose width grows as $(\ln t)^{1/2}$.
For the case $\phi =1$ the exact time-dependent distribution can be obtained  and for the exponential kernel  $\phi(\tau) = \lambda \e^{-\lambda \tau}$ with $\lambda >0$ the exact stationary state can be obtained. 

Interestingly, it was shown in \cite{BEM17} (see also \cite{MUB19} for some rigorous results) that  a whole range of  behaviour  is made possible by choosing different $\phi(\tau)$. These behaviours are classified as follows:

\begin{itemize}
\item  When $\phi(\tau)$ decays faster than $1/\tau$ for large $\tau$, i.e. $\tau\phi(\tau) \to 0$ as $\tau \to\infty$ there is a stationary distribution $p^* (x)$.

\item When $\phi(\tau)$ increases, or decays as, or more slowly than, $1/\tau$ for large $\tau$, i.e., 
$\tau\phi(\tau) > 0$ as $\tau \to\infty$ there is no stationary distribution.
Instead there is a late time behaviour in which the time-dependent distribution takes a Gaussian form with variance $\sigma^2(t$).
The time dependence of the variance exhibits various distinct behaviours depending on $\phi(\tau)$:
\begin{enumerate}
\item  for $\phi(\tau) \sim 1/\tau$, $\sigma^2(t) \sim \ln\ln t$;
 \item for  $ \phi \sim \tau^\alpha$ with $\alpha > -1$, $\sigma^2(t) \sim \ln t$;
\item  for  $\phi(\tau) \sim \exp(a\tau^\beta)$ where $0 < \beta < 1$ and $a$ is a positive constant, $\sigma^2(t) \sim t^\beta$;
\item for $\phi(\tau) \sim \exp(a\tau )$ where $a$ is a positive constant, $\sigma^2(t) \simeq 􏰅 \left(\frac{2a}{a+r}\right) 􏰆Dt$
\item  for $\phi(\tau) \sim \exp(a\tau^\beta )$  where $\beta  > 1$ and $a$ is a positive constant, $\sigma^2(t) =2Dt$.
\end{enumerate}
\end{itemize}
Thus in addition to the logarithmic growth of the variance, which we have seen in the previous subsection, an ultraslow $\ln \ln t$ growth of the variance occurs for $\phi (\tau) \sim 1/\tau$.
Such a double logarithmic growth with time has been reported in data on human mobility \cite{SKWB10}. 

\subsection{Localization-delocalization transition induced by preferential resetting}

Recently, the PVM described above was studied in the presence of a single defect site where the random walk has a finite 
probability to stay~\cite{FCBGM17,BFCGM19}. We recall that, in the absence of the defect site, the walker is always delocalized, i.e, $p(n,t)$ always depends on time $t$
and the variance increases like $\ln t$ at late times. Remarkably, the presence of one single defect site is able to induce a transition from a delocalised to localised phase (where $p(n,t)$ becomes independent of time at late times). More precisely, the model is defined as follows. Again, we consider a single random walker on a $d$-dimensional lattice. At any generic site other than the origin, the walker performs the same dynamics as in the PVM, namely with probability $1-r$ (with $0\leq r \leq 1$), it diffuses to any neighbouring site chosen randomly and, with probability $r$, it relocates to any previously visited site by choosing a past time at random (which is equivalent to choosing a previously visited site with a probability proportional to the number of past visits to that site). The origin is a special site where, with probability $\gamma \in [0,1]$ the walker stays there and with the complementary probability $1-\gamma$, it either diffuses (with probability $(1-\gamma)(1-r)$) or relocates preferentially (with probability $(1-\gamma)r$). The two parameters in this model are thus $\gamma$ and $r$. For $\gamma=0$, it reduces to the PVM described earlier, where the walker is always delocalized. 

\begin{figure}[ht]
  \begin{center}
   \includegraphics[width=0.65\textwidth]{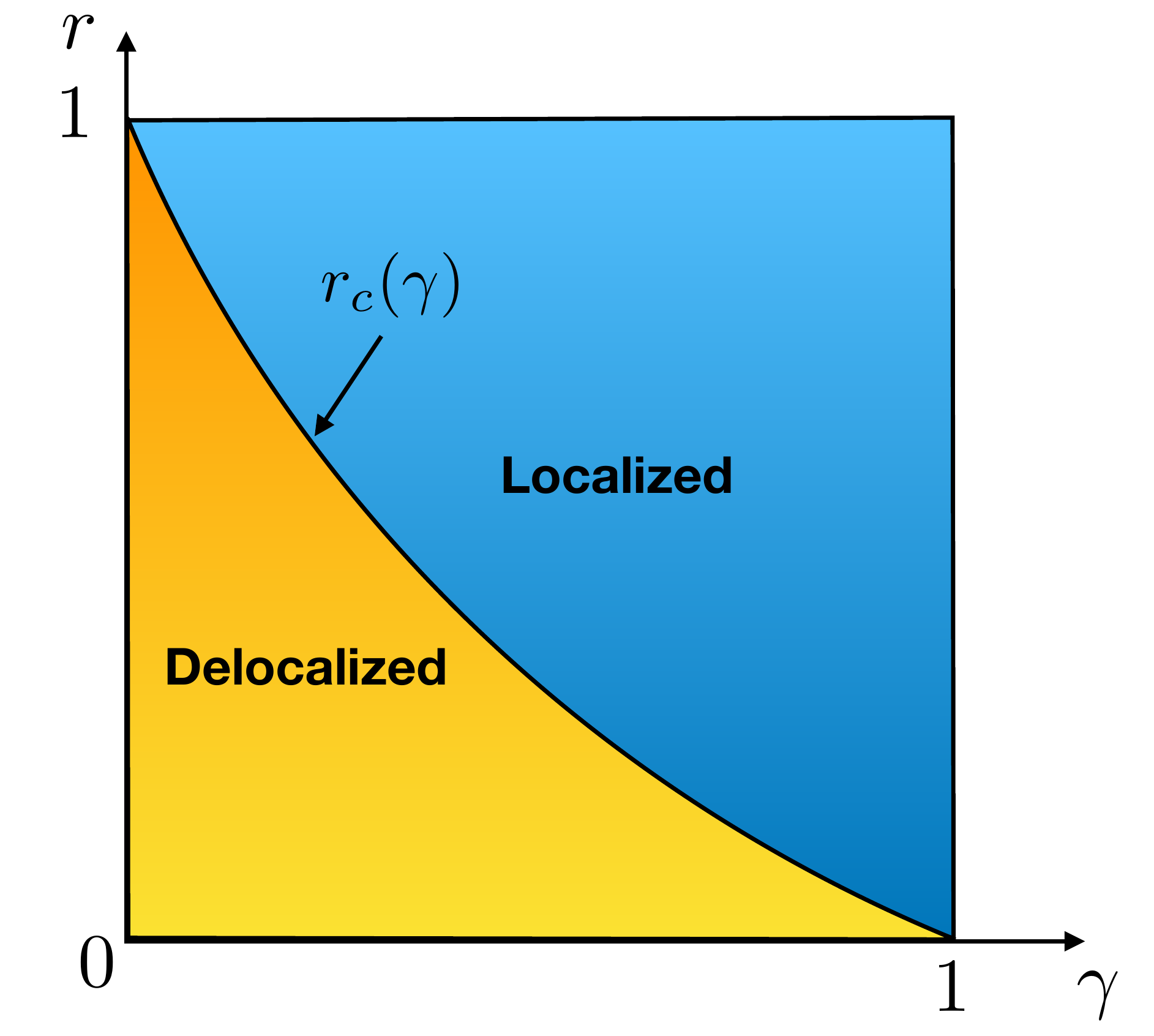}
  \end{center}
  \caption{Phase diagram for the PVM model (for $d > \mu$) with  relocation  probability  $r$ to a previous site,
in the presence of a defect site with a remain probability $\gamma$. 
}
    \label{fig:local}
  \end{figure}
The analysis of this model with a finite $\gamma$ exhibits an interesting phase transition for $d>2$ in the $(\gamma,r)$ plane, across a critical line $r_c(\gamma)$ (see figure \ref{fig:local}). For $r < r_c(\gamma)$, the walker is delocalized and the variance increases with time. In contrast, for $r > r_c(\gamma)$, the walker gets localised, i.e. $p(n,t)$ becomes stationary at late times and the stationary distribution has an exponential tail with a characteristic localization length scale $\xi(r)$ that diverges as one approaches the critical line as $\xi \simeq (r-r_c)^{-\nu}$ where $\nu$ takes the same value as in the self-consistent theory of Anderson localization of waves in random media \cite{FCBGM17,BFCGM19}. The critical value $r_c(\gamma)$ was shown to be related to the probability $P_{\rm no-return}$ of no-return to the origin for the free random walker, i.e, without resetting, via the simple relation
\begin{equation}\label{rc}
r_c(\gamma) = \frac{(1-\gamma) P_{\rm no-return}}{\gamma + (1-\gamma) P_{\rm no-return}} \;.
\end{equation}
It turns out that this relation is quite general and holds for random walks with arbitrary jump length on the lattice, e.g., for L\'evy
flights. For a general random walk with a jump distribution $p(\ell)$, with its Fourier transform given by $\tilde p(k)$, the probability of no-return has a simple general expression
\begin{equation}\label{Pnoret}
P_{\rm no-return} = \left(\frac{1}{(2 \pi)^d} \int_{{\cal B}_d} \frac{d \vec{k}}{1- \hat p(\vec{k})} \right)^{-1} \;,
\end{equation}
where ${\cal B}_d$ is the $d$-dimensional first Brillouin zone. For example, for nearest-neighbour random walk on a $d$-dimensional hyper-cubic lattice, 
\begin{equation}\label{ptildek}
\tilde p(k) =\frac{1}{d} \sum_{i=1}^d \cos k_i \;.
\end{equation}
In this case, $P_{\rm no-return} > 0$ only for $d>2$, i.e. the walk is recurrent for $d \leq 2$ and is transient for $d>2$. For L\'evy flights with L\'evy index $\mu$, such that $p(\ell) \propto \ell^{-1-\mu}$ for large $\ell$ with $0<\mu<2$, $\tilde p(k)$  
behaves as $\tilde p(k) \simeq 1 - |a\,k|^\mu$ as $k \to 0$, where $a$ is a characteristic jump length. In this case, from (\ref{Pnoret}), one finds that $P_{\rm no-return} = 0$ for $d\leq \mu$, i.e., the walk is recurrent. In contrast, for $d>\mu$, $P_{\rm no-return} > 0$ and the L\'evy walk is transient. Therefore, this delocalisation-localisation transition in this resetting model with defect will occur, even for L\'evy flights, as long as $d > \mu$.  For example, if $\mu = 1/2$, this transition will be there even in $d=1$, as was seen in numerical simulations \cite{FCBGM17,BFCGM19}. 
Note that, for a recurrent walk with $d<\mu$, $P_{\rm no-return} = 0$ and hence $r_c(\gamma) = 0$: This indicates that for any finite $r$ the walker is always in the localised phase. Finally, we mention that several variants of this simple model with one defect site were recently studied in Ref. \cite{BFCGM19}.  

To conclude this subsection, we point out that, in the absence of resetting, i.e. $r=0$, for any finite $\gamma$, the walker is always
delocalized (see figure \ref{fig:local}). This is consistent with the well known fact that a single defect is not enough to localise a diffusing particle in dimension
$d>\mu$. However, introducing a finite resetting rate $r>r_c(\gamma)$ can localise the walker in $d>\mu$.

\subsection{Resetting to the past maximum}
\label{sec:pmax}
In this section we consider another example of resetting protocol using the memory of the full history of the process. In this case we consider a random walk for which the resetting move
returns the walker to the previous maximum \cite{MSS15b}.

To be specific  we consider a one-dimensional random walk on the lattice.
At any given time step $n$ if the position $x(n)$ of the walker
is less than the maximum position 
\begin{equation}
m(n) = \mbox{max} \left[x(0) =0, x(1), x(2), \ldots , x(n) \right]
\end{equation}
then in the next time step the position is reset to $m(n)$ with probability $r$
and moves to  either the right or left nearest neighbour site with equal probability
$(1-r)/2$.
If the position is $x(n) = m(n)$  then the walker
moves to the right or left nearest neighbour site with equal probability
$1/2$.

It is useful to consider the distance $y$ of the walker at time $n$ from the maximum
position upto time $n$: $y(n) = m(n) - x(n)$. The master equation of the joint probability $P(y,m,n)$
for the maximum to take value $m$ and the distance from the maximum to take value $y$  obeys, for $y>1$~\cite{MSS15b}
\begin{equation}
P(y,m,n) = \left[ \frac{1-r}{2} + \frac{r}{2}\delta_{y,1}\right] P(y-1,m,n-1)
+ \frac{1-r}{2} P(y+1,m,n-1)
\end{equation}
and for $y=0$
\begin{equation}
P(0,m,n) = \frac{1}{2}  P(0,m-1,n-1)
+ \frac{1-r}{2} P(1,m,n-1) + r \sum_{y=1}^\infty P(y,m,n-1)
\end{equation}
with initial condition $P(y,m,0) = \delta_{y,0}\delta_{m,0}$.

These equations may be solved by generating function techniques\cite{MSS15b}.
The main results of interest for our purposes concern how resetting to the maximum
affects the statistics of the maximum. It turns out that for non zero $r$ the average 
value of the maximum for large $n$ increases ballistically in time as
\begin{equation}
\langle m(n) \rangle \simeq v(r) n
\end{equation}
where the speed $v(r)$ is given by
\begin{equation}
v(r) = \frac{r(1-r)}{r-2 r^2 + \sqrt{r(2-r)}}\;.
\end{equation}
Note that the speed vanishes as $v(r) \simeq \sqrt{r/2}$ signalling a crossover from
ballistic behaviour when $r>0$ to diffusive behaviour $m(n) \sim n^{1/2}$ for $r=0$.

Moreover, the distribution of the distance  from the maximum reaches a stationary state for $y$
held fixed as $n\to \infty$
\begin{equation}
P_y(y,n) \sim \left( \frac{1-r}{1+ \sqrt{r(2-r)}} \right)^y\;.
\end{equation}
Now if one looks at the relaxation to this stationary state one finds a dynamical
transition reminiscent of the relaxation front described in Section~\ref{sec:relss}.
Indeed, for large $n$ the probability  distribution of $y$ obeys a large deviation principle 
\begin{equation}
P_y(y=wn,n) \sim \exp(-nH(w))
\end{equation}
where the rate function takes the form
\begin{equation}
H(w) =
\begin{cases}
& w\ln \left[ \frac{1+ \sqrt{r(2-r)}}{1-r}\right] \quad\mbox{for} \quad w<w^*\\[2ex]
& \frac{w}{2}\ln \left[ \frac{1+w}{1-w}\right] + \ln \left[ \frac{\sqrt{1-w^2}}{1-r}\right]
\quad\mbox{for} \quad w>w^*
\end{cases}
\end{equation}
with $w^* =\sqrt{r(2-r)}$. Thus for $y<y^*= w^*n$ the distribution of $y$ has reached the stationary state but for $y>y^*$ the distribution still depends on time $n$.

\section{Other topics associated to resetting}\label{sec:other}
\setcounter{equation}{0}

In this section we briefly review some exciting new developments that extend the resetting paradigm we have described in previous sections.

\subsection{Thermodynamics of resetting and integral theorems}
The concept of resetting  raises several issues in stochastic thermodynamics, which have been identified and addressed by Fuchs, Goldt and Seifert \cite{FGS16}.
First, resetting to some fixed configuration or finite region of phase space implies  a change in information content 
since information about the preceding state is lost, which in turn
implies a thermodynamic cost. Second, an important question in  biological systems
is the efficiency of computation \cite{BZTL02,MHL12,MHL14}. For a biomolecular search process involving resetting, this  boils down to tensioning the informational efficiency
of the search  against the thermodynamic cost of  resetting.
Finally the  nonequilibrium stationary states generated by resetting  exhibit currents which imply entropy production
and it is of  importance to characterise this. 

In \cite{FGS16}  resetting of colloidal particle in a potential was  considered and used to obtain
a first law of thermodynamics and to  identify  the  thermodynamic work done by resetting.
The resetting entropy production rate was derived for this system with space-dependent resetting rate $r(x)$ to resetting position $X_r$
\begin{equation}
\dot S^{\rm reset} = \int \D x\, r(x) p(x) \ln \frac{p(x)}{p_{X_r}}
\end{equation}
and from this a  second law of thermodynamics including resetting was proposed (see also \cite{BGM19}).

Building on the identification of entropy change due to resetting, Pal and Rahav \cite{PRahav17}  considered how integral fluctuation theorems apply to resetting problems. The integral theorems may be thought of as  generalising the second law
to equalities involving averages over fluctuations \cite{Seifert12}.
They showed how the 
Hatano-Sasa integral fluctuation theorem \cite{HS01}  which pertains to nonequilibrium steady states is also valid
for systems with resetting. Further integral theorems have been considered in \cite{PRahav17,GPP19}.

In \cite{MT17}  the authors considered a probe in contact with a bath held out of equilibrium by a resetting process.
The bath entails particles in a harmonic  potential which are reset to a fixed position with a Poissonian rate.
The probe is coupled to the bath-particles and  for large bath is  governed by an effective Langevin equation which
generates non-Gaussian fluctuations.

\subsection{Large deviations}
As we have already seen in Sections~\ref{sec:relss} and \ref{sec:pmax}, many probabilities associated with stochastic processes obey a large deviation principle.
In particular, additive observables, which are the time integral of a fluctuating quantity  such as time integrated current  or  the area $A_T$ under the space time trajectory,
\begin{equation}
A_T = \frac{1}{T} \int_0^T X_t \D t\;,
\end{equation}
are expected to have a probability distribution which can be written for large
$T$  as
\begin{equation}
P(A_T =a) = \e^{-T I(a) + o(T)}\;.
\end{equation}
This is the large deviation form (often referred to as large deviation principle) and the function $I(a)$ is the rate function or large deviation function.
The rate function is usually determined by considering the generating function
\begin{equation}\label{def_Gk}
{\cal G}(k,t)= \langle  \e^{tk A_t}\rangle
\end{equation}
and finding the asymptotic behaviour for  large $t$
\begin{equation}
 {\cal G}(k,t) \sim \e^{\lambda(k)t}\;.
\label{calGt}
\end{equation}
Then $I(a)$ can be found by the Legendre transform
\begin{equation}
I(a) = \sup_k \left[ ka - \lambda(k)\right],
\end{equation}
see \cite{Touchette} as well as \cite{MS17} for reviews.

Using the renewal approach  Meylahn, Sabhapandit and Touchette \cite{MST15}  showed how the generating function, ${\cal G}_r$, i.e. the equivalent of (\ref{def_Gk}) for an additive observable in a stochastic process subject to Poissonian resetting,
can be written simply in terms of the generating function in the absence of resetting,  ${\cal G}_0$, through the relation in Laplace domain
\begin{equation}
\tilde{\cal G}_r(k,s) =  \frac{\tilde {\cal G}_0(k,s+r)}{1 - r\tilde {\cal G}_0(k,s+r) }  \;.
\label{calGs}
\end{equation}
Then to extract  the large time behaviour (\ref{calGt}) just requires the  knowledge of the poles of the r.h.s. of (\ref{calGs}).
Calculations were carried out explicitly for  an Ornstein Uhlenbeck process under reset.
It was also pointed out that  stochastic observables that do not  ordinarily obey a large deviation principle may acquire one under
resetting. 

Further work \cite{HMMT19} has shown how a variational formula  involving
large deviation rate functions without resetting may be used to obtain the rate function with resetting.
Three examples of additive observables for diffusion with resetting, positive occupation time, area and absolute area, were worked out. 

In \cite{HT17} observables $J_n$ (generalised currents) that
are not reset to zero but retain their value on resetting were considered under discrete time dynamics. It was shown how
phase transitions in the large deviation function may occur  between regimes where the current fluctuations are optimally realised
by a finite frequency of resets  to a  regime where the current  fluctuation is optimally realised by not resetting. Very recently, another functional of Brownian motion with resetting has been studied -- this is the local time spent by the particle at at given position in
space \cite{PCRK19}.

\subsection{Coupling of stochastic process and resetting}

Most of this review has been based on independent processes for the underlying stochastic dynamics and resetting
which has allowed  simple renewal equations to be written. A natural extension  to explore is when
the stochastic dynamics becomes coupled to resetting, e.g. through the dynamics depending on the time to the last reset.
Preliminary examples which have been studied are Brownian particles which are attracted towards each other but reset on contact to an initial separation  \cite{FE17}, and
random walk dynamics where the mean or variance of the microscopic steps of the walk depend on the time since resetting
\cite{HT17}.

\subsection{Quantum dynamics with reset}

The idea of introducing a resetting to an arbitrary classical stochastic process, such as diffusion,  
can be easily generalized to the dynamics of a quantum system. Consider, for simplicity, a generic quantum 
system with a time independent Hamiltonian $H$, prepared in an initial pure state $|\psi_0\rangle$. In the absence
of resetting, this state evolves under the unitary dynamics
\begin{eqnarray}\label{unitary}
|\psi(t)\rangle =\e^{-i H\,t} |\psi(0)\rangle \;.
\end{eqnarray}  
The density matrix
\begin{eqnarray} \label{density_mat}
\hat\rho(t) = |\psi(t) \rangle \langle \psi(t) | 
\end{eqnarray}
evolves via
\begin{eqnarray}\label{dyn_density}
\hat \rho(t) = \e^{i H t} \hat \rho(0) \e^{- i Ht} \quad \, \quad \hat \rho(0) =  |\psi(0) \rangle \langle \psi(0)| \;.
\end{eqnarray}

One can now introduce resetting via the following protocol \cite{MSM18}: the state $|\psi(t)\rangle$ evolves from
time $t$ to $t+\D t$ as follows
\begin{eqnarray}
|\psi(t+\D t)\rangle = 
\begin{cases}
&|\psi(0)\rangle \:, \; \quad \quad \quad \; \quad {\rm with \; proba.} \, r \, \D t \\
&[1-i H \D t] |\psi(t)\rangle \:, \; {\rm with \; proba.} \, 1-r \, \D t \;,
\end{cases}
\end{eqnarray}
where we have set $\hbar = 1$ for convenience. Here $r \geq 0$ denotes the resetting rate with which the
system is projected back to the initial state. Thus, in a small time interval $\D t$, the system
either goes back to its initial state with probability $r\,\D t$, or,
with the complementary probability $(1-r\, \D t)$, it evolves
unitarily with its Hamiltonian $H$. Now the density matrix at time $t$ is denoted
by 
\begin{eqnarray} \label{density_mat_r}
\hat\rho_r(t) = |\psi(t) \rangle \langle \psi(t) | \;.
\end{eqnarray}
Note that for any $r>0$, the
dynamics is a mixture of stochastic and deterministic evolution and
the density matrix in (\ref{density_mat_r}) is stochastic in
the sense that it varies from one realisation of the reset process
to another. Hence, the observed density matrix at time $t$ is
obtained by averaging over all possible reset histories
\begin{equation}
\rho_r(t)= {\mathbb E}\left[{\hat \rho}_r(t)\right]
\label{dens_matrix.1}
\end{equation}
where ${\mathbb E}[\cdot]$ denotes the classical expectation value over all
stochastic evolutions. Our goal is to investigate how a nonzero
$r$ modifies the time evolution of the quantum state, or
equivalently the associated density matrix in (\ref{dens_matrix.1}). Following the same renewal structure that was described 
for the classical systems, one can then write the last renewal equation for the evolution
of the density matrix as
\begin{equation}
\rho_r(t) = \e^{-rt} \e^{-iHt} \hat \rho_0 \e^{iHt}+ \int_0^{t} r \, \e^{-r \tau}\, \e^{-i H\tau} \hat \rho_0 \e^{i H\tau}\, \D \tau  \label{denmat1} \;,
\end{equation}
where $\hat \rho_0 =  |\psi(0) \rangle \langle \psi(0)|$. Here the first term corresponds to no resetting and the second term counts the events following the last resetting before time $t$, that occurs at time $t-\tau$. Now, at long times $t$, the first term vanishes exponentially
and the density matrix $\rho(t)$ approaches a stationary value (as $t\to \infty)$
\begin{eqnarray}
\rho^*  &=&\int_0^{\infty} r \e^{-r \tau}\, \e^{-i H\tau} \hat \rho_0 \e^{i H\tau}\, \D \tau \;.
\label{denmat2}
\end{eqnarray}
where the subscript $^*$ denotes the stationary density matrix. This stationary density matrix was analysed in \cite{MSM18} and it was pointed out that it has non-zero off-diagonal elements in the eigenbasis of $H$. This is at variance with the pure unitary evolution, i.e. $r=0$, where the density matrix, in the eigenbasis of $H$, becomes diagonal. These general results were then applied to various quantum models \cite{MSM18}. For example, the evolution of non-interacting fermions on a one-dimensional lattice, starting from a step initial condition for the density of fermions, and subjected to resetting was shown to lead to a non-trivial steady state density profile. Other models include Dirac fermions and the Bose-Hubbard model in an optical lattice. We refer the readers to Ref. \cite{MSM18} for further details.  

There are of course other interesting questions associated to resetting in quantum systems. For example, 
the spectral properties of quantum systems subjected to resetting have been studied in \cite{RTLG18}. Another
quantum setup where resetting can play an important role is the quantum random walk subjected to perturbations
caused by repeated measurements \cite{DDDS15,DDD15,FKB17,TBK18}. Another situation akin to resetting 
 corresponds to making a stroboscopic series of projective measurements on the initial state.

\subsection{Further topics}

There have been some very recent developments that we have not been able to cover in this review, unfortunately.   
These include, e.g. the connection between home range search and resetting \cite{PKR19}, 
the  interplay between  population dynamics and resetting \cite{MVB18,dSF18}, branching processes and resetting 
 \cite{Eliazar18,PER19}, resetting of the scaled Brownian motion 
with a time-dependent diffusion coefficient, $D(t) \propto t^{\alpha-1}$ with $\alpha >0$
\cite{BCS19a,BCS19b}.

\section{Conclusion}
To summarise, in this review we have attempted to provide a survey
of recent developments in the theory of stochastic processes subjected
to random resettings. The basic idea is very simple and general: 
any stochastic process
evolving under its own natural dynamics is interrupted at random times
and brought back (reset) to a fixed state, say its initial state. 
The intervals between successive reset events are statistically independent
and are drawn from some specified distribution $\psi(t)$. A particularly
simple and illustrative case is the Poissonian resetting where
$\psi(t)=r\, \e^{-r\, t}$ with $r$ denoting the constant reset rate.
One can ask how this stochastically interrupted `reset process' evolves
with time and what are its statistical properties? 

There are two principal effects of resetting that we emphasized in this 
review. First, such `reset' interruptions drive the system to a 
non-trivial nonequilibrium stationary state with a nonzero current in 
the configuration space-- indeed detailed balance is manifestly violated by the 
reset moves. Thus resetting provides a very simple and natural way to 
generate a nonequilibrium stationary state. The second effect of 
resetting concerns the mean time to search or capture a target. We have 
shown that in several models, resetting not only reduces the mean search 
time, but there is typically an optimal resetting rate $r^*$ (for 
Poissonian resetting) at which the mean capture time becomes minimum. 
Thus resetting typically makes the search process efficient.

While the theory of stochastic resetting has seen rather rapid progress in recent years, there have been little progress on the experimental side so far. The recent preliminary results from the experiments in an optimal trap set-up from the group of Ciliberto seem promising~\cite{Bovon2019}. The point about experiments is not just to reproduce the theoretical results (which can be easily done by simulations), but often real experiments require new resetting protocols that have not been theoretically studied. For example, in a theoretical model one often assumes instantaneous resetting which is impossible to achieve experimentally. Thus expermentalists need to devise different types of resetting protocols, which in turn pose interesting theoretical challenges. We hope that this synergy between theory and experiments will advance the field of stochastic resetting even further in the coming years.

Finally, the idea of resetting is so simple and natural that it can be 
used and adapted to ask interesting questions in many different fields,
going beyond classical stochastic processes. For example, we have seen
how resetting in a quantum system leads to a nontrivial steady state 
density matrix with non-zero off-diagonal elements, giving rise to
new non-diagonal ensembles. The idea of resetting has led to interesting
new observations in stochastic thermodynamics, population dynamics, chemical
reactions, just to name a few. We hope that this review will stimulate
further new ideas in this rapidly developing field of research.

\ack
MRE thanks LPTMS, Universit\'e  Paris-Sud for the award of a Visiting Professorship during which this review was completed.
We thank  colleagues for collaborations and useful discussions. They include  O. B\'enichou, B. Besga, R. Blythe, D. Boyer, S. Ciliberto, X. Durang, R. Falcao, A. Falcon-Cortes,  L. Giuggioli, S. Gupta, M. Henkel, A. Kundu, L. Ku\'smierz , C. Maes, K. Mallick, M. Meylahn, D. Mukamel, B. Mukherjee, G. Oshanin, A. Pal,  S. Redner, S. Reuveni, S. Sabhapandit, K. Sengupta,  T. Thiery, H. Touchette, J. Whitehouse.

\appendix

\section{Equivalence of forward and backward renewal equations}
\label{app:renewal} 
In this appendix we show that the last renewal equation and the first renewal equation are equivalent
by showing that they share the same solution.

The last renewal equation (\ref{pt}) reads
\begin{equation}
p( x,t|x_0) = {\rm e}^{-rt}  G_0( x,t| x_0) + r \int_0^t \D \tau \,
{\rm e}^{-r\tau}  G_0( x,\tau| X_r)\;.
\label{ptapp}
\end{equation}
Taking the Laplace transform yields
\begin{equation}
\tilde p( x,s|x_0) =  \tilde G_0( x,r+s| x_0) + r \int_0^\infty \D t  \e^{-st}\int_0^t \D \tau \,
{\rm e}^{-r\tau}  G_0( x,\tau| X_r)\;.
\end{equation}
The final term on the r.h.s. becomes
\begin{eqnarray}
r \int_0^\infty \D \tau  \int_0^\infty \D t' \,
 \e^{-st' -(r+s)\tau}  G_0( x,\tau| X_r)
&=& \frac{r}{s} \tilde G_0( x,r+s| X_r)\;.\nonumber
\end{eqnarray}
Thus the Laplace transform of the solution to (\ref{pt}) is given by
\begin{equation}
\tilde p( x,s|x_0) =   \tilde G_0( x,r+s| x_0)  + \frac{r}{s}\tilde G_0( x,r+s| X_r) \;.
\label{LRapplt}
\end{equation}

Now consider the first renewal equation (\ref{FR})
\begin{eqnarray}
p(x,t|x_0)&=&\e^{-rt}G_0(x,t|x_0) \nonumber \\
&+&r\int_{0}^{t}d\tau_f~\e^{-r\tau_f}~p(x,t-\tau_f|X_r)~,
\end{eqnarray}
Taking the Laplace transform yields
\begin{equation}
\tilde p( x,s|x_0) =  \tilde G_0( x,r+s| x_0) + r \int_0^\infty \D t  \e^{-st}\int_0^t \D \tau \,
{\rm e}^{-r\tau}  p( x,t-\tau| X_r)\;.
\label{FRapplt}
\end{equation}
The final term on the r.h.s. becomes
\begin{eqnarray}
r \int_0^\infty \D \tau  \int_0^\infty \D t' \,
 \e^{-st' -(r+s)\tau}  p( x, t'| X_r)\nonumber 
&=& \frac{r}{r+s} \tilde p( x,s| X_r)
\end{eqnarray}
so that the Laplace transform of the solution to the first renewal equation obeys
\begin{equation}
\tilde p( x,s|x_0) =   \tilde G_0( x,r+s| x_0)  + \frac{r}{r+s}\tilde p( x,s| X_r) \;.
\label{FRapplt2}
\end{equation}
Assuming that the solutions of the first renewal and last renewal equations are the same
we may subtract (\ref{LRapplt}) from (\ref{FRapplt2}) to obtain
\begin{equation}
\frac{r}{s}\tilde G_0( x,r+s| X_r)  = \frac{r}{r+s} \tilde p(x,s|X_r)
\end{equation}
which is indeed consistent with the solution
(\ref{LRapplt}).

\end{document}